\documentclass[aps,prb,superscriptaddress,longbibliography,twocolumn,usenames,dvipsnames]{revtex4-2}
\usepackage[english]{babel}
\usepackage{amsmath,amssymb,bbm,graphicx,color,comment}%txfonts
\usepackage[bookmarks=true,colorlinks,citecolor=blue,urlcolor=blue]{hyperref}

\usepackage{dsfont}
\usepackage{soul}
\usepackage{subfigure}
\usepackage{extarrows}
\usepackage{float}
\usepackage{txfonts}
\usepackage{nicematrix}
\usepackage{mathtools}
\usepackage[]{xcolor} 
\usepackage{mathrsfs}

\usepackage[export]{adjustbox}
\usepackage{adjustbox}

\usepackage{dcolumn}   % Align table columns on decimal point
\usepackage{bm}        % bold math
\usepackage{amsfonts}  % Common math fonts
\usepackage{filecontents}
\usepackage{lineno}

\usepackage{pifont}

\DeclareMathOperator{\diag}{diag}

\renewcommand{\vec}[1]{\bm{#1}}
\AtBeginEnvironment{pmatrix}{\setlength{\arraycolsep}{4pt}}

\definecolor{mygreen}{rgb}{0,0.6,0}

\usepackage{times}

\newcommand{\angstr}{\mathring{\mathrm{A}}}

\begin{document}

\title{Altermagnetic splitting of magnons in hematite (\texorpdfstring{$\alpha-$Fe$_2$O$_3$)}{}}%

\author{Rhea Hoyer}
\affiliation{Department of Physics, Johannes Gutenberg University Mainz, 55128 Mainz, Germany}

\author{P. Peter Stavropoulos }
\affiliation{Institut für Theoretische Physik, Goethe-Universität Frankfurt, 60438 Frankfurt am Main, Germany}

\author{Aleksandar Razpopov}
\affiliation{Institut für Theoretische Physik, Goethe-Universität Frankfurt, 60438 Frankfurt am Main, Germany}

\author{Roser Valentí}
\affiliation{Institut für Theoretische Physik, Goethe-Universität Frankfurt, 60438 Frankfurt am Main, Germany}

\author{Libor Šmejkal}
 \affiliation{Max Planck Institute for the Physics of Complex Systems, N\"othnitzer Str. 38, 01187 Dresden, Germany}
\affiliation{Max Planck Institute for Chemical Physics of Solids, N\"othnitzer Str. 40, 01187 Dresden, Germany} 
\affiliation{Department of Physics, Johannes Gutenberg University Mainz, 55128 Mainz, Germany}
\affiliation{Institute of Physics, Czech Academy of Sciences, Cukrovarnická 10, 162 00 Praha 6, Czech Republic}

\author{Alexander Mook}
\affiliation{Department of Physics, Johannes Gutenberg University Mainz, 55128 Mainz, Germany}

\begin{abstract}
We develop a four-sublattice spin-wave theory for the $g$-wave altermagnet candidate hematite ($\alpha$-Fe$_2$O$_3$), considering both its easy-axis phase below and its weak ferromagnetic phase above the Morin temperature. A key question is whether the defining altermagnetic feature---magnon spin splitting (also called chirality or polarization splitting) due to nonrelativistic time-reversal symmetry breaking---remains intact when relativistic corrections, which contribute to hematite’s magnetic order, are included. Using a detailed symmetry analysis supported by density functional theory, we show that capturing the magnon splitting within a Heisenberg model requires exchange interactions extending at least to the 13\textsuperscript{th} neighbor. We find an altermagnetic band splitting of approximately $2\,$meV, which contrasts with the total band width of about $100\,$meV. The splitting scales as $k^4$ in the long-wavelength limit ($k$ is the crystal momentum) and exhibits complex direction dependence. While three of the four expected altermagnetic nodal surfaces align with crystallographic mirror planes, the fourth deviates from planarity, leading to twelve instead of six nodes in the $k_z = 0$ plane. To evaluate the experimental observability of this splitting, we analyze relativistic corrections to the magnon spectrum in both magnetic phases. We show that spin-orbit coupling---manifesting as magnetocrystalline anisotropies and the Dzyaloshinskii-Moriya interaction (DMI)---does not obscure the key altermagnetic features. In the easy-axis phase, DMI introduces small spectral corrections on the order of $100\,\mu$eV. In the easy-plane phase, while DMI induces the well-documented weak ferromagnetic moment due to spin canting, its effect on the magnon spectrum is negligibly small, on the order of $25\,\mu$eV. The dominant relativistic effect arises from easy-plane anisotropy, which splits magnon modes at the Brillouin zone center and suppresses their spin expectation value. However, this effect remains weaker than the altermagnetic splitting and rapidly diminishes away from the zone center. Our analysis suggests that nonrelativistic altermagnetic splitting dominates at energies above $\sim30\,$meV, where the magnon spin polarization nearly recovers its quantized value. These findings indicate that inelastic neutron scattering can directly probe altermagnetic magnon splitting in hematite. We also discuss implications for magnon transport, particularly magnonic contributions to the thermal Hall effect (which requires spin-orbit coupling) and to spin splitter effects (which do not). Notably, we predict a third-order nonlinear magnon spin splitter effect: when a temperature gradient is applied along a direction in the $ab$ plane that does not coincide with a mirror plane, a spin current emerges along the $c$ axis. This result suggests that the $g$-wave magnon spin splitting in hematite enables transverse heat-to-spin conversion without requiring an external magnetic field.
\end{abstract}

\date{\today}
\maketitle

%%%%%%%%%%%%%%%%%%%%%%%%%%%%%%%%%%%%%%%%%%%%%%%%%%%%%%%%%%%%%%%%%%%%
%
% INTRODUCTION
%
%%%%%%%%%%%%%%%%%%%%%%%%%%%%%%%%%%%%%%%%%%%%%%%%%%%%%%%%%%%%%%%%%%%%
\section{Introduction}
%\pps{About the figures: In the figures I see "Momentum $k$", "Energy $\varepsilon$", "Splitting $\Delta\varepsilon$" and so on written explicitly as x- or y-axis label. I think the qualifiers "Momentum", "Energy", "Splitting" are completely redundant and makes the plot busy to the eye. Also, when plotting between momenta points $\Gamma$-$X$-... The labeling of the x-axis is complacently unnecessary. On a personal preference, momentum points $\Gamma$, $X$,... should be bold, because the momentum point is really a vector, but this is neither here nor there. One more point, in the first few figures, some of the fonts are incredibly small. I think especially for the axis labels in the figures the fonts should be big enough not to dominate the figure but also not require one to zoom in completely to resolve it.}
The iron oxide $\alpha-$Fe$_2$O$_3$, known as \textit{hematite}, has great technological relevance \cite{Mirzaei2015, Ahmad2016, Tamirat2016, Wan2023ReviewHematite}. In the context of magnetism, it holds a place as archetypical weak ferromagnet---a magnet with an almost compensated antiparallel order but a small canting of the magnetic moments which gives rise to a finite net magnetization. This canting arises from the relativistic antisymmetric exchange, which is also referred to as Dzyaloshinskii-Moriya interaction \cite{dzyaloshinskyThermodynamicTheoryWeak1958,MoriyaAnisotropicSuperex1960}. Below the Morin transition temperature of $T_{\text{M}} \approx 250$ K \cite{morinMagneticSusceptibilityFe1950,neelNewResultsAntiferromagnetism1953,velikov1969antiferromagnetic, ellistonAntiferromagneticResonanceMeasurements} hematite undergoes the first-order spin flop transition from the high-temperature easy-plane phase with a weak ferromagnetic moment to the low-temperature easy-axis phase without a weak ferromagnetic moment \cite{morinMagneticSusceptibilityFe1950, hillNeutronDiffractionStudy2008}. The magnetic long-range order in the high-temperature weak ferromagnetic phase is stable up to the Néel temperature $T_\text{N} \approx 955$ K \cite{morinMagneticSusceptibilityFe1950,irshinskii1980moessbauer,hillNeutronDiffractionStudy2008}. These magnetic properties have made hematite an exciting material platform for spintronics applications \cite{Baltz2018, Ross2020, Meer2023}. For example, it supports an electrically switchable magnetic order \cite{Cheng2020, Cogulu2021} and spin transport over long distances \cite{Lebrun2018hematitelongdistance, Ross2019}.

%polarized neutron scattering to study the domains in hematite \cite{nathansPolarizedNeutronStudyHematite1964}

Recently, hematite was identified as an altermagnet \cite{SmejkalBeyondConv2022,SmejkalChiralMagnons2023,chenUnconventionalMagnonsCollinear2025}. Altermagnets are compensated collinear magnets, which---in contrast to antiferromagnets---break time-reversal symmetry in momentum space. This is because where antiferromagnets have a sublattice transposing symmetry involving inversion or translation, altermagnets require (possibly nonsymmorphic) rotations or reflections \cite{SmejkalBeyondConv2022}. As a result, the electronic band structure in altermagnets lacks Kramers degeneracy and exhibits a spin splitting \cite{SmejkalBeyondConv2022, SmejkalEmerging2022, Krempask2024, Lee2024}. In contrast to ferromagnets, this spin splitting has an even parity beyond $s$-wave, implying that the splitting has either $d$-wave, $g$-wave, or $i$-wave character \cite{SmejkalBeyondConv2022}. As these partial waves have nodal planes \cite{Mazin2021}, the spin splitting in the electronic band structure is highly anisotropic and direction dependent, giving rise to several electronic transport effects with great potential in spintronics, for example, and an anomalous Hall effect \cite{Smejkal2020CrystalHall}, giant magnetoresistance \cite{Smejkal2022GMR}, and spin-polarized currents \cite{Naka2019, GonzalezHernandez2021}. %\am{\@ Libor, you had a comment here, but I did not quite catch it. Could you check if it is resolved?}

Hematite is an insulator and, as such, does not naturally support electronic transport effects; doping is required to render it conductive \cite{Galindez-Ruales2023}. However, spin transport remains possible through collective magnetic excitations known as magnons, the quanta of spin waves. In altermagnets, the magnon band structure is expected to exhibit spin splitting \cite{Naka2019, SmejkalChiralMagnons2023, gohlkeSpuriousSymmetryEnhancement2023, Gomonay2024} (sometimes also called chirality \cite{Odashima2013} or polarization \cite{Nambu2020} splitting).
%, or handedness \cite{Onoda2008}}
 This splitting has been experimentally observed in MnTe via inelastic neutron scattering \cite{Liu2024MnTeMagnonSplit}, and is even slightly larger than what was predicted from first-principles \cite{SmejkalChiralMagnons2023}. In hematite, first-principles calculations of Heisenberg exchange parameters predict a $g$-wave magnon band splitting of up to $3\,$meV \cite{SmejkalChiralMagnons2023}. Notably, these calculations neglect relativistic effects, raising the question of whether such corrections could obscure the magnon spin splitting.

To address this question, a microscopic understanding of the magnon band structure across the entire Brillouin zone is required. Here, we develop a spin-wave theory model to systematically study hematite’s magnon excitations. We first analyze how magnon spin splitting arises in the nonrelativistic limit and find that exchange interactions up to the 13\textsuperscript{th} neighbor shell must be included to recover the characteristic $g$-wave splitting. This occurs because a Heisenberg model truncated at the twelfth shell exhibits higher symmetries than the underlying lattice. For a general discussion on the minimum required neighbor shell to construct a Heisenberg model without artificial symmetry enhancements, we refer the reader to Ref.~\cite{Dagnino2024}. Although the necessity for far-neighbor Heisenberg exchange interactions is not necessarily a general feature of all altermagnets, it has already been pointed out for the rutiles \cite{SmejkalChiralMagnons2023, gohlkeSpuriousSymmetryEnhancement2023} and also for MnTe \cite{McClarty2024Landau, Liu2024MnTeMagnonSplit}. In hematite, while the 13\textsuperscript{th}-neighbor exchange interactions are naturally weaker than those between the nearest neighbors, first-principles calculations suggest they remain sufficiently strong to induce a splitting of approximately $2\,$meV. 

Next, we incorporate relativistic corrections---including easy-axis anisotropy, easy-plane anisotropy, and Dzyaloshinskii-Moriya interactions (DMI)---and examine their effects on the magnon band structure and magnon spin polarization. The magnitudes of these relativistic terms are estimated by fitting spin-wave gaps and canting angles to experimental data \cite{andersonMagneticResonanceFe1954,kumagaiFrequencyDependenceMagnetic1955,fonerLowTemperatureAntiferromagneticResonance1965,ellistonAntiferromagneticResonanceMeasurements,velikov1969antiferromagnetic,morrishCantedAntiferromagnetismHematite1994,morrishCantedAntiferromagnetismHematite1994,hillNeutronDiffractionStudy2008,chouHighResolutionTerahertzOptical2012,Schönfeld2025} and by referring to previous first-principles calculations \cite{danneggerMagneticPropertiesHematite2023}. While relativistic corrections are essential in determining the ground state, we find that in both the easy-axis and easy-plane weak ferromagnetic phases, their impact on the magnon spectrum is negligible away from the Brillouin zone center. The most pronounced effect arises from easy-plane anisotropy in the weak ferromagnetic phase, which hybridizes magnon modes near the zone center and destroys the magnon spin polarization. However, this hybridization rapidly diminishes with increasing momentum, allowing the altermagnetic splitting---scaling with the fourth power of momentum---to dominate. Consequently, although magnon spin polarization is suppressed near the zone center, it quickly recovers its nearly quantized value at higher momenta. These findings suggest that relativistic effects will not obscure the nonrelativistic altermagnetic splitting in hematite.

%\pps{I think this sentence was meant to be the last sentence of the previous paragraph, and not the first of this paragraph: } 
Finally, we discuss spectroscopic and transport experiments that could detect the predicted $g$-wave magnon spin splitting. A symmetry analysis of magnetic point groups suggests that hematite is compatible with a thermal Hall effect in the weak ferromagnetic phase. Furthermore, inspired by recent work on nonlinear electron spin splitters \cite{Ezawa2025}, we propose a nonlinear magnon spin splitter effect driven by a temperature gradient. Specifically, we predict that a spin current along the $c$ axis emerges in response to a temperature gradient in the basal $ab$ plane, provided the gradient is not aligned with a mirror plane of hematite.

The rest of the paper is organized as follows: In Sec.~\ref{sec:model}, we discuss important symmetries, outlining the crystal structure in Sec.~\ref{sec:crystal_struct}, with a focus on the shortest bonds accounting for the altermagnetic nature of hematite in Sec.~\ref{sec:am_properties}. In Sec.~\ref{sec:nonrel}, we discuss the nonrelativistic physics and the resulting magnon spectrum with time-reversal symmetry breaking and magnon spin splitting. In Sec.~\ref{sec:rel}, we investigate relativistic physics and introduce the contributions from spin-orbit coupling. We then study the easy-axis phase (EAP) in Sec.~\ref{sec:EAP}, where we introduce the easy-axis anisotropy as well as the DMI. We then shift to the easy-plane or weak ferromagnetic phase (WFP) in Sec.~\ref{sec:WFP}, introducing an effective easy-plane anisotropy in Sec.~\ref{sec:WFP_anisotropy}, the effect of the DMI in Sec.~\ref{sec:WFP_dmi}, and the influence of altermagnetism in Sec.~\ref{sec:WFP_full}. We further investigate implications for experiments in Sec.~\ref{sec:symmetry} and conclude in Sec.~\ref{sec:conclusion}. In the Appendix we show the symmetries of the space group in Sec.~\ref{ap:symm}, followed by density functional theory (DFT) data in Sec.~\ref{ap:DFT}. We close with pointing out that order-by-quantum disorder considerations cause a triaxial anisotropy in the system in Sec.~\ref{sec:order_by_disorder}.
%%%%%%%%%%%%%%%%%%%%%%%%%%%%%%%%%%%%%%%%%%%%%%%%%%%%%%%%%%%%%%%%%%%%
%
% Model
%
%%%%%%%%%%%%%%%%%%%%%%%%%%%%%%%%%%%%%%%%%%%%%%%%%%%%%%%%%%%%%%%%%%%%
\section{Symmetries}
\label{sec:model}
%\pps{this section is called "Model", but no model is presented actually. The explicit models are given in equations in the latter sections "Nonrelativistic Physics" and "Relativistic Physics". It seems to me this section is more about symmetry. Maybe you want to name this section "Symmetries", and then subsection A is "Crystal structure" and subsection B is "Altermagnetisim".}
\subsection{Crystal structure}
\label{sec:crystal_struct}
\begin{figure}
    \centering       
    \includegraphics[width=\columnwidth]{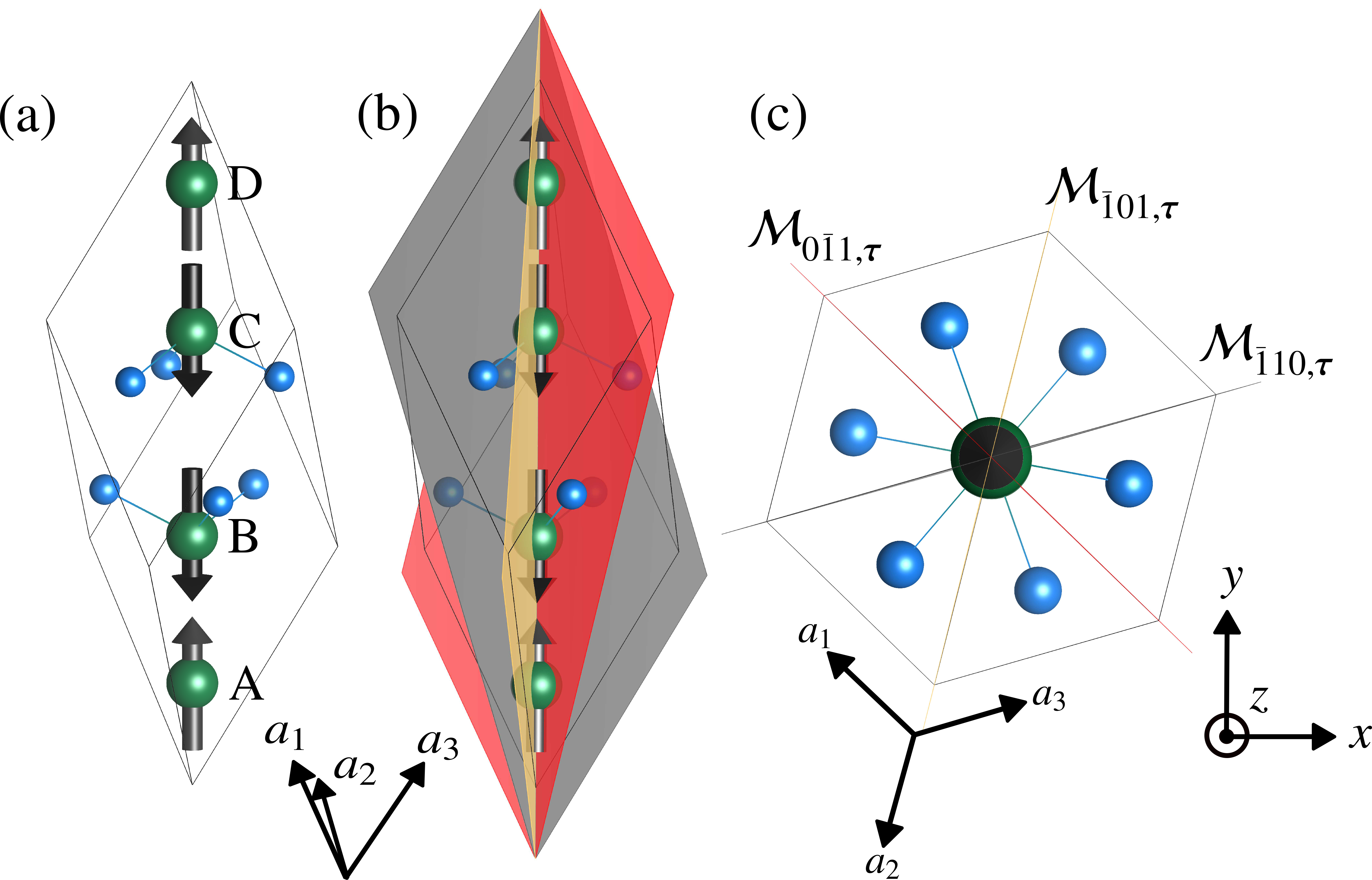}
       \caption{
       Rhombohedral unit cell of hematite in the ground state (easy-axis) phase with magnetic iron (Fe) atoms in green, and nonmagnetic oxygen (O) atoms in blue.  (a) Side view with the iron atoms denoted A, B, C, and D from bottom to top.  (b) Same view as (a) with the three unitary glide-mirrors $\mathcal{M}_{0\overline{1}1, \bm{\tau}}$ (red), $\mathcal{M}_{\overline{1}01, \bm{\tau}}$ (yellow), and $\mathcal{M}_{\overline{1}10, \bm{\tau}}$ (gray), with $\bm{\tau} = \left(\frac{1}{2}, \frac{1}{2}, \frac{1}{2} \right) $.  (c) Top view of the unit cell with the mirrors denoted.
       }
       \label{fig:unit_cell}
\end{figure}

Hematite has the space group R$\overline{3}$c (No. 167) \cite{brock2016internationalTabCryst} in Hermann Mauguin notation, meaning that it is centrosymmetric and has a rhombohedral (trigonal) primitive unit cell with three mirrors. The space group, after factorizing out the translational group of the Bravais lattice, has twelve coset representatives, which are tabulated in Tab.~\ref{tab:symmetries} in Appendix~\ref{ap:symm}. The primitive unit cell in the easy-axis phase is shown in Fig.~\ref{fig:unit_cell} (generated with the program VESTA \cite{Momma2008VESTAAT}). The four magnetic iron (Fe) atoms are colored green, and the six non-magnetic oxygen (O) atoms are colored blue. We label the iron atoms A, B, C, and D from bottom to top [cf. Fig.~\ref{fig:unit_cell}(a)]. 
The three mirrors shown in Fig.~\ref{fig:unit_cell}(b) are unitary glide-mirrors $\mathcal{M}_{0\overline{1}1, \bm{\tau}}$ (red), $\mathcal{M}_{\overline{1}01, \bm{\tau}}$ (yellow), and $\mathcal{M}_{\overline{1}10, \bm{\tau}}$ (gray), with $\bm{\tau} = \left(\frac{1}{2}, \frac{1}{2}, \frac{1}{2} \right) $, where $\bm{\tau}$ is given in fractional lattice coordinates. The lattice vector $\bm{a}_1$/$\bm{a}_2$/$\bm{a}_3$ lies inside the glide-mirror $\mathcal{M}_{0\overline{1}1, \bm{\tau}}$/$\mathcal{M}_{\overline{1}01, \bm{\tau}}$/$\mathcal{M}_{\overline{1}10, \bm{\tau}}$ [cf.~Fig.~\ref{fig:unit_cell}(c)].

%\pps{\textbf{CURRENT}: 
%We build the primitive lattice vectors $\bm{a}_i$ from the conventional hexagonal unit cell with lattice constants 
%$a_\text{c} = 5.0342\,\angstr$ 
%and $c_\text{c} = 13.7519\,\angstr$ 
%as given in Ref.~\cite{pailheInvestigationNanocrystallizedAFe22008} as follows

%\textbf{SUGGESTION}:
The primitive lattice vectors $\bm{a}_i$ are determined from the conventional lattice vectors $\bm{a}^c_i$, %\am{I'm confused by this notation. Should there be transpositions?}\pps{I think (from looking at other eq in the manuscript, for example eq(4a)) when you write vectors, they are to be understood as column vectors. In that case in eq(1-3) the $\bm{a}$ and $\bm{a}^c$ do need to be transposed. I think there are two ways to amend this (see below and chose one which you prefer):
\begin{align}
    \left(\begin{array}{c}
         \bm{a}^c_1 \ \ \bm{a}^c_2\ \ \bm{a}^c_3
    \end{array} \right) = 
    \left(
    \begin{array}{rrr}
         \frac{\sqrt{2 + \sqrt{3}}}{2}a_c & -\frac{\sqrt{2 - \sqrt{3}}}{2}a_c & 0 \\
         -\frac{\sqrt{2 - \sqrt{3}}}{2}a_c &\frac{\sqrt{2 + \sqrt{3}}}{2}a_c & 0 \\
         0 & 0 & c_c \\
    \end{array}
\right),
\end{align}
by
\begin{equation}
    \left(\begin{array}{c}
         \bm{a}_1 \ \ \bm{a}_2\ \ \bm{a}_3
    \end{array} \right) = 
\left(
    \begin{array}{c}
         {\bm{a}^c_1} \ \ {\bm{a}^c_2}\ \ {\bm{a}^c_3}
    \end{array}
\right)\left(
    \begin{array}{rrr}
         -1/3 & -1/3 & 2/3 \\
         1/3 &-2/3 & 1/3 \\
         1/3 & 1/3 & 1/3 \\
    \end{array}
\right)
,
\end{equation}
with unit cell lengths $a_c\! = \!|\bm{a}^c_1| \!=\!|\bm{a}^c_2| \!=\! 5.0342\,\angstr$, 
and $c_c\!=\!|\bm{a}^c_3| \!=\! 13.7519\,\angstr$,
as reported in Ref.~\cite{pailheInvestigationNanocrystallizedAFe22008}.
We find 
\begin{align}
    \left(\begin{array}{c}
         \bm{a}_1 \ \ \bm{a}_2\ \ \bm{a}_3
    \end{array} \right) = 
    \begin{pNiceMatrix}
        - \frac{a_\text{c}}{\sqrt{6}} & - \frac{a_\text{c}}{2 \sqrt{2}} + \frac{a_\text{c}}{2 \sqrt{6}} &  \frac{a_\text{c}}{2 \sqrt{2}} + \frac{a_\text{c}}{2 \sqrt{6}}\\
        \frac{a_\text{c}}{\sqrt{6}} & - \frac{a_\text{c}}{2 \sqrt{2}} - \frac{a_\text{c}}{2 \sqrt{6}} & \frac{a_\text{c}}{2 \sqrt{2}} - \frac{a_\text{c}}{2 \sqrt{6}}\\
        \frac{c_\text{c}}{3} & \frac{c_\text{c}}{3} &  \frac{c_\text{c}}{3}
    \end{pNiceMatrix}.
\end{align}

We set the real space positions of the iron atoms (in fractional coordinates)  %\am{From which paper did you get these numbers?} \rh{the $\delta$s are from \cite{pailheInvestigationNanocrystallizedAFe22008} as cited below. I remember asking Peter about the $1/3$ etc., and I don't remember the exact answer.}
%\pps{Yes these are my definitions. Experimentalist crystalographers do not report data like this. The $\delta_{\mathrm{Fe}}$ and $\delta_{\mathrm{O}}$ are such that the crystallographic data matches Ref.44 specifically Table 1. I think the exact position formulas are not needed, definitely not main text information. It suffice to say "the crystal structure is as reported in Ref.44" or "we use the crystalographic information from Ref.44" or something like this, plus mention the Wyckoff positions of Fe (which is important as the magnetic site Fe will transform as Wyckoff 12c, i.e., subgroup $C_3$, under crystallographic symmetry operations). If needed the exact position formulas that follow can be tabulated and sent to the appendices.}
\begin{subequations}
    \begin{align}
    \bm{r}_\text{A}^\text{Fe} = \left( \frac{1}{6} - \delta_\text{Fe}, \frac{1}{6} - \delta_\text{Fe}, \frac{1}{6} - \delta_\text{Fe}\right)^\mathrm{T}, \\
    \bm{r}_\text{B}^\text{Fe} = \left( \frac{1}{3} + \delta_\text{Fe},  \frac{1}{3} + \delta_\text{Fe},  \frac{1}{3} + \delta_\text{Fe}\right)^\mathrm{T}, \\
    \bm{r}_\text{C}^\text{Fe} = \left( \frac{2}{3} - \delta_\text{Fe},  \frac{2}{3} - \delta_\text{Fe},  \frac{2}{3} - \delta_\text{Fe}\right)^\mathrm{T}, \\
    \bm{r}_\text{D}^\text{Fe} = \left( \frac{5}{6} + \delta_\text{Fe}, \frac{5}{6} + \delta_\text{Fe}, \frac{5}{6} + \delta_\text{Fe}\right)^\mathrm{T}, 
    \end{align}
\end{subequations}
where $\delta_\text{Fe} = 0.35498 - \frac{1}{3}$ \cite{pailheInvestigationNanocrystallizedAFe22008}. They occupy the Wyckoff position 4c. The position of the oxygen atoms is given by
\begin{subequations}
    \begin{align}
    \bm{r}_1^\text{O} = \left( \frac{7}{12} - \delta_\text{O}, \frac{11}{12} + \delta_\text{O}, \frac{1}{4} \right)^\mathrm{T}, \\
    \bm{r}_2^\text{O} = \left( \frac{1}{4},  \frac{7}{12} - \delta_\text{O}, \frac{11}{12} + \delta_\text{O}\right)^\mathrm{T}, \\
    \bm{r}_3^\text{O} = \left( \frac{11}{12} + \delta_\text{O},  \frac{1}{4},  \frac{7}{12} - \delta_\text{O}\right)^\mathrm{T}, \\
    \bm{r}_4^\text{O} = \left( \frac{5}{12} + \delta_\text{O}, \frac{1}{12} - \delta_\text{O}, \frac{3}{4} \right)^\mathrm{T}, \\
    \bm{r}_5^\text{O} = \left(\frac{3}{4}, \frac{5}{12} + \delta_\text{O}, \frac{1}{12} - \delta_\text{O} \right)^\mathrm{T}, \\
    \bm{r}_6^\text{O} = \left(\frac{1}{12} - \delta_\text{O}, \frac{3}{4},  \frac{5}{12} + \delta_\text{O} \right)^\mathrm{T},
    \end{align}
\end{subequations}
where $\delta_\text{O} = \frac{1}{3} - 0.3082$ \cite{pailheInvestigationNanocrystallizedAFe22008}. They occupy the Wyckoff position 6e. % with the letter ``e''.\rh{6e in primitive unit cell}
The center of inversion is situated in the middle of the bond connecting $\bm{r}_\text{B}^\text{Fe}$ to $\bm{r}_\text{C}^\text{Fe}$ (Wyckoff position 2b), as seen in Fig.~\ref{fig:unit_cell}(a). 

\subsection{Altermagnetism}
\label{sec:am_properties}
As shown in Fig.~\ref{fig:unit_cell}, the magnetic moments of iron ions A and D are parallel but opposite to those of B and C. In the low-temperature easy-axis phase, they are aligned (or antialigned) with the $z$ direction. In the high-temperature weak ferromagnetic easy-plane phase, the moments lie in the $ab$ basal plane and exhibit a small canting within the plane. We will explore relativistic effects in Sec.~\ref{sec:rel}. Here, let us assume the nonrelativistic limit, such that the absolute orientation of the spins in real space does not matter. Their relative orientation and collinear order, however, remain important, as they are defined in spin space.

\begin{figure}
    \centering       
    \includegraphics[width=\columnwidth]{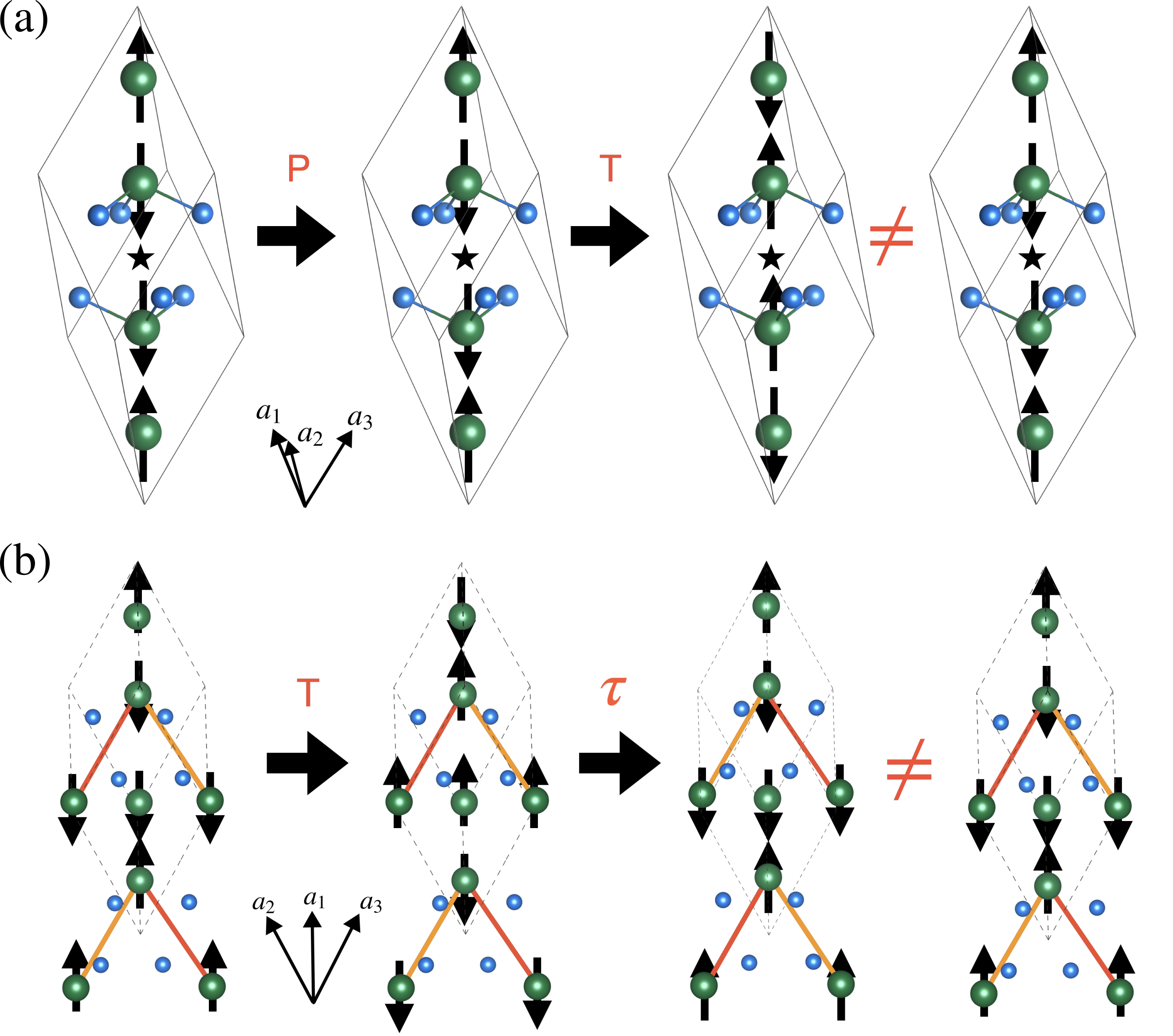}
       \caption{%\pps{The lattice vectors and "star center of inversion" are way to small fonts.}
       Absence of antiferromagnetism in hematite in the nonrelativistic limit.  (a) Hematite does not possess $PT$ symmetry, where $P$ denotes inversion symmetry and $T$ time-reversal symmetry. Inversion $P$ with respect to the center of inversion (marked by a star) is a symmetry of the crystal. The magnetic moments, being axial vectors, only change position but do not get reversed under $P$. Applying time reversal $T$ flips all spins. The resulting configuration is not identical to the original crystal.  (b) Hematite does not posses $T \vec{\tau}$ symmetry, where $\vec{\tau}$ denotes translation. Time reversal $T$ flips all spins and a translation by $\vec{\tau} =(1/2, 1/2, 1/2)$ (given in fractions of the lattice vectors) shifts the atoms along the $c$ direction by half a unit cell. The resulting configuration is not identical to the original crystal. The difference arises because of the blue oxygen atoms. In the context of the effective Heisenberg spin model, which does not explicitly account for nonmagnetic ions, the symmetry-breaking due to the presence of these oxygen atoms enters via the 13\textsuperscript{th}-neighbor exchange interactions. These come with two different values due to the symmetry-inequivalent bonds (marked by different colors).
    }
       \label{fig:no_AFM}
\end{figure}

%First, note that hematite is not an antiferromagnet because its oppositely aligned sublattices are neither related by a $PT$ nor a $T \vec{\tau}$ symmetry \cite{SmejkalBeyondConv2022}. 
First, note that hematite is neither a $PT$, nor a $T \vec{\tau}$ antiferromagnet because its oppositely aligned sublattices are neither related by a $PT$ nor a $T \vec{\tau}$ symmetry \cite{SmejkalBeyondConv2022}. Here, $P$ denotes inversion symmetry, $T$ denotes time-reversal symmetry, and $\vec{\tau}$ denotes a translation. As shown in Fig.~\ref{fig:no_AFM}(a), $P$ is a good symmetry because hematite is centrosymmetric. A subsequent action of $T$ flips the spins, resulting in a configuration that is not identical to the original one. Similarly, as shown in Fig.~\ref{fig:no_AFM}(b), a time-reversed order cannot be brought to the original configuration by a translation. Here, the difference becomes apparent only by considering the oxygen atoms.  

%Hematite does neither possess PT, nor T$\tau$ symmetry, where P denotes parity, T time-reversal and $\tau$ a translation. As we can see in Fig.~\ref{fig:no_AFM}(a), parity P does nothing to the unit cell (second column), since there is a center of inversion (marked by a star). Applying time reversal T flips all the spins, and the result in the third column is not identical with the starting point (column four). Thus, PT does not leave the unit cell invariant and is therefore not a symmetry of the system. Figure~\ref{fig:no_AFM}(b) shows the same view of the crystal as in Fig.~\ref{fig:13th}NN(c). Applying time reversal T to the shown atoms flips all the spins (column two). Applying a half-unit cell translation $\tau$ shifts the lower red and orange bonds including the oxygen atoms up by $\tau = (1/2, 1/2, 1/2)$ in lattice coordinates. The result in the third column does not agree with the starting point (fourth column). We conclude that hematite is not T$\tau$-symmetric either. Therefore, hematite is no PT or T$\tau$ antiferromagnet.

\begin{figure}
    \centering       
    \includegraphics[width=\columnwidth]{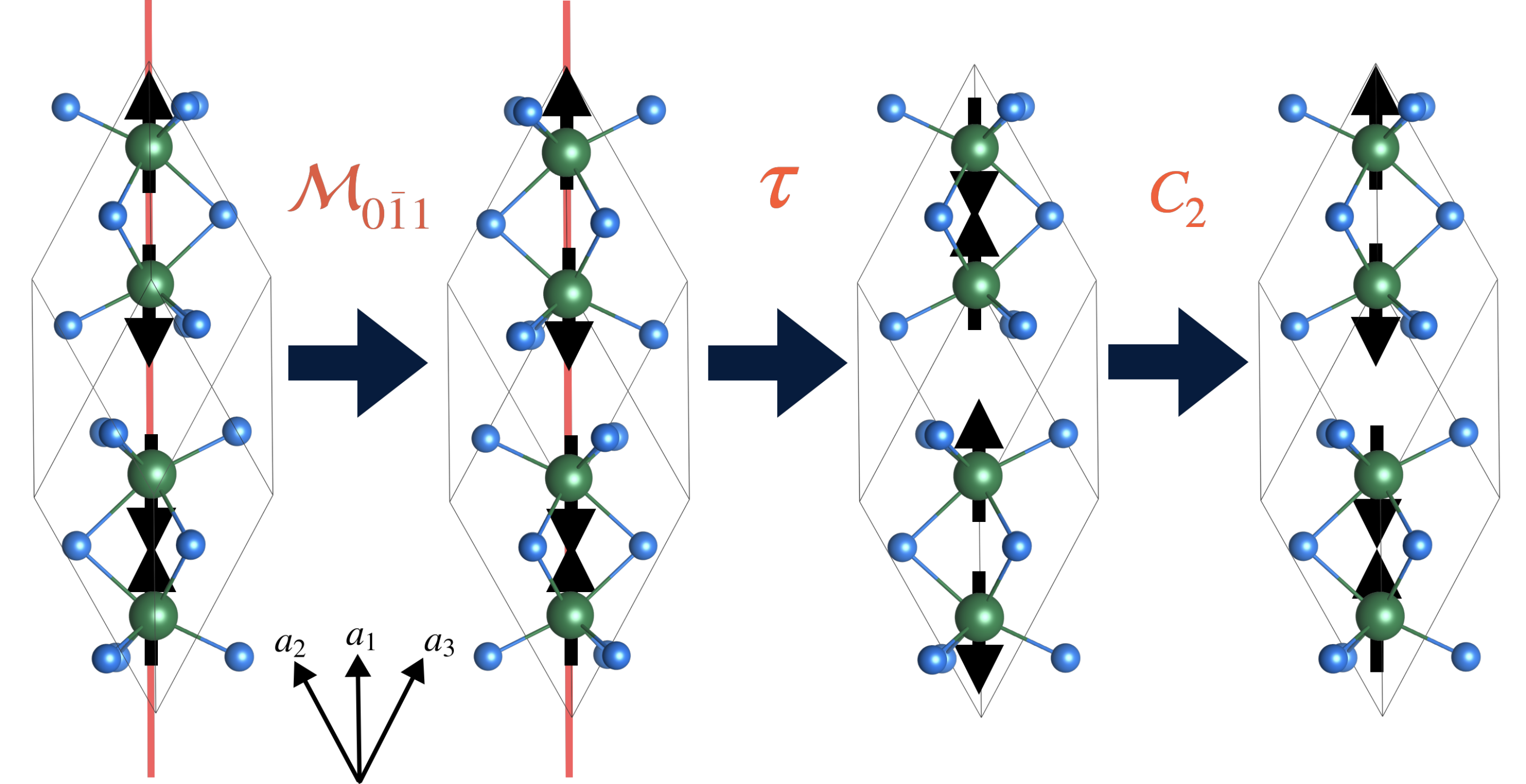}
    \caption{
       Sublattice transposing symmetry responsible for altermagnetism in hematite. The panels show a step-by-step application of the symmetry operation $[C_2 || \mathcal{M}_{0 \overline{1} 1} | \vec{\tau}]$ with $\vec{\tau} = (\frac{1}{2},\frac{1}{2},\frac{1}{2})$. First, the mirror $\mathcal{M}_{0 \overline{1} 1}$, indicated by the orange line, is applied in direct space, leading to a reversal of the oxygen cages. Second, a translation by $\vec{\tau}$ maps the atomic positions onto the original positions, but also moves the spins. Finally, a $C_2$ rotation in spin space reverses the direction of the spin, mapping back onto the original configuration.
    }
       \label{fig:but_altermagnetism}
\end{figure}

%\am{\@ Libor: Could you please double check this paragraph and the argument?}
Instead of inversion or translation symmetries, mapping the two oppositely aligned sublattices onto each other requires, e.g., one of the glide-mirrors indicated in Fig.~\ref{fig:unit_cell}(b,c) followed by a reversal of the spins. Specifically, since hematite belongs to the spin Laue group ${}^{1}\overline{3} {}^{2}\text{m}$ \cite{SmejkalBeyondConv2022}, it supports the sublattice transposition symmetry $[C_2 || \mathcal{M}_{0 \overline{1} 1} | \vec{\tau}]$ with $\vec{\tau} = (\frac{1}{2},\frac{1}{2},\frac{1}{2})$, where the left of $||$ denotes the operation in spin space and the right denotes the operation in direct space (real space). The successive application of the mirror, translation, and spin rotation contained in this operation is visualized in Fig.~\ref{fig:but_altermagnetism}.
%The spin space group of hematite reads ${}^{1}\overline{3} /{}^{2}\text{m} = [ E || \overline{3}] + [C_2|| \overline{3}\text{m} - \overline{3} | \bm{\tau}]$, where the left of $||$ denotes the operation in spin space and the right denotes the operation in real space.

The spin Laue group ${}^{1}\overline{3} {}^{2}\text{m}$ renders hematite a (bulk-type) $g$-wave altermagnet \cite{SmejkalBeyondConv2022}. We are guaranteed to find four nodal surfaces in reciprocal space, where the spin splitting is zero. Three nodal surfaces are vertical \textit{planes}, symmetry-protected by the three glide mirrors in the system that connect opposite spin sublattices [cf.~Fig.~\ref{fig:unit_cell}(b)]. The fourth nodal surface is not forced to be planar, since the system lacks any additional mirror symmetry (connecting opposite spin sublattices) with an out-of-plane rotation axis. Note the difference to the $g$-wave altermagnet MnTe, which is also bulk-type $g$-wave but belongs to another spin Laue group \cite{SmejkalBeyondConv2022}, and supports four planar nodal surfaces \cite{Liu2024MnTeMagnonSplit, Lee2024}.

As we have seen, hematite's altermagnetic nature arises solely due to the presence of oxygen ions, which break the $T \vec{\tau}$ symmetry. In an effective Heisenberg spin model that does not explicitly include nonmagnetic ions, the influence of oxygen can only be captured by identifying magnetic bonds that exist in two symmetry-inequivalent versions. The key question, then, is how far one must extend the bond analysis to encounter the first such inequivalent pair. In other words, what is the closest shell of magnetic ions that correctly implements the lattice symmetries?

This type of problem was addressed in a general setting in Ref.~\cite{Dagnino2024}. Recent studies have demonstrated that in RuO$_2$ the relevant bonds appear at the 6th shell \cite{SmejkalChiralMagnons2023}, whereas in MnTe, they emerge at the 10th and 11th shells \cite{Liu2024MnTeMagnonSplit}. These findings highlight that capturing the altermagnetic properties of magnons sometimes requires extending the analysis beyond the first few neighbors. The minimal distance at which such symmetry-inequivalent bonds appear is fundamentally a property of the underlying lattice. As we show in the case of hematite, it is the 13\textsuperscript{th} neighbors, which are indicated by different colors in Fig.~\ref{fig:no_AFM}(b) to highlight the breaking of $T \vec{\tau}$ symmetry.

\begin{figure*}
    \centering       
    \includegraphics[width=13cm]{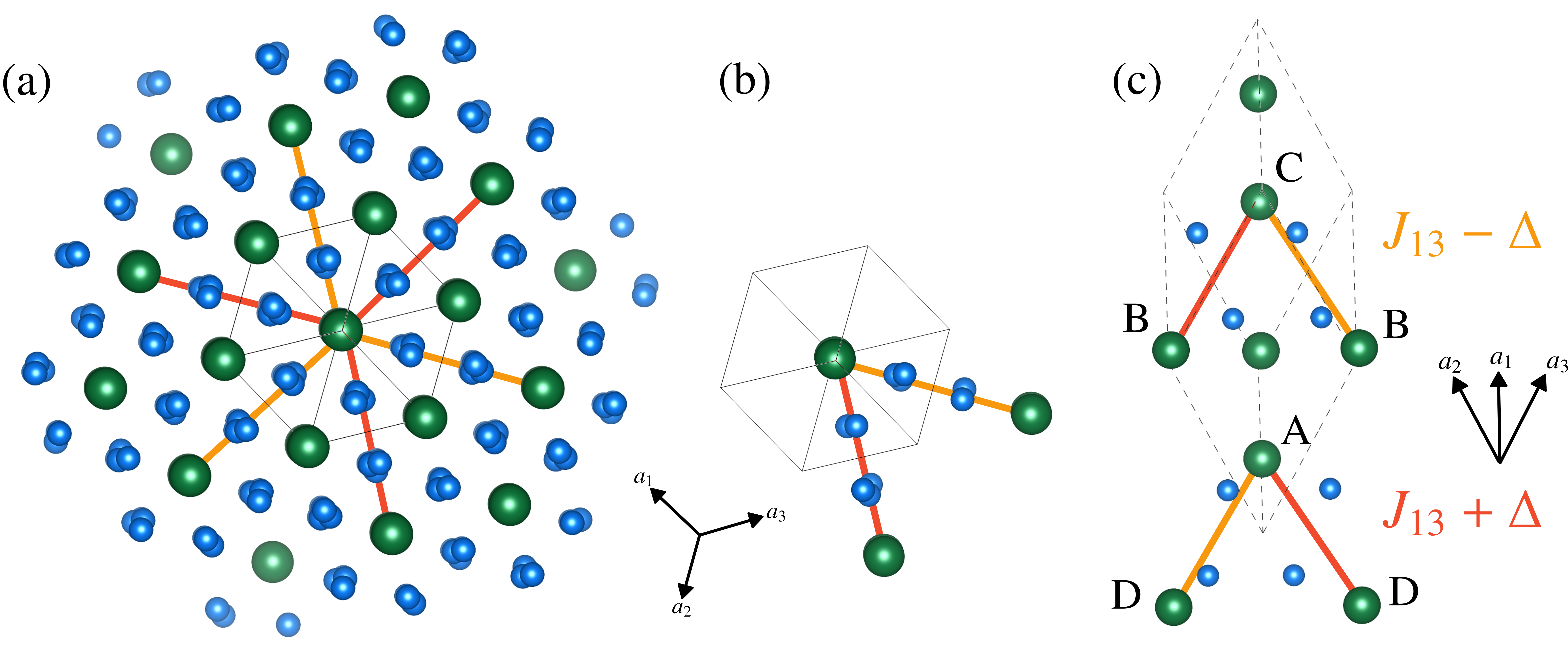}
       \caption{
       (a) Crystal structure of hematite beyond the first unit cell seen from above along the $c$ axis. The first unit cell is outlined in black. The red and orange lines mark the direction of the 13\textsuperscript{th} neighbor bonds (only for the bond connecting iron atoms C and B).  (b) Same as (a) but all atoms deleted but for two symmetry-inequivalent directions in the 13\textsuperscript{th} shell.  (c) The inequivalent 13\textsuperscript{th} neighbor bonds---or ``altermagnetic bonds''---are shown from a side view (perpendicular to the $c$ axis). Here, $J_{13}$ denotes the isotropic part of the exchange interaction and $\Delta$ the anisotropic part or the ``altermagnetic exchange''. The indicated bonds correspond to the directions shown in (b).
       }
       \label{fig:13thNN}
\end{figure*}

In Fig.~\ref{fig:13thNN}, we present a detailed analysis of why the 13\textsuperscript{th} bond plays a crucial role. Viewed from above in Fig.~\ref{fig:13thNN}(a,b), the 13\textsuperscript{th} bond---henceforth referred to as the ``altermagnetic bond''---connects iron atoms of the first unit cell to those in adjacent unit cells. Each iron atom is linked by six such bonds, shown in red and orange, at a distance of $6.42278 \angstr$. These bonds connect atom A to D (C to B), meaning they couple parallel-oriented spins. A key distinction arises from the positions of the blue oxygen atoms, which differ between the red and orange bonds. Importantly, no symmetry operation of the crystal maps an orange bond onto a red bond, leading to two inequivalent Heisenberg exchange constants. We denote the exchange interaction as $J_{13} + \Delta$ for red bonds and $J_{13} - \Delta$ for orange bonds, where $\Delta$, the ``altermagnetic exchange'', quantifies the symmetry-induced inequivalence. (We note in passing that further-neighbor bonds up to the 26\textsuperscript{th} shell that are ``altermagnetic bonds'' are $J_{17}$, $J_{21}$, $J_{23}$, and $J_{26}$.)

The physical consequence of a finite $\Delta$ is an intuitive spin splitting of magnons. Consider the limiting case where all exchange interactions connecting A to B or C, and D to B or C vanish. In this scenario, hematite effectively decomposes into two decoupled ferromagnets. Maintaining the antiparallel alignment of these sublattices, the magnons in each ferromagnet carry opposite spin (or chirality). Since the A-D sublattice ferromagnet features orange $13$th bonds in directions where the B-C sublattice ferromagnet contains red $13$th bonds, magnons of opposite spin acquire different group velocities along these directions, resulting in spin splitting.

%We also emphasize that the eighth-neighbor bonds, which are also along the directions marked in Fig.~\ref{fig:13thNN}(a) and (b), can all be mapped onto each other. Hence, they must have the same exchange constant. This makes the 13\textsuperscript{th} neighbor bonds the firsts to capture the altermagnetic property and leads to the characteristic spin (or chirality) splitting of magnon bands in the magnon dispersion relation [cf. Fig.~\ref{fig:disperion_nonrelativistic}].

%%%%%%%%%%%%%%%%%%%%%%%%%%%%%%%%%%%%%%%%%%%%%%%%%%%%%%%%%%%%%%%%%%%%
%
% Nonrelativistic physics
%
%%%%%%%%%%%%%%%%%%%%%%%%%%%%%%%%%%%%%%%%%%%%%%%%%%%%%%%%%%%%%%%%%%%%
\section{Nonrelativistic physics}
\label{sec:nonrel}
We first look into the nonrelativistic physics and therefore disregard the DMI and anisotropies derived from spin-orbit coupling (SOC). Two independent DFT calculations without SOC---that of Ref.~\cite{SmejkalChiralMagnons2023} and our own shown in Appendix~\ref{ap:DFT}---confirm the 13\textsuperscript{th} iron-iron bond to be the shortest bond with two symmetry-inequivalent versions responsible for the altermagnetism. We extracted exchange parameters $J_{i,j}$ from both \textit{ab initio} studies by mapping onto a Heisenberg model $\sum_{i,j} J_{i,j} \vec{S}_i \cdot \vec{S}_j$, and studied the magnon spectrum for both sets of parameters. While the overall band width and main features in the dispersion roughly agree with those found in inelastic neutron scattering experiments \cite{samuelsenInelasticNeutronScattering1970}, we have decided to take the DFT calculations only as a guide for the magnitude of the 13\textsuperscript{th} neighbor exchange interactions. To describe the material as well as possible, we do not rely on DFT data for exchange interactions of shorter bonds and instead opt for ``parameter set 1'' reported in Ref.~\cite{samuelsenInelasticNeutronScattering1970}, which was obtained by fitting linear spin-wave dispersions to inelastic neutron scattering data. This set of parameters includes the exchanges of the first five shells.
Therefore, the effective spin Hamiltonian 
\begin{align}
\mathcal{H} = \mathcal{H}^\text{ISO} + \mathcal{H}^\text{AM}
\label{eq:ham_EAP_iso_am}
\end{align}
is split into an isotropic Heisenberg exchange part $\mathcal{H}^\text{ISO}$ and an altermagnetic term $\mathcal{H}^\text{AM}$, where
\begin{align}
\label{eq:ham_iso}
   \mathcal{H}^\text{ISO} & = \sum_{r=1}^{5} \sum_{\langle i, j \rangle_r} J_r \bm{S}_{i} \cdot \bm{S}_{j} + \sum_{\langle i, j \rangle_{13}} J_{13} \bm{S}_{i} \cdot \bm{S}_{j}, 
\end{align}
%\begin{align}
%\label{eq:ham_iso}
%   \mathcal{H}^\text{ISO} & = \sum_{r=1}^{5} \sum_{\langle i, j \rangle_r} J_{ij} \bm{S}_{i} \cdot \bm{S}_{j} + \sum_{\langle i, j \rangle_{13}} J_{ij} \bm{S}_{i} \cdot \bm{S}_{j}, 
%\end{align}
and 
\begin{align}
   \mathcal{H}^\text{AM} & = \Delta  \left( \sum_{\langle i, j \rangle_{13}^1}  \bm{S}_{i} \cdot \bm{S}_{j} - \sum_{\langle i, j \rangle_{13}^2}   \bm{S}_{i} \cdot \bm{S}_{j} \right).
   \label{eq:ham_am}
\end{align}
We list the parameters used in the nonrelativistic limit in Tab.~\ref{tab:params_nonrel}, where we chose $J_1$ to $J_5$ to come from Ref.~\cite{samuelsenInelasticNeutronScattering1970} because they were obtained by fitting to neutron scattering data. The parameters $J_{13}$ and $\Delta$ are obtained from a fit to our own DFT calculations, described in Appendix~\ref{ap:DFT}, using a reduced set of exchange parameters: $J_1$ to $J_5$, $J_{13}$, and $\Delta$. The numbers from another fit including all exchanges up to the 13\textsuperscript{th} neighbor ($J_{13} \approx 0.21\,$meV and $\Delta \approx 0.11\,$meV) are very similar, while the numbers from the data in the Supplemental Material of Ref.~\cite{SmejkalChiralMagnons2023} ($J_{13} \approx 0.04\,$meV and $\Delta \approx 0.15\,$meV) differ for $J_{13}$, but match quite well for $\Delta$. In this data set, exchange constants for a very large number of neighbors were considered, which leads to a substantial difference in $J_{13}$.
%specifically, from \mkch{the Supplemental Material of} Ref.~\cite{SmejkalChiralMagnons2023}, but the numbers of our own DFT calculation in Appendix~\ref{ap:DFT} are very similar ($J_{13} = 0.21\,$meV and $\Delta = 0.11\,$meV in a fit including all exchanges up to the 13\textsuperscript{th} neighbor, and $J_{13} = 0.14\,$meV and $\Delta = 0.15\,$meV in a fit including a reduced set of parameters). 
We emphasize that we have used a parametrization where the spins do not have unit length but the spin quantum number $S=5/2$ ($\hbar = 1$). Furthermore, the sum over $\langle i, j \rangle_r$ is a sum over unique bonds belonging to the $r$\textsuperscript{th} shell (i.e., ions $i$ and $j$ are $r$-th nearest neighbors).

\begin{table}[t]
\centering
\caption{Parameters in the nonrelativistic limit. We use DFT calculations for $J_{13}$ and $\Delta$, and  ``parameter set 1'' reported in Ref.~\cite{samuelsenInelasticNeutronScattering1970} for $J_1$ to $J_5$. We have converted the exchanges to our convention which does not normalize the exchange with respect to the spin length. Additionally, we assumed a spin length of $5/2$ ($\hbar = 1$).}
\label{tab:params_nonrel}
\begin{tabular}{ p{1.3cm}|p{1.5cm}|p{1.cm}} 
 \toprule
 \centering Parameter & \centering Value & \centering Unit
  \arraybackslash \\
 \hline
 \centering $S$ & \centering $5/2$ & \centering 
   \arraybackslash  \\
  \centering $J_1$ & \centering $-0.982$ & \centering meV
  \arraybackslash \\
  
    \centering  $J_2$ & \centering $-0.169$ & \centering meV
  \arraybackslash \\
 
 \centering $J_3$ & \centering $5.452$ & \centering meV
   \arraybackslash \\
 
 \centering $J_4$ & \centering $4.008$ & \centering meV
  \arraybackslash  \\

  \centering $J_5$ & \centering $0.337$ & \centering meV
  \arraybackslash  \\

  \centering $J_{13}$ & \centering 
  %$0.04000$ 
 $0.143$ & \centering meV
  \arraybackslash  \\

  \centering $\Delta$ & \centering 
  %$0.15000$ 
  $0.148$ & \centering meV
  \arraybackslash  \\
  
 \botrule
\end{tabular}
\end{table}

%We set $J_1, J_2 <0$ and $J_3, J_4, J_5, J_{13} > 0$. 
%to avoid frustration. \am{Würde ich weglassen. Die $J$'s sind was sie sind. (Sind die von Samuelson unfrustriert?) Etwas Frustration täte der Ordnung auch keinen Abbruch.} 
%The addition (subtraction) of $\Delta$ to $J_{13}$ makes $\left( J_{13} + \Delta \right) > 0$ [$\left( J_{13} - \Delta \right) < 0$].
% \subsection{Spin-to-boson transformation}

Since hematite is known for supporting both collinear as well as canted spin textures, we use a general orthonormal basis $\{\hat{\bm{x}}^{\alpha}, \hat{\bm{y}}^{\alpha}, \hat{\bm{z}}^{\alpha}\}$, where
\begin{subequations}
\begin{align}
    \hat{\bm{x}}^{\alpha} & = \left(\cos \theta_{\alpha} \cos \phi_{\alpha}, \cos \theta_{\alpha} \sin \phi_{\alpha}, -\sin \theta_{\alpha} \right)^{\mathrm{T}}, \\
    \hat{\bm{y}}^{\alpha} & = \left(- \sin \phi_{\alpha}, \cos \phi_{\alpha}, 0 \right)^{\mathrm{T}}, \\
    \hat{\bm{z}}^{\alpha}  & = \left(\sin \theta_{\alpha} \cos \phi_{\alpha}, \sin \theta_{\alpha} \sin \phi_{\alpha}, \cos \theta_{\alpha} \right)^{\mathrm{T}}, \label{eq:N}
\end{align}
\label{eq:coordinate-system}
\end{subequations}
and $\alpha \in \{\text{A}, \text{B}, \text{C}, \text{D} \}$ stands for the corresponding sublattice. Within this general ansatz, the four sublattice moments point in a generic direction. We write the spins as
\begin{align}
    \bm{S}_i^{\alpha} = \widetilde{S}^{\alpha, x}_i \hat{\bm{x}}^{\alpha} +  \widetilde{S}^{\alpha, y}_i \hat{\bm{y}}^{\alpha} +  \widetilde{S}^{\alpha, z}_i \hat{\bm{z}}^{\alpha}.
\end{align}
However, it is known that the sublattices A and D (B and C) are parallel, such that $\theta_\text{D} \equiv \theta_\text{A}$ and $\phi_\text{D} \equiv \phi_\text{A}$ ($\theta_\text{C} \equiv \theta_\text{B}$ and $\phi_\text{C} \equiv \phi_\text{B}$). The ground state is collinear and in the relativistic limit, there is no preference of the Néel vector direction. We therefore choose the Néel vector $\bm{N} = \bm{M}_{\text{A}} - \bm{M}_{\text{B}}$ to be parallel to $\hat{\bm{z}}$,
%\am{We have to write here that it is collinear and we choose it to point along $z$ becuase in the nonrelativistic limit, there is no preference.} 
%The Néel vector $\bm{N} = \bm{M}_{\text{A}} - \bm{M}_{\text{B}}$ is therefore parallel to $\hat{\bm{z}}$
%($\bm{N} \parallel \hat{\bm{z}}$)
where $\bm{M}_{\text{A}}$ ($\bm{M}_{\text{B}}$) corresponds to the magnetization of sublattice A (B).% \am{Hier könnte man sich fragen, warum du so komplizierst ansetzt, obwohl du danach sofort sagst, dass die Ordnung kollinear ist, was die Winkel einfach macht. Das würde ich einfach etwas anders schreiben. We make the general ansatz, allowing for all sublattice moments to point in a general direction. However, it is known that ... such that ...}

We employ the Holstein-Primakoff \cite{holsteinprimakoff1940} transformation up to leading order in the spin quantum number $S$,
\begin{subequations}
    \begin{align}
        \widetilde{S}^{\alpha, x}_i & \approx \frac{\sqrt{2S}}{2} \left( \gamma_i + \gamma_i^\dagger \right), \\
        \widetilde{S}^{\alpha, y}_i & \approx  \frac{\sqrt{2S}}{2 \mathrm{i}} \left( \gamma_i - \gamma_i^\dagger  \right), \\
        \widetilde{S}^{\alpha, z}_i &= S - \gamma_i^\dagger \gamma_i,
    \end{align}
\end{subequations}
to rewrite the Hamiltonian $\mathcal{H}$ in terms of bosonic creation (annihilation) operators $\gamma_i^\dagger$ ($\gamma_i$), with $\gamma \in \{a, b, c, d\}$, where $a_i$/$b_i$/$c_i$/$d_i$ corresponds to a boson of sublattice A/B/C/D. The creation and annihilation operators obey the bosonic commutation relations $[\gamma_i, \gamma_j^\dagger] = \delta_{ij}$. We expand the Hamiltonian with respect to the number of bosons 
\begin{align}
    \mathcal{H} = \mathcal{H}_0 + \mathcal{H}_1 + \mathcal{H}_2 + \mathcal{H}_3 + \ldots \, , 
\end{align}
where $\mathcal{H}_0 \equiv E_0$ denotes the classical ground state energy, $\mathcal{H}_1 = 0$ as we expand around a stable ground state, $\mathcal{H}_2$ describes the free, noninteracting theory of magnons, and $\mathcal{H}_3$ and above cover interactions between magnons. We will restrict ourselves to the harmonic (noninteracting) Hamiltonian piece $\mathcal{H}_2$. 

We Fourier transform the bosonic operators
\begin{align}
    \gamma_{i}^\dagger = \frac{1}{\sqrt{N}} \sum_{\bm{k}} \mathrm{e}^{-\mathrm{i} \bm{k} \cdot \bm{r}_{i}} \gamma_{\bm{k}}^\dagger, \qquad \gamma_{i} = \frac{1}{\sqrt{N}} \sum_{\bm{k}} \mathrm{e}^{\mathrm{i} \bm{k} \cdot \bm{r}_{i}} \gamma_{\bm{k}},
    \label{eq:FT}
\end{align}
and find
\begin{subequations}
\begin{align}
	\mathcal{H}_2^{\text{ISO}} & =  \sum_{\bm{k}} \mathcal{A}_{\bm{k}}^{\text{ISO}}  \left(a_{\bm{k}}^{\dagger} a_{\bm{k}} + b_{\bm{k}}^{\dagger} b_{\bm{k}} + c_{\bm{k}}^{\dagger} c_{\bm{k}} + d_{\bm{k}}^{\dagger} d_{\bm{k}}\right)  \nonumber \\
 & \qquad
 + \mathcal{D}_{\bm{k}}  \left(a_{\bm{k}}^\dagger d_{\bm{k}}  +  c_{\bm{k}}^\dagger b_{\bm{k}} \right)  + \mathrm{H.c.} \nonumber \\
 & \qquad + \frac{\mathcal{B}_{\bm{k}}}{2}  \left( a_{-\bm{k}} b_{\bm{k}}  +  c_{-\bm{k}} d_{\bm{k}} + a_{\bm{k}}^{\dagger} b_{-\bm{k}}^{\dagger} + c_{\bm{k}}^{\dagger} d_{-\bm{k}}^{\dagger} \right)  + \mathrm{H.c.}  \nonumber \\
    & \qquad + \frac{\mathcal{C}_{\bm{k}}}{2}  \left(a_{-\bm{k}} c_{\bm{k}}  + b_{-\bm{k}} d_{\bm{k}} + a_{\bm{k}}^\dagger c_{-\bm{k}}^\dagger  + b_{\bm{k}}^\dagger d_{-\bm{k}}^\dagger\right) + \mathrm{H.c.} ,  \\ 
 \mathcal{H}_2^\text{AM} & =  \sum_{\bm{k}}  \Delta_{\bm{k}} \left( a_{\bm{k}}^\dagger d_{\bm{k}} -  c_{\bm{k}}^\dagger b_{\bm{k}} \right) + \mathrm{H.c.}\, ,
 \end{align}
 \end{subequations}
with the components given by
\begin{subequations}
\label{eq:nonrel_componenets}
\begin{align}
    \mathcal{A}_{\bm{k}}^{\text{ISO}} &= 
     S \left[ J_1 - 3 J_2  + 3 J_3 + 6 J_4 - J_5 - 6 J_{13}\right], \\
    \mathcal{D}_{\bm{k}} & = 
     S \left( J_2 \sum_{j=1}^3 \mathrm{e}^{\mathrm{i} \bm{k} \cdot \bm{\delta}_{2_j}} +  J_5  \mathrm{e}^{\mathrm{i} \bm{k} \cdot \bm{\delta}_5}+  J_{13} \sum_{j=1}^6 \mathrm{e}^{\mathrm{i} \bm{k} \cdot \bm{\delta}_{13_j}} \right), 
\end{align}
\begin{align}
    \mathcal{B}_{\bm{k}} & = 
     S \left( J_1  \mathrm{e}^{\mathrm{i} \bm{k} \cdot \bm{\delta}_1} +  J_3  \sum_{j=1}^3 \mathrm{e}^{\mathrm{i} \bm{k} \cdot \bm{\delta}_{3_j}} \right), \\
    \mathcal{C}_{\bm{k}} & = 
     2 S  J_4 \sum_{j=1}^3 \cos( \bm{k} \cdot \bm{\delta}_{4_j}), \\
      \Delta_{\bm{k}} & =  S \Delta \left( \sum_{j=1}^3 \mathrm{e}^{\mathrm{i} \bm{k} \cdot \bm{\delta}_{13_j}} - \sum_{j = 4}^6 \mathrm{e}^{\mathrm{i} \bm{k} \cdot \bm{\delta}_{13_j}} \right) \label{eq:delta} .
\end{align}
\end{subequations}
We can read off from these components that the nonrelativistic dispersion relation does not depend on the Néel vector orientation, as expected for the SO(3) symmetric Heisenberg model. In Eq.~\eqref{eq:nonrel_componenets}, we define the distances between neighbors %\am{Rhea, now that we have changed the symbols for the symmetry elements in the Table in the Appendix, we also have to change them here.}
\begin{subequations}
    \begin{align}
        \bm{\delta}_1 &=  \bm{r}_\text{B}^\text{Fe} - \bm{r}_\text{A}^\text{Fe}, 
    \end{align}
    \begin{align}
        \bm{\delta}_{2_1} & = \left(\bm{r}_\text{D}^\text{Fe} - \bm{r}_\text{A}^\text{Fe} \right) - \bm{a}_2 - \bm{a}_3, \nonumber \\
         \bm{\delta}_{2_2}  &= \mathcal{C}_{3} \bm{\delta}_{2_1}, \quad \bm{\delta}_{2_3}  = \mathcal{C}_{3}^{-1} \bm{\delta}_{2_1}, 
    \end{align}
    \begin{align}
        \bm{\delta}_{3_1} & = \left(\bm{r}_\text{B}^\text{Fe} - \bm{r}_\text{A}^\text{Fe} \right) - \bm{a}_3, \nonumber \\
         \bm{\delta}_{3_2} & = \mathcal{C}_{3} \bm{\delta}_{3_1}, \quad
        \bm{\delta}_{3_3} =\mathcal{C}_{3}^{-1} \bm{\delta}_{3_1}, 
    \end{align}
    \begin{align}
        \bm{\delta}_{4_1} & = \left(\bm{r}_\text{C}^\text{Fe} - \bm{r}_\text{A}^\text{Fe} \right) - \bm{a}_2 - \bm{a}_3, \nonumber \\
         \bm{\delta}_{4_2} & = \mathcal{C}_{3} \bm{\delta}_{4_1}, \quad
        \bm{\delta}_{4_3}  =\mathcal{C}_{3}^{-1} \bm{\delta}_{4_1}, 
        \label{eq:delta_4}
    \end{align}
    \begin{align}
        \bm{\delta}_5 &=  \bm{r}_\text{B}^\text{Fe} - \bm{r}_\text{C}^\text{Fe}, 
    \end{align}
    \begin{align}
        \bm{\delta}_{13_1} & = \left(\bm{r}_\text{D}^\text{Fe} - \bm{r}_\text{A}^\text{Fe} \right) - \bm{a}_2 - 2 \bm{a}_3, \quad 
         \bm{\delta}_{13_2} = \mathcal{C}_{3} \bm{\delta}_{13_1}, \nonumber \\
        \bm{\delta}_{13_3}  &=\mathcal{C}_{3}^{-1} \bm{\delta}_{2_1}, \quad  \bm{\delta}_{13_4}  =\mathcal{M}_{\overline{1}10, \bm{\tau}} \bm{\delta}_{13_1}, \nonumber \\
        \bm{\delta}_{13_5}  &=\mathcal{M}_{\overline{1}01, \bm{\tau}} \bm{\delta}_{13_1}, \quad \bm{\delta}_{13_6} = \mathcal{M}_{0\overline{1}1, \bm{\tau}}  \bm{\delta}_{13_1}.
    \end{align}
\end{subequations}

\begin{figure*}
    \centering       
    \includegraphics[width=\textwidth]{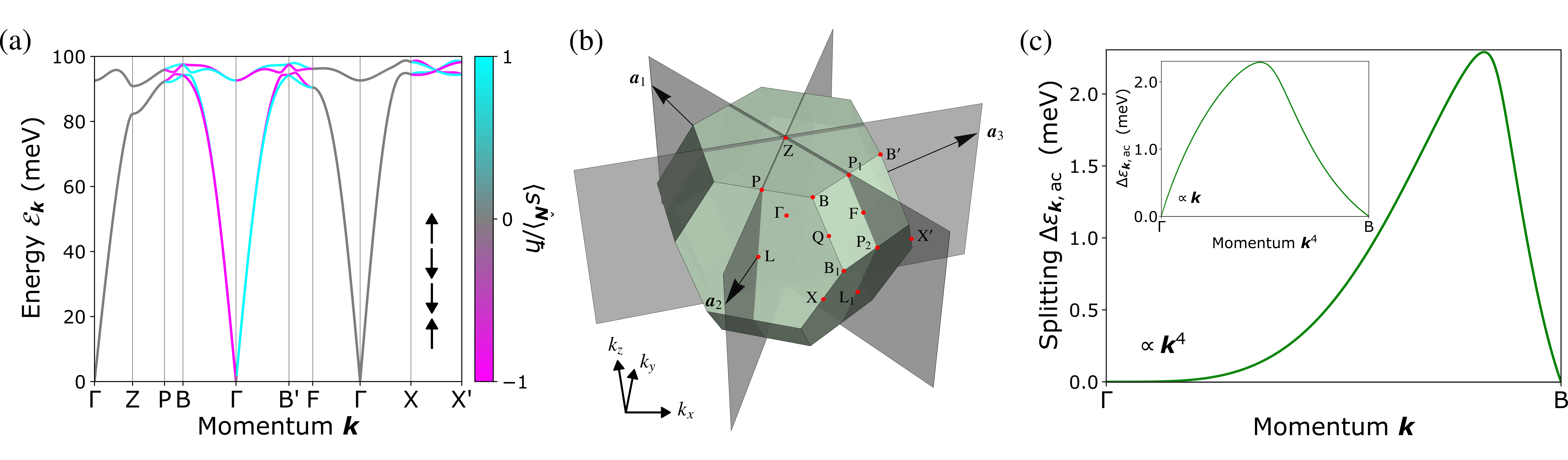}
       \caption{
       (a) Magnon dispersion relation along a path connecting high-symmetry points in the first Brillouin zone of hematite. The color bar shows the magnon spin expectation value $\langle s^{\hat{\bm{N}}} \rangle$ along the Néel vector direction. The parameters used are tabulated in Table~\ref{tab:params_nonrel}.  (b) First Brillouin zone of hematite with indicated high-symmetry points, as well as three gray-colored nodal planes. The fourth nodal \textit{surface} is not drawn, but cuts the Brillouin zone horizontally.  (c) Splitting $\Delta \varepsilon _{\bm{k},\text{ac}}$ of the acoustic magnon modes along $\Gamma$-B. Close to $\Gamma$, the splitting is proportional to $\bm{k}^4$.
       }
       \label{fig:disperion_nonrelativistic}
\end{figure*}

We further write
\begin{subequations}
 \begin{align}
    \mathcal{H}_2 &= \mathcal{H}_2^{\text{ISO}} + \mathcal{H}_2^\text{AM}, \\
	 & = \frac{1}{2} \sum_{\bm{k}} \bm{\Psi}_{\bm{k}}^{\dagger} \left(H_{\bm{k}}^\text{ISO} + H_{\bm{k}}^\text{AM} \right) \bm{\Psi}_{\bm{k}} , \\
    & = \frac{1}{2} \sum_{\bm{k}} \bm{\Psi}_{\bm{k}}^{\dagger} H_{\bm{k}} \bm{\Psi}_{\bm{k}},
\end{align}
\end{subequations}
with 
\begin{align}
\bm{\Psi}_{\bm{k}}^{\dagger} = \left(a_{\bm{k}}^{\dagger}, b_{\bm{k}}^{\dagger}, c_{\bm{k}}^{\dagger}, d_{\bm{k}}^{\dagger}, a_{-\bm{k}}, b_{-\bm{k}}, c_{-\bm{k}}, d_{-\bm{k}}\right).
\label{eq:psi}
\end{align}

The Hamilton kernel $H_{\bm{k}}$ reads
\begin{widetext}
\begin{align} 
   H_{\bm{k}} = H_{\bm{k}}^\text{ISO} + H_{\bm{k}}^\text{AM}
   = \begin{pmatrix}
     \mathcal{A}_{\bm{k}}^{\text{ISO}} & 0 & 0 &  \mathcal{D}_{\bm{k}}  & 0 & \mathcal{B}_{\bm{k}}  & \mathcal{C}_{\bm{k}} & 0 \\[4pt]
     0 & \mathcal{A}_{\bm{k}}^{\text{ISO}} &  \mathcal{D}_{\bm{k}}^*  & 0 & \mathcal{B}_{\bm{k}}^* & 0 & 0 & \mathcal{C}_{\bm{k}} \\[4pt]
      0 & \mathcal{D}_{\bm{k}}   &  \mathcal{A}_{\bm{k}}^{\text{ISO}} &  0 & \mathcal{C}_{\bm{k}} & 0 & 0 & \mathcal{B}_{\bm{k}} \\[4pt]
      \mathcal{D}_{\bm{k}}^* & 0 & 0  &  \mathcal{A}_{\bm{k}}^{\text{ISO}} & 0 & \mathcal{C}_{\bm{k}} & \mathcal{B}_{\bm{k}}^* & 0 \\[4pt]
    0 & \mathcal{B}_{\bm{k}}  & \mathcal{C}_{\bm{k}} & 0 & \mathcal{A}_{\bm{k}}^{\text{ISO}} & 0 & 0 &  \mathcal{D}_{\bm{k}} \\[4pt]
    \mathcal{B}_{\bm{k}}^*  & 0 & 0 & \mathcal{C}_{\bm{k}}  & 0 & \mathcal{A}_{\bm{k}}^{\text{ISO}} &  \mathcal{D}_{\bm{k}}^* & 0 \\[4pt]
     \mathcal{C}_{\bm{k}}  & 0 & 0 & \mathcal{B}_{\bm{k}}  & 0 &  \mathcal{D}_{\bm{k}}  & \mathcal{A}_{\bm{k}}^{\text{ISO}} & 0 \\[4pt]
     0 & \mathcal{C}_{\bm{k}}  & \mathcal{B}_{\bm{k}}^* & 0 &  \mathcal{D}_{\bm{k}}^*  & 0 & 0 & \mathcal{A}_{\bm{k}}^{\text{ISO}} 
    \end{pmatrix} 
    + 
    \begin{pmatrix}
     0 & 0 & 0 &   \Delta_{\bm{k}} & 0 & 0  & 0 & 0 \\[4pt]
     0 & 0 &   \Delta_{\bm{k}}^* & 0 & 0 & 0 & 0 & 0 \\[4pt]
      0 &  \Delta_{\bm{k}}  &  0 &  0 & 0 & 0 & 0 & 0\\[4pt]
       \Delta_{\bm{k}}^* & 0 & 0  &  0 & 0 & 0 & 0 & 0 \\[4pt]
    0 & 0  & 0 & 0 & 0 & 0 & 0 &   \Delta_{\bm{k}} \\[4pt]
    0  & 0 & 0 & 0 & 0 & 0 &  \Delta_{\bm{k}}^* & 0 \\[4pt]
     0  & 0 & 0 & 0 & 0 & \Delta_{\bm{k}} & 0 & 0 \\[4pt]
     0 & 0 & 0 & 0 &   \Delta_{\bm{k}}^* & 0 & 0 & 0 
    \end{pmatrix}.
    \label{eq:ham_kernel_nonrel}
\end{align}   
\end{widetext}
We note in passing that the kernel can be rearranged into a block-diagonal form with the block bases $\widetilde{\bm{\Psi}}_{\bm{k}, 1}^{\dagger} = \left(a_{\bm{k}}^{\dagger}, d_{\bm{k}}^{\dagger},b_{-\bm{k}}, c_{-\bm{k}}\right)$, and $\widetilde{\bm{\Psi}}_{\bm{k}, 2}^{\dagger} = \left(b_{\bm{k}}^{\dagger}, c_{\bm{k}}^{\dagger}, a_{-\bm{k}}, d_{-\bm{k}}\right)$, signaling 
%\am{No, nothing in spurious here. The Hamiltonian has SO(3), which break down to SO(2) due to spontaneous long-range order in the ground state. The SO(2) makes the spin along the order a good quantum number. Magnon-magnon interactions will not change that. Only when you have the DMI in the system, the Hamiltonian is no longer SO(3) symmetric and you would expect spin not be a good quantum number. Then, $H_2$ still is block diagonal but that is a spurious symmetry of $H_2$ and magnon-magnon interactions would correct it.} 
spin conservation in the system, making spin (or the magnon ``chirality'' or ``polarization'') a good quantum number. 
However, we keep the $\bm{\Psi}_{\bm{k}}^{\dagger}$ basis and perform a paraunitary Bogoliubov transformation \cite{COLPA1978327}
\begin{align}
    \mathcal{E}_{\bm{k}} & \equiv T_{\bm{k}}^\dagger H_{\bm{k}} T_{\bm{k}} \nonumber \\
    &= \diag (\varepsilon_{1, \bm{k}}, \varepsilon_{2, \bm{k}}, \varepsilon_{3, \bm{k}}, \varepsilon_{4, \bm{k}}, \varepsilon_{1, -\bm{k}}, \varepsilon_{2, -\bm{k}}, \varepsilon_{3, -\bm{k}}, \varepsilon_{4, -\bm{k}}),
\end{align}
such that
\begin{align}
    \mathcal{H}_2 = \sum_{\bm{k}} \sum_{n = 1}^4 \varepsilon_{n\bm{k}} \left( \beta^\dagger_{n\bm{k}} \beta_{n\bm{k}} + \frac{1}{2} \right),
\end{align}
where $\beta_{n\bm{k}}$ denote the magnon normal modes with energies $\varepsilon_{n\bm{k}}$.
The paraunitary matrix $T_{\bm{k}}$ obeys \cite{COLPA1978327}
\begin{subequations}
\begin{align}
    T_{\bm{k}}^\dagger G T_{\bm{k}} &= G, \\
    G &= \diag (1, 1, 1, 1, -1, -1, -1, -1).
\end{align}
\end{subequations}

\begin{figure}[H]
    \centering          
    \includegraphics[width=\columnwidth]{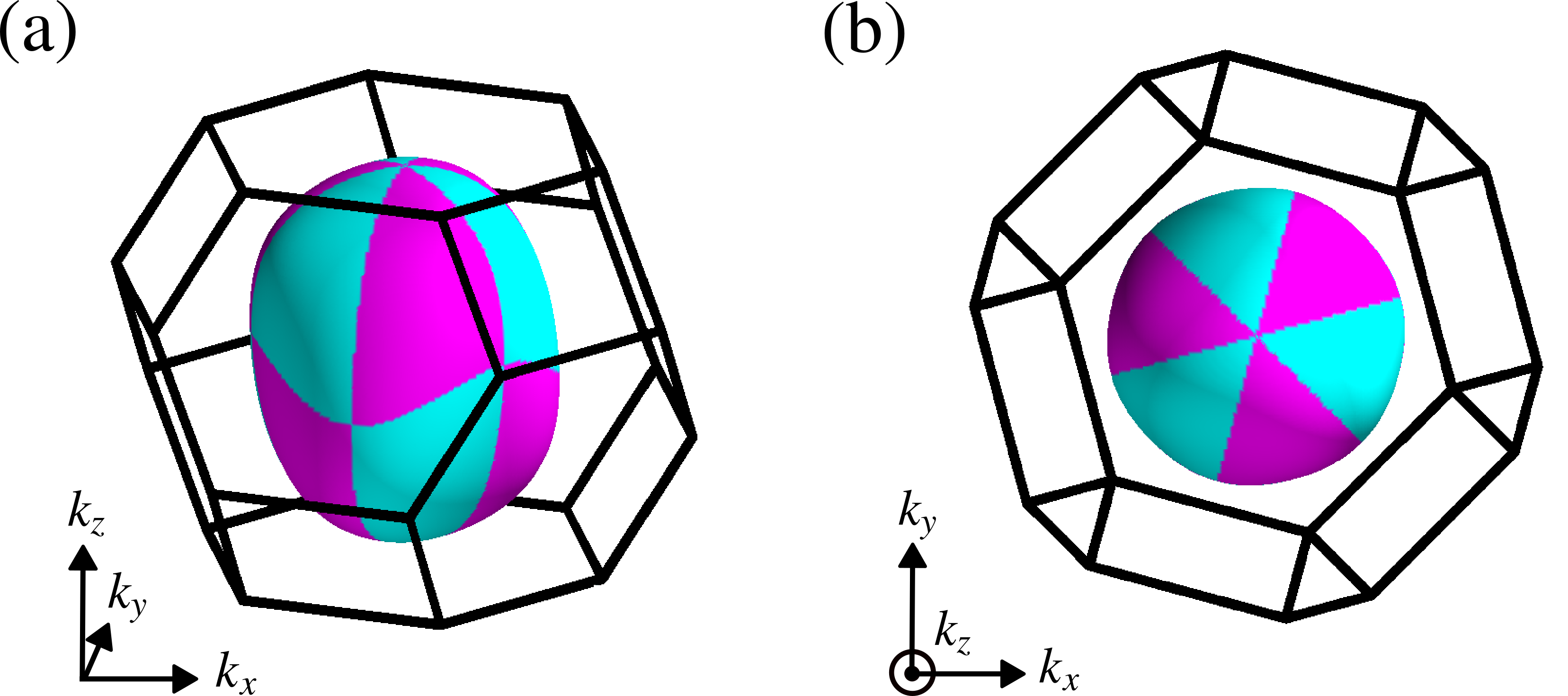} %trim={0.8cm 1.3cm 2cm 1.5cm},clip,
       \caption{
       Isoenergy surface of the acoustic magnon bands at $80$ meV in hematite showing the bulk $g$-wave spin splitting, viewed from the side (a) and from the top (b). The colors indicate the magnon spin expectation value and are chosen according to the color bar in Fig.~\ref{fig:disperion_nonrelativistic}(a). The regions of spin split magnons are clearly visible, as well as the nodal surface cutting the Brillouin zone horizontally in (a), and the nodal planes cutting the Brillouin zone vertically in (b).
       }
       \label{fig:iso_nonrel_surf}
\end{figure}

\begin{figure*}
    \centering          
    \includegraphics[width=11cm]{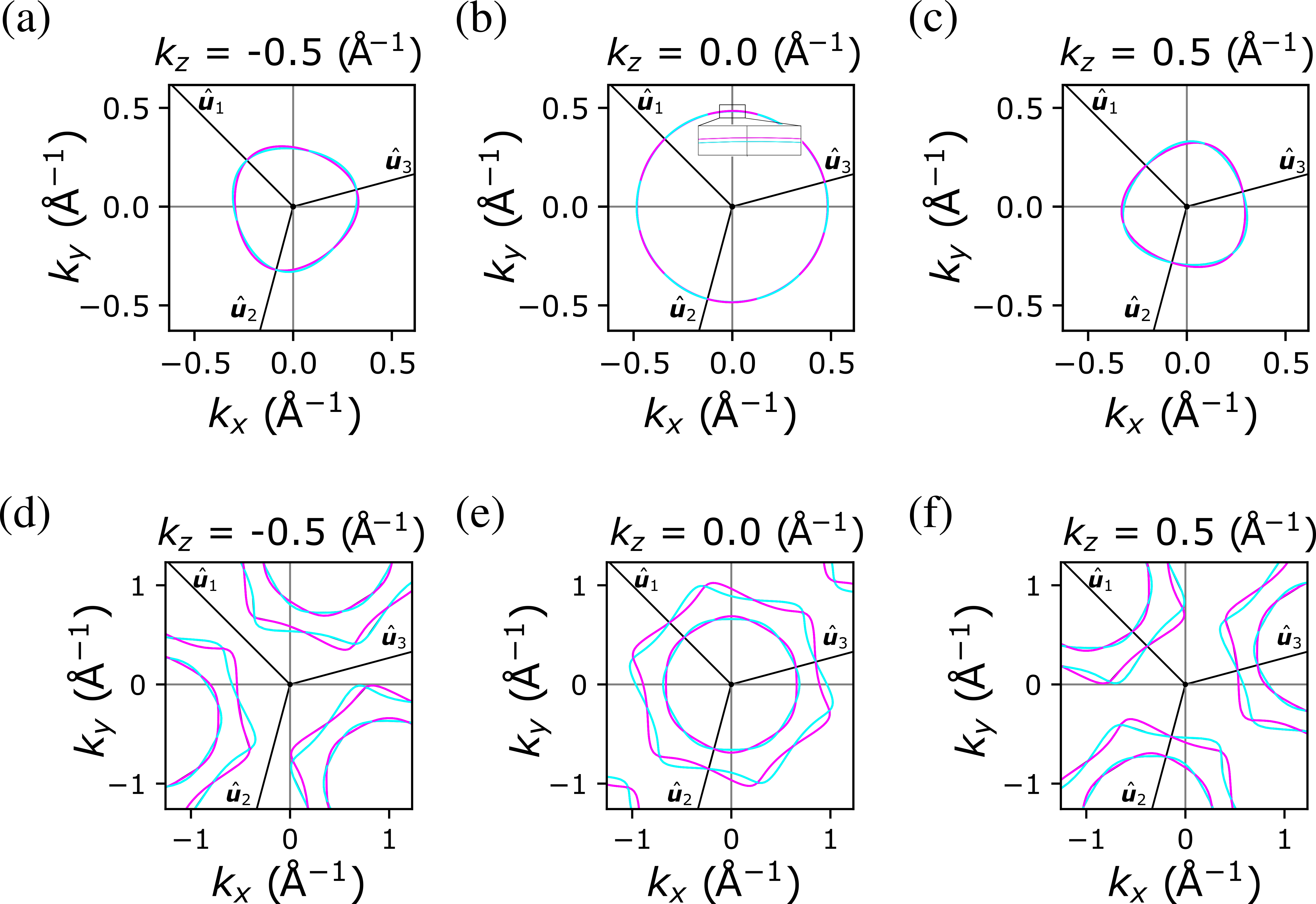}
       \caption{
       Isoenergy cuts at constant-$k_z$ planes of the magnon bands in hematite.  (a-c) Cuts at $80$ meV---which can be considered horizontal slices through the isoenergy surface in Fig.~\ref{fig:iso_nonrel_surf} at $k_z = -0.5$ $\angstr^{-1}$  (a), $k_z = 0.0$ $\angstr^{-1}$ (b), and $k_z=0.5$ $\angstr^{-1}$(c). The inset in (d) shows that not the entire $k_z = 0.0$ $\angstr^{-1}$ plane is degenerate.  (d-f) Cuts at $93$ meV showing the acoustic and optical magnon bands at $k_z = -0.5$ $\angstr^{-1}$ (d), $k_z = 0.0$ $\angstr^{-1}$ (e), and $k_z=0.5$ $\angstr^{-1}$ (f) within and beyond the first Brillouin zone.
       The colors indicate the magnon spin expectation value and are chosen according to the color bar in Fig.~\ref{fig:disperion_nonrelativistic}(a).
       }
       \label{fig:iso_nonrel}
\end{figure*}

We show the magnon dispersion relation in Fig.~\ref{fig:disperion_nonrelativistic}(a) along a path connecting high-symmetry points in the first Brillouin zone (BZ) [cf.~Fig.~\ref{fig:disperion_nonrelativistic}(b)]. Along a path lying in one of the three gray-colored nodal planes [cf. Fig.~\ref{fig:disperion_nonrelativistic}(b)], e.g, $\Gamma$-Z-P or F-$\Gamma$, or in the fourth nodal \textit{surface} 
%\am{If this is the first time the nodal surface is mentioned, it is a bit too sudden. We would need to write a precursor somewhere that altermagnets have nodal planes/surfaces and would have to say right away that hematite has both.} 
[cf. Section~\ref{sec:am_properties}, not shown in Fig.~\ref{fig:disperion_nonrelativistic}(b)], e.g., $\Gamma$-X, the two acoustic and the two optical magnon bands are
%doubly \am{I think ``doubly'' is too much because of two modes anyway.}
degenerate, respectively. The color reflects the spin expectation value along the Néel vector direction $\langle s^{\hat{\bm{N}}} \rangle$. As spin is a good quantum number, the magnon bands carry a momentum-independent spin expectation value (either magenta or cyan).
%, which is zero for degenerate bands. \am{Strictly speaking, it is not zero because due to the block diagonal nature, the two blocks do not know of each other and the spin is a good quantum number despite the degeneracy. I would write: ``
We emphasize that spin is a good quantum number  also along the direction with degenerate bands, but we use gray color to indicate the degeneracy. The paths $\Gamma$-Z and F-$\Gamma$-X in Fig.~\ref{fig:disperion_nonrelativistic}(a) reproduce the inelastic neutron scattering data in Fig.~4 (bottom panel) of Ref.~\cite{samuelsenInelasticNeutronScattering1970}, which also shows doubly degenerate bands. The paths described in Ref.~\cite{samuelsenInelasticNeutronScattering1970}, which are labeled there $\Sigma$-$\Gamma$-D and $\Gamma$-Z, correspond to our spin degenerate paths X-$\Gamma$-F and $\Gamma$-Z, respectively.
 
Along a path that is not in a nodal plane or surface, e.g., P-B-$\Gamma$-B$'$-F, or X-X$'$, the acoustic and optical magnon bands are spin (or chirality) split, as we expect from magnons in altermagnets \cite{Naka2019, SmejkalBeyondConv2022, SmejkalChiralMagnons2023}. The splitting 
\begin{align}
    \Delta \varepsilon _{\bm{k},\text{ac}} \equiv \varepsilon_{3,\bm k} - \varepsilon_{4,\bm k}
    \label{eq:splitting_ac}
\end{align}
of the acoustic magnon bands along $\Gamma$-B is proportional to $\bm{k}^4$ near $\Gamma$, as shown in Fig.~\ref{fig:disperion_nonrelativistic}(c). This long-wavelength splitting is expected for $g$-wave altermagnets \cite{SmejkalBeyondConv2022}.

The path X-X$'$ lies in the $k_z = 0$ $\angstr^{-1}$ plane and we see spin splitting of the magnon bands with a nodal point in the middle [cf.~Fig.~\ref{fig:disperion_nonrelativistic}(a)]. This nodal point is in one of the vertical nodal planes [cf.~Fig.~\ref{fig:disperion_nonrelativistic}(b)]. The spin splitting displays the nonplanar nature of the fourth nodal surface. If it was planar, this path would be degenerate. Another way to demonstrate the shape of the nodal surface is to study an isoenergy surface of the magnon bands. In Fig.~\ref{fig:iso_nonrel_surf}, we show the three-dimensional acoustic magnon isoenergy surface at $80$ meV. %The bulk $g$-wave nature of hematite becomes apparent from the four nodal surfaces with alternating spin polarization of the magnon bands in the sectors between them \cite{SmejkalBeyondConv2022}. 
The bulk $g$-wave nature of hematite becomes apparent from the four nodal surfaces with alternating magnon spin in the sectors between them \cite{SmejkalBeyondConv2022}. The colors represent the magnon spin expectation value along the Néel vector direction [according to the color bar in Fig.~\ref{fig:disperion_nonrelativistic}(a)]. In Fig.~\ref{fig:iso_nonrel_surf}(a), we see the isoenergy surface from the side, which reveals the nodal surface cutting the BZ horizontally, whereas the top view in Fig.~\ref{fig:iso_nonrel_surf}(b) makes the three vertical nodal planes apparent. The vertical nodal planes are symmetry-protected by the three glide-mirrors in the system, whereas the horizontal nodal surface is not. What leads to the nodal surface are the two-fold rotations $\mathcal{C}_{2\langle 0\overline{1}1 \rangle, \bm{\tau}}$,$\mathcal{C}_{2\langle  \overline{1}01 \rangle, \bm{\tau}}$, and $\mathcal{C}_{2\langle  \overline{1}10 \rangle, \bm{\tau}}$; the nodal surface is therefore not forced to be planar \cite{SmejkalBeyondConv2022}.  Note that this observation sets hematite apart from MnTe \cite{Liu2024MnTeMagnonSplit}. 

We show different constant-$k_z$ cuts of the isoenergy surface at $80$ meV in Fig.~\ref{fig:iso_nonrel}(a-c), and at $93$ meV in Fig.~\ref{fig:iso_nonrel}(d-f). In all panels we indicate the in-plane components of the lattice vectors $\hat{\bm{u}}_i$, with $i \in \{1, 2, 3\}$ in black.
%In Fig.~\ref{fig:iso_nonrel}(a)-(c), we show different constant-$k_z$ cuts of the isoenergy surface. 
The isoenergy lines in the $k_z = -0.5$ $\angstr^{-1}$ cut in Fig.~\ref{fig:iso_nonrel}(a) reveal the $\mathcal{C}_3$ symmetry of the system. The nodal points are in accordance with the nodal planes shown in Fig.~\ref{fig:disperion_nonrelativistic}(b). In Fig.~\ref{fig:iso_nonrel}(b), we show a cut at $k_z = 0.0$ $\angstr^{-1}$, where the inset shows a spin split region along this isoenergy line, revealing that not the whole plane of $k_z = 0.0$ $\angstr^{-1}$ is spin degenerate. Again, this shows the nonplanar shape of the horizontal nodal surface. In Fig.~\ref{fig:iso_nonrel}(c), we show a cut at $k_z = 0.5$ $\angstr^{-1}$, which, if we compare with Fig.~\ref{fig:iso_nonrel}(a), displays the inversion symmetry in the system. The cuts at $93$ meV in Fig.~\ref{fig:iso_nonrel}(d-f)
%, ...} In Fig.~\ref{fig:iso_nonrel}(d)-(f) we show cuts at $k_z=-0.5$ (d), $k_z=0.0$ (e), and $k_z=0.5$ (f), 
additionally reveal the optical magnon bands together with the acoustic ones in the first BZ, extending into the second BZ. The splitting of the magnon bands is more pronounced than in Fig.~\ref{fig:iso_nonrel}(a-c), as expected from Fig.~\ref{fig:disperion_nonrelativistic}(a). 
%Because of this bigger splitting at this energy level, it \am{
%\am{Maybe drop because it comes again in the Discussion:} This splitting is likely to be resolved in inelastic neutron scattering. Specifically the chirality, or spin polarization of the modes can be made visible with polarized neutron scattering, as Ref.~\cite{McClartyAltermagNeutrons2025} predicts.

%In Fig.~\ref{fig:iso_nonrel}(e) we cut the isoenergy surface at $k_z=0.0$ and the nonplanar nature of the horizontal nodal surface is clearly showcased. In Fig.~\ref{fig:iso_nonrel}(f) the cut is made at $k_z=0.5$ highlighting the $\mathcal{C}_3$ symmetry and inversion in the system. 

%%%%%%%%%%%%%%%%%%%%%%%%%%%%%%%%%%%%%%%%%%%%%%%%%%%%%%%%%%%%%%%%%%%%
%
% Relativistic physics
%
%%%%%%%%%%%%%%%%%%%%%%%%%%%%%%%%%%%%%%%%%%%%%%%%%%%%%%%%%%%%%%%%%%%%
\section{Relativistic physics}
\label{sec:rel}

%\subsection{Hamiltonian}
%We now delve into relativistic physics and look into the easy-axis phase (EAP), which is the ground state \am{Careful: Ground state implies $T=0$.} The easy-axis single-ion anisotropy (SIA) dominates over the Dzyaloshinskii-Moriya interaction (DMI), the two-ion anisotropy, and the dipole-dipole interaction, ordering the moments out-of-plane. \am{You already give here the info what is the case and then revert back to having a look at what is symmetry allowed. I would do symmetry first to motivate what you neglect.} 

We now delve into relativistic physics and investigate the symmetry of the full exchange Hamiltonian $\mathcal{H}^{\text{EX}}$ for the first five nearest neighbor bonds. The full exchange Hamiltonian can be written as
\begin{align}
   \mathcal{H}^{\text{EX}} = \frac{1}{2} \sum_{i,j} \bm{S}_i^\text{T} \mathcal{J}_{i,j} \bm{S}_j,
\end{align}
where the exchange matrix $\mathcal{J}_{i,j}$ encompasses the isotropic Heisenberg exchange $J_{i,j}$, the antisymmetric exchange (DMI) $\bm{D}_{i,j}$, and the symmetric exchange $\Gamma_{i,j}$. The symmetries of the space group (cf. Tab.~\ref{tab:symmetries}) dictate the shape of the exchange matrices. Selecting a representative bond for the first five nearest-neighbor shells, we find 
\begin{subequations}
\begin{align}
        \mathcal{J}_{1} &= \begin{pmatrix}
            J_1^x & D_1^z & 0 \\[4pt]
            -D_1^z & J_1^x & 0 \\[4pt]
            0& 0& J_1^z
        \end{pmatrix}, \label{eq:calJ_1} \\
         \mathcal{J}_{2} &= \begin{pmatrix}
            J_2^x & \Gamma_2^z & \Gamma_2^y \\[4pt]
            \Gamma_2^z & J_2^y &  \Gamma_2^x \\[4pt]
             \Gamma_2^y & \Gamma_2^x & J_2^z
        \end{pmatrix}, \label{eq:calJ_2}
    \end{align}
\begin{widetext}
    \begin{align}
        \mathcal{J}_{3} &= \begin{pmatrix}
            J_3^x & D_3^z + \Gamma_3^z & \left(-2 + \sqrt{3} \right) D_3^x + \left( 2 + \sqrt{3}\right) \Gamma_3^x \\[4pt]
            -D_3^z + \Gamma_3^z & J_3^x - 2 \sqrt{3}  \Gamma_3^z &  D_3^x + \Gamma_3^x \\[4pt]
             \left(2 - \sqrt{3} \right) D_3^x + \left( 2 + \sqrt{3}\right) \Gamma_3^x  & -D_3^x + \Gamma_3^x & J_3^z
        \end{pmatrix}, \label{eq:calJ_3}
    \end{align}
\end{widetext}
    \begin{align}
         \mathcal{J}_{4} &= \begin{pmatrix}
            J_4^x & D_4^z + \Gamma_4^z & - D_4^y + \Gamma_4^y \\[4pt]
            -D_4^z + \Gamma_4^z & J_4^y  &  D_4^x + \Gamma_4^x \\[4pt]
             D_4^y + \Gamma_4^y & -D_4^x + \Gamma_4^x & J_4^z
        \end{pmatrix}, \label{eq:calJ_4}
    \end{align}
    \begin{align}
         \mathcal{J}_{5} &= \begin{pmatrix}
            J_5^x &0 & 0 \\[4pt]
            0 & J_5^x  &  0\\[4pt]
            0 & 0 & J_5^z
        \end{pmatrix}. \label{eq:calJ_5}
\end{align}
\end{subequations}
%\rh{which agrees with Ref.~\cite{karaki2023materialSearch} (except $\mathcal{J}_3$). \textbf{Comment:} Here, the difference is that the paper uses a specific in-plane setting, which then has the magnetic space group C2'/c', which has only one glide-mirror left. The symmetries are therefore different. This is also why their Wyckoff positions are different (depends on the space group).}

The rich structure of these general symmetry-allowed matrices reflects the low symmetry of the bonds. The first and fifth bonds have the highest symmetry, allowing only for XXZ-type interactions and, in case of the first bond, for one component of the DMI. 
The midpoint of the second bond is a center of inversion and hence, the second bond does not support DMI, but only the traceless symmetric $\Gamma$ terms.
The third and fourth bonds allow for both the DMI and the $\Gamma$ terms.

\begin{figure}
    \centering          
    \includegraphics[width=\columnwidth]{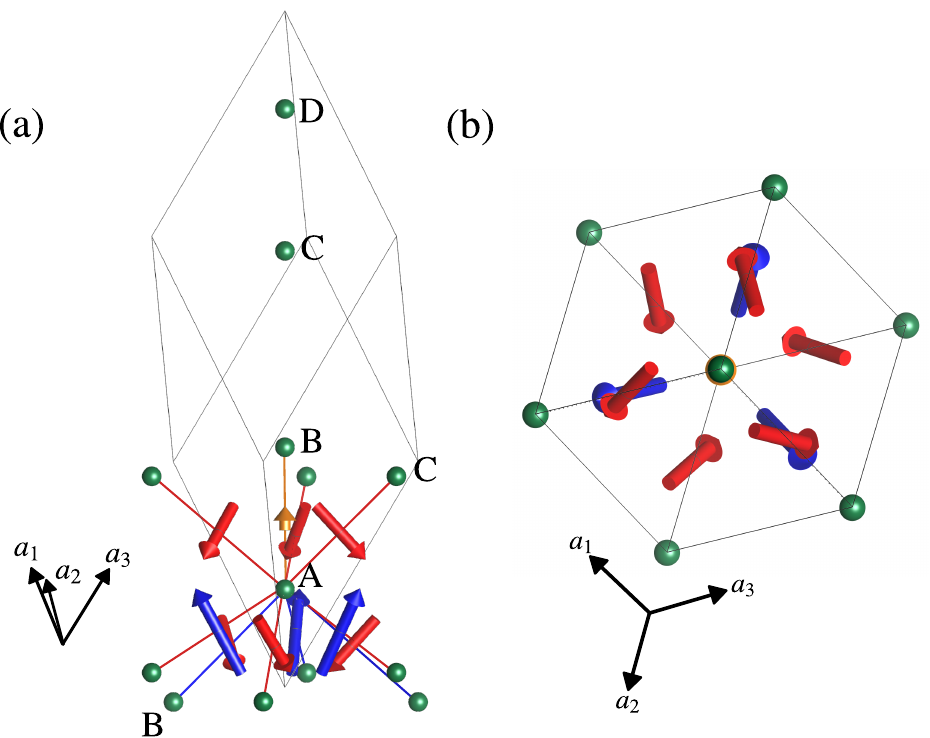}
       \caption{
       Dzyaloshinskii-Moriya interaction (DMI) vectors for the first, third and fourth nearest neighbor shell, defined from iron atom A outwards. (a) Side view with the orange/blue/red bonds connecting first/third/fourth nearest neighbors. (b) View from above along the $c$ axis.
       }
       \label{fig:DMI_vecs}
\end{figure}

Hematite is known to have small relativistic effects, as also confirmed by DFT calculations \cite{danneggerMagneticPropertiesHematite2023}. The DMI is derived in first order from SOC and is known to be important for the canting in the weak-ferromagnetic phase. We therefore build the symmetry-allowed local DMI vectors for the first four neighbor shells according to the rules in Ref.~\cite{MoriyaAnisotropicSuperex1960}. As an example, we consider DMI unit vectors from iron atom A outwards to the first, third and fourth nearest neighbor bonds in Fig.~\ref{fig:DMI_vecs}. According to the rules in Ref.~\cite{MoriyaAnisotropicSuperex1960}, we find that $\bm{D}^{(1)}$, shown along the orange bond in Fig.~\ref{fig:DMI_vecs}(a), only has a $z$ component. This is in accordance with Eq.~\eqref{eq:calJ_1}. The midpoint between iron atoms A and D is a center of inversion, hence $\bm{D}^{(2)} \equiv 0$, which agrees with Eq.~\eqref{eq:calJ_2}. For the third and fourth nearest neighbor shell, we construct the DMI vector directions by assuming that the DMI is mediated only through superexchange with the nearest oxygen atom. This approximation likely does not paint the full picture, but is sufficient to capture all relevant effects. The result for $\bm{D}^{(3)}_{\bm{\delta}_{3_1}}$ does agree with the restriction in Eq.~\eqref{eq:calJ_3}. %and with the DMI vectors found in Ref.~\cite{danneggerMagneticPropertiesHematite2023}. 
These third-neighbor shell DMI vectors are shown along the blue bonds in Fig.~\ref{fig:DMI_vecs}(a,b). For the fourth nearest neighbor shell, there are no restrictions on the shape of the DMI vector in Eq.~\eqref{eq:calJ_4}, hence, the microscopic environment is of importance. 
From these unit vectors, we extract their azimuthal and polar angles $\phi_{\text{D}_i}$ and $\theta_{\text{D}_i}$, respectively. We base the parameters $D_1, D_3, D_4$ (tabulated in Tab.~\ref{tab:dmi}) on the \textit{ab initio} calculations of Ref.~\cite{danneggerMagneticPropertiesHematite2023}, which provide the $z$ component of the net DMI vector per neighbor shell. We take these values $D_{1,\text{ref}}^z$, $D_{3,\text{ref}}^z$, and $D_{4,\text{ref}}^z$ [from Fig.~3(b) in Ref.~\cite{danneggerMagneticPropertiesHematite2023}] and divide by the number of bonds in each shell to get the $z$ component per bond. We then solve the equality $D^{z,\text{ref}}_i = D_i \cos \theta_{\text{D}_i}$ for $D_i$. We now find $\bm{D}_i = D_i \left(\sin \theta_{\text{D}_i} \cos \phi_{\text{D}_i},  \sin \theta_{\text{D}_i} \sin \phi_{\text{D}_i},  \cos \theta_{\text{D}_i} \right)$ for one DMI vector per neighbor shell and generate the rest from the matrix representations of the symmetry elements $\mathcal{C}_3$ and $\mathcal{C}_3^{-1}$ in Tab.~\ref{tab:symmetries} [cf. Eqs.~\eqref{eq:dmi}]. 
To fit the found parameters $D_1, D_3, D_4$ to the experimental canting angle in the weak ferromagnetic phase \cite{morrishCantedAntiferromagnetismHematite1994,hillNeutronDiffractionStudy2008}, we have to multiply them by a factor of approximately $12$.

%We use the parameters extracted from Fig.~(3) of Ref.~\cite{danneggerMagneticPropertiesHematite2023} to find their ratio per bond \mkch{$D_{1,\text{ref}}^z:D_{3,\text{ref}}^z/3:D_{4,\text{ref}}^z/6$ to be approximately $0.013:1:-0.45$.} We then convert those parameters to our convention and, while \mkch{roughly} keeping their ratio \mkch{($D_1: D_3: D_4 = 0.012:1:-0.53$)}, fit them to the experimental canting angle in the weak ferromagnetic phase \cite{morrishCantedAntiferromagnetismHematite1994,hillNeutronDiffractionStudy2008}. 

%$0.0045:1:-1.02$. We then convert those parameters to our convention and, while keeping their ratio, fit them to the experimental canting angle in the weak ferromagnetic phase \cite{morrishCantedAntiferromagnetismHematite1994,hillNeutronDiffractionStudy2008}. 

The approximation to restrict ourselves to the DMI mediated by superexchange from the nearest oxygen atom results in DMI vectors along the red vectors in Fig.~\ref{fig:DMI_vecs}(a,b), which show a slight difference from the DMI vectors predicted by \textit{ab initio} methods in Ref.~\cite{danneggerMagneticPropertiesHematite2023} (apart from a global sign change due to a difference in the definition of the Hamiltonian). However, the DMI vectors we find do not violate any rule in Ref.~\cite{MoriyaAnisotropicSuperex1960}. We therefore proceed with the found vectors and provide them here for reference:
\begin{subequations}
\label{eq:dmi}
\begin{align}
    \bm{D}^{(1)} & = D_1 \left( 0, 0,1 \right)^{\mathrm{T}}, 
\end{align}
\begin{align}
 \bm{D}^{(3)}_{\bm{\delta}_{3_1}}  &= \mathcal{C}_{3}^{-1} \bm{D}^{(3)}_{\bm{\delta}_{3_2}}, \nonumber \\
    \bm{D}^{(3)}_{\bm{\delta}_{3_2}} & = D_3 \left( -0.486952, -0.130478, 0.863628 \right)^{\mathrm{T}}, \nonumber \\
    \bm{D}^{(3)}_{\bm{\delta}_{3_3}} & = \mathcal{C}_{3} \bm{D}^{(3)}_{\bm{\delta}_{3_2}},  
\end{align}
and 
\begin{align}
 \bm{D}^{(4)}_{\bm{\delta}_{4_1}} & = \mathcal{C}_{3} \bm{D}^{(4)}_{\bm{\delta}_{4_3}}, \quad
    \bm{D}^{(4)}_{\bm{\delta}_{4_2}}  =\mathcal{C}_{3}^{-1}  \bm{D}^{(4)}_{\bm{\delta}_{4_3}}, \nonumber \\
    \bm{D}^{(4)}_{\bm{\delta}_{4_3}} & = D_4 \left(-0.49938, -0.46367, 0.73187 \right)^{\mathrm{T}}, \nonumber \\
    \bm{D}^{(4)}_{- \bm{\delta}_{4_1}} & = D_4 \left(0.46369, 0.49936, 0.73187 \right)^{\mathrm{T}}, \nonumber \\
    \bm{D}^{(4)}_{- \bm{\delta}_{4_2}}  & = \mathcal{C}_{3} \bm{D}^{(4)}_{- \bm{\delta}_{4_1}} , \quad
    \bm{D}^{(4)}_{-\bm{\delta}_{4_3}}  = \mathcal{C}_{3}^{-1} \bm{D}^{(4)}_{-\bm{\delta}_{4_1}},
    \label{eq:dmi_4} 
\end{align}
\end{subequations}

For Eq.~\eqref{eq:dmi_4} we emphasize that $\bm{D}^{(4)}_{\bm{\delta}_{4_i}}$ corresponds to the DMI vector associated with the bond $\bm{\delta}_{4_i}$ [cf. Eq.~\eqref{eq:delta_4}], and $\bm{D}^{(4)}_{-\bm{\delta}_{4_i}}$ is associated with the bond $-\bm{\delta}_{4_i}$. %Furthermore, we stress that applying the mirrors $\mathcal{M}_{\overline{1}01, \bm{\tau}}$, $\mathcal{M}_{0\overline{1}1, \bm{\tau}}$, and $\mathcal{M}_{\overline{1}10, \bm{\tau}}$ to the DMI vector $\bm{D}^{(4)}_{-\bm{\delta}_{4_3}}$ in Eq.~\eqref{eq:dmi_4} means that we simply multiply the corresponding matrix in Tab.~\ref{tab:symmetries} in Cartesian coordinates \mkch{by} the DMI vector $\bm{D}^{(4)}_{-\bm{\delta}_{4_3}}$, which we treat as a polar vector in this regard. The facts that \mkch{generally} DMI vectors are axial vectors and that the DMI is an antisymmetric exchange are already taken into account in this calculation.

%%%%%%%%%%%%%%%%%%%%%%%%%%
\begin{table}[b]
\centering
\caption{Dzyaloshinskii-Moriya interaction parameters extracted from Ref.~\cite{danneggerMagneticPropertiesHematite2023} and adapted to fit the canting angle in the weak ferromagnetic phase from experiments \cite{morrishCantedAntiferromagnetismHematite1994,hillNeutronDiffractionStudy2008}.}
\label{tab:dmi}
\begin{tabular}{ p{1.3cm}|p{1.8cm}|p{1.cm}} 
 \toprule
 \centering Parameter & \centering Value & \centering Unit
  \arraybackslash \\
 \hline
  
    \centering  $D_1$ & \centering 
    %$0.00025$ 
    %$0.0015$
    $0.003$
    & \centering meV
  \arraybackslash \\
 
 \centering $D_3$ & \centering 
 %$0.02171$ 
 %$0.13026$
 $0.261$
 & \centering meV
   \arraybackslash \\
 
 \centering $D_4$ & \centering 
 %$-0.01139$ 
 %$-0.06834$
 $-0.137$
 & \centering meV
  \arraybackslash  \\
  
 \botrule
\end{tabular}
\end{table}

The traceless symmetric exchange terms encoded in the $\Gamma_{i,j}$ terms are second order in SOC, and hence negligible. Additionally, anisotropies and dipolar interactions are known to be important in hematite \cite{danneggerMagneticPropertiesHematite2023}. We therefore implement the single-ion anisotropy (SIA) in the easy-axis phase, and, for simplicity, merge the two-ion anisotropy and the dipolar interactions into an effective easy-plane anisotropy in the weak ferromagnetic phase. We emphasize that our goal is not to develop a spin model that describes hematite across all temperatures, as was done in Ref.~\cite{danneggerMagneticPropertiesHematite2023}, as this would require inclusion of magnon-magnon interactions to capture the Morin transition. Instead, we aim to reproduce the qualitatively most important features of the harmonic magnon band structure in the two magnetic phases of hematite.

%\am{Say that relativistic effects are weak. Since Gamma terms are second order in SOC, we drop them. DMI is first order, and known to be important for the ordering of the material, so we keep it. Dipolar interactions and anisotropy are also known to be important, so we merge them into an effective anisotropy.}

%\am{Hematite is known to have small relativistic effects, ...}
%We know that the relativistic effects in hematite are in general very small, so we do not implement the $\Gamma$ matrices, 
%but merge the two-ion anisotropy and dipole-dipole interaction into an effective in-plane anisotropy affecting $J_i^z$, for $i \in \{1, 2, 3\}$, as we will discuss for the weak ferromagnetic phase (WFP) in Sec.~\ref{sec:WFP_anisotropy}.

%\am{When all this is discussed, say that you will build up the level of complexity, starting with EAP.}

%%%%%%%%%%%%%%%%%%%%%%%%%%%%%%%%%%%%%%%%%%%%%%%%%%%%%%%%%%%%%%%%%%%%
%
% Easy-axis phase
%
%%%%%%%%%%%%%%%%%%%%%%%%%%%%%%%%%%%%%%%%%%%%%%%%%%%%%%%%%%%%%%%%%%%%
\subsection{Easy-axis phase}
\label{sec:EAP}
In the easy-axis phase, we incorporate the single-ion anisotropy $ \mathcal{H}^\text{SIA}$, as well as the DMI $\mathcal{H}^\text{DMI}$ into the spin Hamiltonian 
\begin{align}
    \mathcal{H}' = \mathcal{H} + \mathcal{H}^\text{SIA} + \mathcal{H}^\text{DMI}, %= \mathcal{H}_{\text{ISO}} + \mathcal{H}_\text{AM} + \mathcal{H}_\text{SIA} + \mathcal{H}_\text{DMI},
\end{align}
where we define
\begin{subequations}
\begin{align}
   \mathcal{H}^\text{SIA} & = d_2 \sum_i \left(S_i^z\right)^2, \label{eq:anisoEAP}\\ 
   \mathcal{H}^\text{DMI} &= \sum_{r=1}^4 \sum_{\langle i, j \rangle_r } \bm{D}_{ij} \cdot  \left( \bm{S}_{i} \times \bm{S}_{j} \right),\label{eq:HDMI}
\end{align}
\end{subequations}
and take $\mathcal{H}$ as in Eq.~\eqref{eq:ham_EAP_iso_am}.
The effective single-ion anisotropy $d_2$ is fitted to match the spin wave gap $\Delta_{\text{gap,EAP}}$ in Ref.~\cite{chouHighResolutionTerahertzOptical2012} in the presence of DMI, so that $\Delta_{\text{gap,EAP}} = 0.1381$ meV [cf. Tab.~\ref{tab:easy_axis}]. We implement the DMI as shown in Sec.~\ref{sec:rel}.

%%%%%%%%%%%%%%%%%%%%%%%%%%
\begin{table}[b]
\centering
\caption{Magnitude of the effective single-ion anisotropy $d_2$. The parameter is made to fit the experimental spin wave gap in Ref.~\cite{chouHighResolutionTerahertzOptical2012} in the presence of the Dzyaloshinskii-Moriya interaction.}
\label{tab:easy_axis}
\begin{tabular}{ p{1.3cm}|p{1.8cm}|p{1.cm}} 
 \toprule
 \centering Parameter & \centering Value & \centering Unit
  \arraybackslash \\
 \hline % -0.00005490
  \centering $d_2$ & \centering $-5.490\cdot 10^{-5}$& \centering meV
  \arraybackslash \\
 \botrule
\end{tabular}
\end{table}

\begin{figure*}
    \centering       
    \includegraphics[width=14cm]{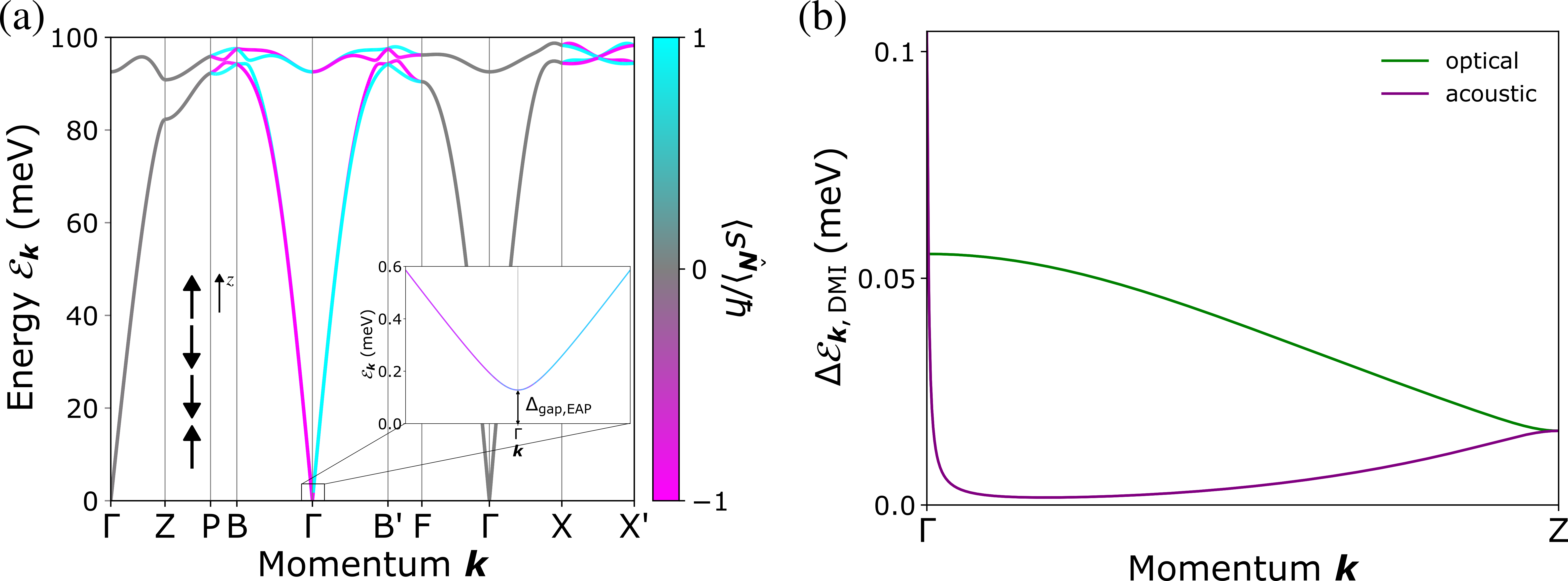}
       \caption{
       (a) Magnon dispersion relation in the easy-axis phase of hematite with Dzyaloshinskii-Moriya interaction (DMI). The color bar indicates the magnon spin expectation value $\langle s^{\hat{\bm{N}}} \rangle$. The inset shows the gap $\Delta_\text{gap,EAP}$ opened by the easy-axis anisotropy at $\Gamma$. Here, the modes are still doubly degenerate.  (b) Energy difference between the magnon dispersion relation in the easy-axis phase without DMI and with DMI, $\Delta \varepsilon_{\bm{k},\text{DMI}} = \varepsilon_{\bm{k}}(D_i = 0) - \varepsilon_{\bm{k}}(D_i)$, with $i \in \{1, 3, 4 \}$.
       The parameters used are listed in Tab.~\ref{tab:easy_axis} and \ref{tab:dmi}.
       }
       \label{fig:bands_easy_axis}
\end{figure*}

We write the harmonic Hamiltonian as
\begin{align}
	\mathcal{H}_2' & = \mathcal{H}_2 + \mathcal{H}_2^\text{SIA} + \mathcal{H}_2^\text{DMI}, \\
    & = \frac{1}{2} \sum_{\bm{k}} \bm{\Psi}_{\bm{k}}^{\dagger} H_{\bm{k}}' \bm{\Psi}_{\bm{k}},
\end{align}
where the new Hamilton kernel $H_{\bm{k}}'$ is given by
\begin{widetext}
\begin{align} 
    H_{\bm{k}}' = H_{\bm{k}} + \begin{pmatrix}
     \mathcal{A}_{\bm{k}}^{\text{SIA}} & 0 & 0 &  0 & 0 &  - \mathrm{i} \mathcal{B}'_{\bm{k}} &  - \mathrm{i} \mathcal{C}'_{\bm{k}}& 0 \\[4pt]
     0 & \mathcal{A}_{\bm{k}}^{\text{SIA}}&  0 & 0 &  - \mathrm{i} (\mathcal{B}'_{\bm{k}})^* & 0 & 0 & - \mathrm{i} \mathcal{C}'_{\bm{k}}\\[4pt]
      0 &0  & \mathcal{A}_{\bm{k}}^{\text{SIA}} &  0 &  - \mathrm{i} (\mathcal{C}'_{\bm{k}})^*& 0 & 0 &  - \mathrm{i} \mathcal{B}'_{\bm{k}}\\[4pt]
      0 & 0 & 0  &  \mathcal{A}_{\bm{k}}^{\text{SIA}} & 0 &  - \mathrm{i} (\mathcal{C}'_{\bm{k}})^*&  - \mathrm{i} (\mathcal{B}'_{\bm{k}})^*& 0 \\[4pt]
    0 &  \mathrm{i} \mathcal{B}'_{\bm{k}} &  \mathrm{i} \mathcal{C}'_{\bm{k}}& 0 & \mathcal{A}_{\bm{k}}^{\text{SIA}} & 0 & 0 & 0 \\[4pt]
     \mathrm{i} (\mathcal{B}'_{\bm{k}})^* & 0 & 0 &  \mathrm{i} \mathcal{C}'_{\bm{k}} & 0 & \mathcal{A}_{\bm{k}}^{\text{SIA}} &  0 & 0 \\[4pt]
      \mathrm{i} (\mathcal{C}'_{\bm{k}})^* & 0 & 0 &  \mathrm{i} \mathcal{B}'_{\bm{k}} & 0 &  0 &\mathcal{A}_{\bm{k}}^{\text{SIA}} & 0 \\[4pt]
     0 & \mathrm{i} (\mathcal{C}'_{\bm{k}})^* &  \mathrm{i} (\mathcal{B}'_{\bm{k}})^*& 0 &  0 & 0 & 0 & \mathcal{A}_{\bm{k}}^{\text{SIA}}
    \end{pmatrix},
\end{align}   
\end{widetext}
where $H_{\bm{k}}$ corresponds to Eq.~\eqref{eq:ham_EAP_iso_am}.
The additional components read
\begin{align}
    \mathcal{A}_{\bm{k}}^{\text{SIA}} & = - 2 S d_2, \\\
     \mathcal{B}'_{\bm{k}} & = S \hat{\bm{z}} \cdot \left( \bm{D}^{(1)} \mathrm{e}^{\mathrm{i} \bm{k} \cdot \bm{\delta}_{1}} + \sum_{j=1}^3 \bm{D}^{(3)}_{\bm{\delta}_{3_j}}\mathrm{e}^{\mathrm{i} \bm{k} \cdot \bm{\delta}_{3_j}}\right), \\
    \mathcal{C}'_{\bm{k}} & = S \hat{\bm{z}} \cdot  \sum_{j=1}^3 \left(\bm{D}^{(4)}_{\bm{\delta}_{4_j}} \mathrm{e}^{\mathrm{i} \bm{k} \cdot \bm{\delta}_{4_j}} + \bm{D}^{(4)}_{- \bm{\delta}_{4_j}} \mathrm{e}^{- \mathrm{i} \bm{k} \cdot \bm{\delta}_{4_j}} \right).
\end{align}

We use the paraunitary diagonalization as in the previous section and find the dispersion relation in Fig.~\ref{fig:bands_easy_axis}(a). The Goldstone modes of Fig.~\ref{fig:disperion_nonrelativistic}(a) are now lifted with the gap size $\Delta_\text{gap,EAP}$ because of the single-ion anisotropy, as shown in the inset.

In Fig.~\ref{fig:bands_easy_axis}(b), we show the difference of the acoustic and optical magnon dispersion relation between the cases without, and with DMI along the path $\Gamma$-Z, i.e., 
\begin{align}
\Delta \varepsilon_{\bm{k},\text{DMI}} = \varepsilon_{\bm{k}} (D_i=0) - \varepsilon_{\bm{k}} (D_i),
\label{eq:dmi_diff}
\end{align}
with $i \in \{1, 3, 4 \}$. The changes in the dispersion relation caused by the DMI are largest at the $\Gamma$ point ( $\sim 0.1$ meV). %\am{Maybe drop: This is beyond the resolution of neutron scattering \cite{kellerNeutronSpinEchoInstrumentation2022}.}
%, e.g., neutron spin echo \cite{MesotNeutronSpinEcho2006, ...}. 
Furthermore, the DMI cannot lead to any canting in the easy-axis phase, since the sum of the DMI vectors for each considered shell of neighbors only shows a finite $z$ component,
\begin{align}
    \bm{D}^{(1)} = \begin{pmatrix}
        0 \\ 0 \\ D_1
    \end{pmatrix}, \, \sum_{j=1}^3 \bm{D}^{(3)}_{\bm{\delta}_{3_j}} = \begin{pmatrix}
        0 \\ 0 \\ \mathcal{D}_3
    \end{pmatrix}, \,
    \sum_{j=1}^3 \left( \bm{D}^{(4)}_{\bm{\delta}_{4_j}} + \bm{D}^{(4)}_{-\bm{\delta}_{4_j}}\right) = \begin{pmatrix}
        0 \\ 0 \\ \mathcal{D}_4
    \end{pmatrix},
\end{align}
that is, a net component parallel to the magnetic order.

We point out a particular defect of the DMI within the harmonic theory, which goes under the name of ``spurious symmetries'' \cite{gohlkeSpuriousSymmetryEnhancement2023}. The DMI explicitly breaks the SO(3) symmetry of the spin Hamiltonian, and therewith also the SO(2) symmetry in the collinearly ordered state. Consequently, spin is not a good quantum number. However, within the harmonic theory, only the DMI components parallel to the Néel vector enter, such that the block diagonal structure of the Hamilton kernel is kept and the magnon spin is spuriously conserved. Thus, the harmonic magnon theory is blind to the breaking of SO(2) symmetry. Only magnon-magnon interactions, appearing to leading order as cubic interactions proportional to the DMI vector component perpendicular to the Néel vector, can lift the spurious SO(2) symmetry---already at zero temperature \cite{chernyshevDampedTopologicalMagnons2016,mookInteractionStabilizedTopologicalMagnon2021,gohlkeSpuriousSymmetryEnhancement2023}. However, since hematite has a large spin quantum number $S$, and the self-energies associated with cubic magnon-magnon interactions are an order of $1/S$ smaller than the bare energies, it is safe to neglect DMI-induced many-body effects. Thus, for all relevant purposes, the magnon spin is effectively conserved in the easy-axis phase of hematite.
%The DMI also does not influence the spin polarization of the magnons in the harmonic theory, as we see in the following. The Hamiltonian is now no longer SO(3) [but SO(2)] symmetric because of the DMI, and we expect spin to not be a good quantum number anymore. However, the Hamilton kernel $\mathcal{H}_{\bm{k}}'$ can still be arranged to be block-diagonal, meaning that spin is still a good quantum number. This implies that spin conservation is a spurious symmetry of the harmonic theory in the easy-axis phase, which would be corrected by including magnon-magnon interactions.

In short, relativistic corrections to the magnon spectrum in the easy-axis phase are restricted to a region close to $\Gamma$ and the splitting they give rise to is orders of magnitude smaller than the predicted splitting due to altermagnetism.
%} therefore do not obscure the altermagnetic splitting. \am{
We conclude that relativistic corrections will not obscure the detection of altermagnetic magnon splitting in the easy-axis phase of hematite. %This makes the splitting well suited to be measured in neutron scattering (e.g., polarized inelastic neutron scattering \cite{McClartyAltermagNeutrons2025}, neutron spin echo \cite{MesotNeutronSpinEcho2006,BayrakciNeutronSpinEcho2006,kellerNeutronSpinEchoInstrumentation2022}).
%\am{The Keller article is a great intro do spin echo}.

%%%%%%%%%%%%%%%%%%%%%%%%%%%%%%%%%%%%%%%%%%%%%%%%%%%%%%%%%%%%%%%%%%%%
%
% Easy-plane (weak ferromagnetic) phase
%
%%%%%%%%%%%%%%%%%%%%%%%%%%%%%%%%%%%%%%%%%%%%%%%%%%%%%%%%%%%%%%%%%%%%
\subsection{Easy-plane (weak ferromagnetic) phase}
\label{sec:WFP}
Above the Morin transition temperature $T_\text{M}$, in the weak ferromagnetic phase, the easy-plane two-ion anisotropy, as well as the dipole-dipole interaction force the spins to lie fully in-plane \cite{danneggerMagneticPropertiesHematite2023}. This leads to the polar angles $\theta_\text{A} \equiv \theta_\text{B} = \pi/2$.
On inclusion of the DMI, the ground state is not collinear anymore. The DMI induces a weak ferromagnetic moment \cite{dzyaloshinskyThermodynamicTheoryWeak1958,MoriyaAnisotropicSuperex1960} by canting the spins with a canting angle we define to be $|\delta \phi |= |\phi_\text{A} -\phi_\text{B} - \pi|$, where $\phi_\text{A}$ ($\phi_\text{B}$) is the azimuthal angle of sublattices A and D (B and C). Experiments find the canting angle $|\delta \phi/2| = 0.0554(8)^\circ$ at $295\,$K \cite{morrishCantedAntiferromagnetismHematite1994,hillNeutronDiffractionStudy2008}.
In our model, we introduce an effective in-plane two-ion anisotropy, the DMI, and the altermagnetic splitting one after the other to investigate their effect on the magnon dispersion. We neglect any distortion of the lattice due to magnetoelastic effects coming from the weak ferromagnetic moment on the grounds of weak spin-orbit coupling.

\subsubsection{Influence of easy-plane anisotropy}
\label{sec:WFP_anisotropy}
For simplicity, we introduce an effective easy-plane anisotropy $\delta J$, which includes the two-ion anisotropy and mimicks the dipole-dipole interaction. We incorporate it into the exchange interactions $J_1$, $J_2$, and $J_3$, reducing their $z$ component by $\delta J$, and therefore forcing the spins in-plane. To achieve this, we define the term %H_{\text{aniso}}$ 
\begin{align}
\label{eq:aniso_WFP}
    \mathcal{H}^{\text{ANISO}} =& -  \delta J \sum_{r=1}^{3} \sum_{\langle i, j \rangle_r} S_i^z S_j^z.
\end{align}
%\begin{align}
%\label{eq:aniso_WFP}
%    H_{\text{aniso}} =& \sum_{r=1}^{3} \sum_{\langle i, j \rangle_r}  \bm{S}_i^\text{T} \mathscr{J}_{ij} \bm{S}_j,
%\end{align}
%with $\mathscr{J}_{r} = \mathrm{diag}(0, 0, - \delta J)$. 
We first investigate $\widetilde{\mathcal{H}} = \mathcal{H}^{\text{ISO}} + \mathcal{H}^\text{ANISO}$ with $\mathcal{H}^\text{ISO}$ defined in Eq.~\eqref{eq:ham_iso}, but neglect the interactions responsible for the altermagnetic splitting. We study the general, azimuthal angle-dependent harmonic Hamiltonian, which is given by
%, and write the harmonic Hamiltonian as
%\begin{subequations}
%\begin{align}
%	H_2 & = \sum_{\bm{k}} A_{\bm{k}}  \left(a_{\bm{k}}^{\dagger} a_{\bm{k}} + b_{\bm{k}}^{\dagger} b_{\bm{k}} + c_{\bm{k}}^{\dagger} c_{\bm{k}} + d_{\bm{k}}^{\dagger} d_{\bm{k}}\right)  \nonumber \\
% & \quad + \frac{B_{\bm{k}}}{2}  \left(a_{\bm{k}}^\dagger b_{\bm{k}}  + c_{\bm{k}}^\dagger d_{\bm{k}}  + b_{-\bm{k}}^\dagger a_{-\bm{k}} + d_{-\bm{k}}^\dagger c_{-\bm{k}}\right)  + \mathrm{H.c.} \nonumber \\ 
%&+ \frac{C_{\bm{k}}}{2}  \left(a_{\bm{k}}^\dagger c_{\bm{k}}  + b_{\bm{k}}^\dagger d_{\bm{k}} + c_{-\bm{k}}^\dagger a_{-\bm{k}}  + d_{-\bm{k}}^\dagger b_{-\bm{k}}\right)  + \mathrm{H.c.} \nonumber \\ 
%& + \frac{D_{\bm{k}}}{2} \left( a_{\bm{k}}^\dagger d_{\bm{k}} + d_{-\bm{k}}^\dagger a_{-\bm{k}} + c_{\bm{k}}^\dagger b_{\bm{k}} + b_{-\bm{k}}^\dagger c_{-\bm{k}}  \right)  + \mathrm{H.c.} \nonumber \\
% &+ \frac{\widetilde{B}_{\bm{k}}}{2}  \left( a_{-\bm{k}} b_{\bm{k}}  +  c_{-\bm{k}} d_{\bm{k}} + a_{\bm{k}}^{\dagger} b_{-\bm{k}}^{\dagger} + c_{\bm{k}}^{\dagger} d_{-\bm{k}}^{\dagger} \right)  + \mathrm{H.c.}, \nonumber \\
% &+ \frac{\widetilde{C}_{\bm{k}}}{2} \left(a_{-\bm{k}} c_{\bm{k}}  + b_{-\bm{k}} d_{\bm{k}} + a_{\bm{k}}^\dagger c_{-\bm{k}}^\dagger  + b_{\bm{k}}^\dagger d_{-\bm{k}}^\dagger\right)+ \mathrm{H.c.} \nonumber  \\ 
%  &+ \frac{\widetilde{D}_{\bm{k}}}{2}   \left(a_{-\bm{k}} d_{\bm{k}} + a_{\bm{k}}^\dagger d_{-\bm{k}}^\dagger + c_{-\bm{k}} b_{\bm{k}} + c_{\bm{k}}^\dagger b_{-\bm{k}}^\dagger \right) + \mathrm{H.c.}
\begin{align}
	\widetilde{\mathcal{H}}_2&= \frac{1}{2} \sum_{\bm{k}} \bm{\Psi}_{\bm{k}}^{\dagger} \widetilde{H}_{\bm{k}} \bm{\Psi}_{\bm{k}},
    \label{eq:harm_ham_wfp}
\end{align}
with $ \bm{\Psi}_{\bm{k}}^{\dagger}$ as in Eq.~\eqref{eq:psi}. 
The Hamilton kernel reads
\begin{widetext}
\begin{align} 
    \widetilde{H}_{\bm{k}} = \begin{pmatrix}
     A_{\bm{k}} & B_{\bm{k}} & C_{\bm{k}} &  D_{\bm{k}}  &  0 & \widetilde{B}_{\bm{k}}  & \widetilde{C}_{\bm{k}} & \widetilde{D}_{\bm{k}} \\[4pt]
     B_{\bm{k}}^* & A_{\bm{k}} &  D_{\bm{k}}^* & C_{\bm{k}}  & \left(\widetilde{B}_{\bm{k}}\right)^*  & 0 & \left(\widetilde{D}_{\bm{k}}\right)^*& \widetilde{C}_{\bm{k}} \\[4pt]
      C_{\bm{k}}^* & D_{\bm{k}}  &  A_{\bm{k}} & B_{\bm{k}}  & \left(\widetilde{C}_{\bm{k}}\right)^* & \widetilde{D}_{\bm{k}}  & 0 & \widetilde{B}_{\bm{k}} \\[4pt]
      D_{\bm{k}}^*  & C_{\bm{k}}^* & B_{\bm{k}}^*  &  A_{\bm{k}} & \left(\widetilde{D}_{\bm{k}}\right)^* & \left(\widetilde{C}_{\bm{k}}\right)^* & \left(\widetilde{B}_{\bm{k}}\right)^* & 0 \\[4pt]
    0 & \widetilde{B}_{\bm{k}}  & \widetilde{C}_{\bm{k}} & \widetilde{D}_{\bm{k}} & A_{\bm{k}} & B_{\bm{k}} & C_{\bm{k}}  &  D_{\bm{k}}  \\[4pt]
    \left(\widetilde{B}_{\bm{k}}\right)^*  & 0 & \left(\widetilde{D}_{\bm{k}}\right)^* & \widetilde{C}_{\bm{k}}  & B_{\bm{k}}^* & A_{\bm{k}} &  D_{\bm{k}}^*& C_{\bm{k}}  \\[4pt]
     \left(\widetilde{C}_{\bm{k}}\right)^*  & \widetilde{D}_{\bm{k}}  & 0 & \widetilde{B}_{\bm{k}}  & C_{\bm{k}}^* &  D_{\bm{k}}  & A_{\bm{k}} & B_{\bm{k}} \\[4pt]
     \left(\widetilde{D}_{\bm{k}}\right)^* & \left(\widetilde{C}_{\bm{k}}\right)^* & \left(\widetilde{B}_{\bm{k}}\right)^* & 0 &  D_{\bm{k}}^*  & C_{\bm{k}}^*  & B_{\bm{k}}^* & A_{\bm{k}} 
    \end{pmatrix},
    \label{eq:ham_kernel_WFP}
\end{align}
\end{widetext}
with its components given by
\begin{subequations}
    \begin{align}
         A_{\bm{k}} &= -S \cos (\phi_\text{A} - \phi_\text{B}) \left( J_1 + 3 J_3 + 6 J_4 \right) \nonumber \\
         & \quad + 3 J_2 + J_5 + 6 J_{13},
         \label{eq:Ak_WFP_1}
    \end{align}
    \begin{align}
    B_{\bm{k}} & =  \frac{S}{2} \left\{
   \left[J_1 -\delta J + J_1  \cos (\phi_\text{A} - \phi_\text{B}) \right] \mathrm{e}^{\mathrm{i} \bm{k} \cdot \bm{\delta}_1}  \right. \nonumber \\
   & \left. +  \left[J_3  -\delta J + J_3 \cos (\phi_\text{A} - \phi_\text{B}) \right] \sum_{j=1}^3 \mathrm{e}^{\mathrm{i} \bm{k} \cdot \bm{\delta}_{3_j}} \right\},  
   \end{align}
   \begin{align}
C_{\bm{k}} & =  S 
     \left[ \cos (\phi_\text{A} - \phi_\text{B}) +1 \right] J_4 \sum_{j=1}^3 \cos( \bm{k} \cdot \bm{\delta}_{4_j}),  \\
     D_{\bm{k}} &= S \left[ \left( J_2 - \frac{\delta J}{2}  \right)\sum_{j=1}^3 \mathrm{e}^{\mathrm{i} \bm{k} \cdot \bm{\delta}_{2_j}} +  J_5  \mathrm{e}^{\mathrm{i} \bm{k} \cdot \bm{\delta}_5}+  J_{13} \sum_{j=1}^6 \mathrm{e}^{\mathrm{i} \bm{k} \cdot \bm{\delta}_{13_j}} \right], 
     \end{align}
     \begin{align}
    \widetilde{B}_{\bm{k}} & = \frac{S}{2} \left\{
     \left[J_1 -\delta J - J_1  \cos (\phi_\text{A} - \phi_\text{B}) \right] \mathrm{e}^{\mathrm{i} \bm{k} \cdot \bm{\delta}_1}  \right. \nonumber \\
    &  \left. \quad+  \left[J_3  -\delta J - J_3 \cos (\phi_\text{A} - \phi_\text{B}) \right] \sum_{j=1}^3 \mathrm{e}^{\mathrm{i} \bm{k} \cdot \bm{\delta}_{3_j}} \right\}, 
    \end{align}
    \begin{align}
    \widetilde{C}_{\bm{k}} & = S \left\{
      \left[ 1 - \cos (\phi_\text{A} - \phi_\text{B}) \right] J_4 \sum_{j=1}^3 \cos( \bm{k} \cdot \bm{\delta}_{4_j}) \right\} , \\
    \widetilde{D}_{\bm{k}} & =  - \frac{S}{2} \delta J \sum_{j=1}^3 \mathrm{e}^{\mathrm{i} \bm{k} \cdot \bm{\delta}_{2_j}}.
    \end{align}
\end{subequations}
One can read off the dependence on the azimuthal angles $\phi_\text{A}$ and $\phi_\text{B}$. This will become of interest as soon as we introduce the DMI into the system. However, as long as the sublattices are perfectly collinear, i.e., if $\phi_\text{B} = \phi_\text{A} + \pi$, we find $\cos (\phi_\text{A} - \phi_\text{B}) = -1$ and the expressions simplify considerably. 

\begin{figure}
    \centering       
    \includegraphics[width=7cm]{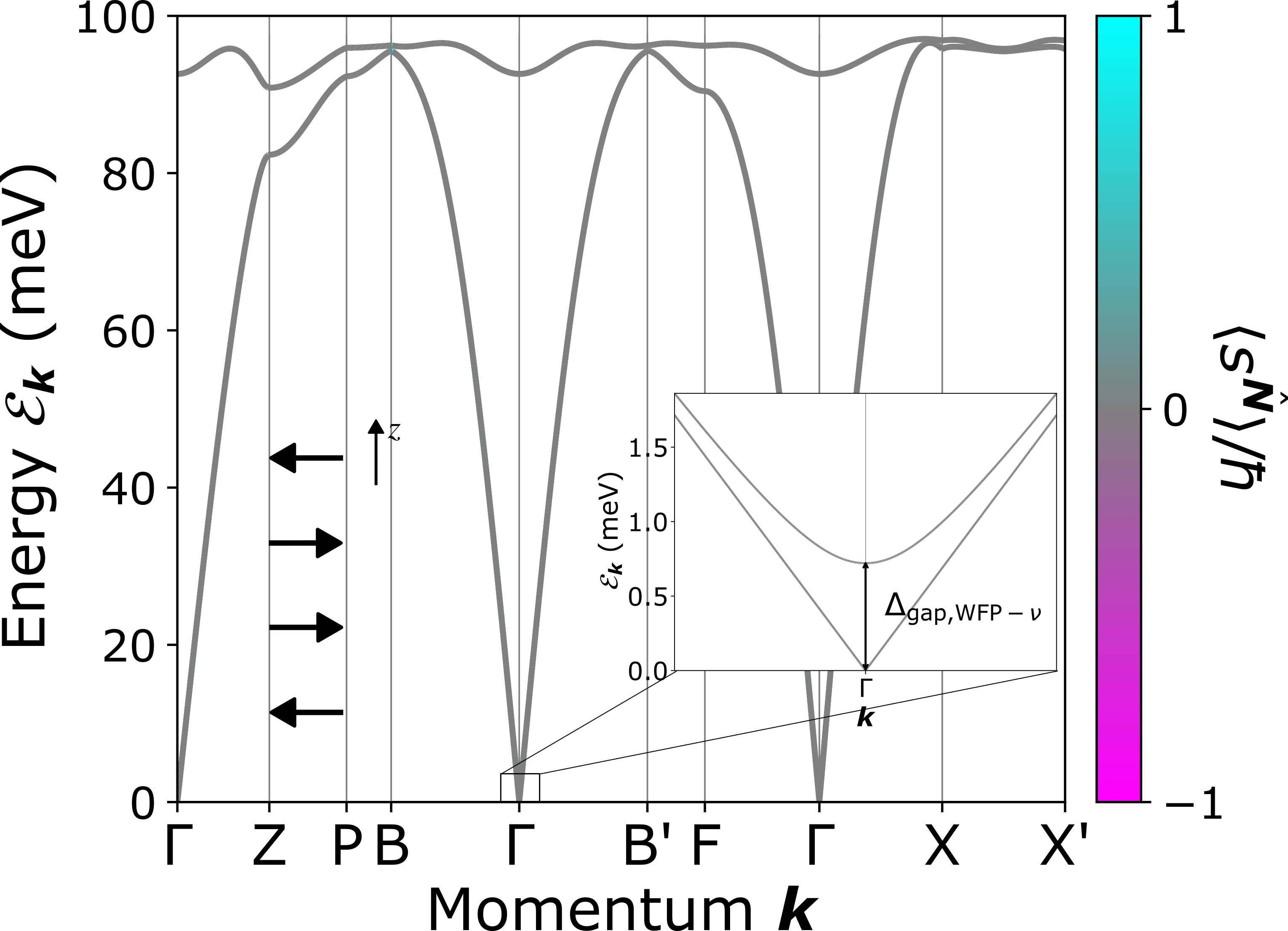}
       \caption{
       Magnon dispersion relation in the weak ferromagnetic phase (WFP) of hematite with easy-plane anisotropy but neglecting altermagnetic splitting and the Dzyaloshinskii-Moriya interaction. The acoustic modes split into the lower quasi-ferromagnetic and upper quasi-antiferromagnetic modes (see inset) with a gap of $\Delta_\text{gap,WFP} - \nu$. The color indicates the spin expectation value $\langle s^{\hat{\bm{N}}} \rangle$.
       We use the parameters in Tab.~\ref{tab:params_nonrel} (but set $\Delta = 0$) and Tab.~\ref{tab:easy_plane}.
       }
       \label{fig:bands_WFP_no_dmi_no_am}
\end{figure}

We Bogoliubov diagonalize and examine the magnon dispersion in Fig.~\ref{fig:bands_WFP_no_dmi_no_am}. The collinear ground state spontaneously breaks the SO(2) symmetry of the Hamiltonian by choosing a Néel vector direction in-plane, which leads to one Goldstone mode 
%from the easy-axis phase being lifted 
and no spin conservation. Hence, spin is not a good quantum number anymore and the spin polarization of the magnons along the Néel vector direction is mostly zero, i.e., the dispersion is gray. 
At $\Gamma$ the acoustic magnon modes split into the quasi-ferromagnetic (lower) and quasi-antiferromagnetic (upper) modes [cf. inset in Fig.~\ref{fig:bands_WFP_no_dmi_no_am}] with a gap of $\Delta_\text{gap,WFP} - \nu$ between them. 
To match antiferromagnetic resonance experiments \cite{fonerLowTemperatureAntiferromagneticResonance1965,ellistonAntiferromagneticResonanceMeasurements,velikov1969antiferromagnetic,Schönfeld2025} we fix the value of $\delta J$ (cf.~Tab.~\ref{tab:easy_plane}) such that $\Delta_\text{gap,WFP} \sim 180$~GHz $\approx 0.7444$ meV at $300$ K in the presence of DMI. We find the quasi-ferromagnetic mode to be the Goldstone mode, which would be lifted by including a triaxial anisotropy \cite{besserMagnetocrystallineAnisotropyPure1967} in the Hamiltonian
%or by considering quantum and thermal fluctuations (more to this in the Appendix.~\ref{sec:order_by_disorder}) \am{Drop the lifting by fluctuations part, because it is wrong here. This only happens when the Goldstone mode is not a true but a pseudo-Goldstone mode. It only is pseudo in the presence of DMI.} 
to match the experimental gap \cite{andersonMagneticResonanceFe1954,kumagaiFrequencyDependenceMagnetic1955,morrishCantedAntiferromagnetismHematite1994,Schönfeld2025}.

%%%%%%%%%%%%%%%%%%%%%%%%%%
\begin{table}[b]
\centering
\caption{Additional parameter in the weak ferromagnetic limit, chosen to match antiferromagnetic resonance experiments \cite{Schönfeld2025}.}
\label{tab:easy_plane}
\begin{tabular}{ p{1.3cm}|p{1.5cm}|p{1.cm}} 
 \toprule
 \centering Parameter & \centering Value & \centering Unit
  \arraybackslash \\
 \hline
  \centering $\delta J$ & \centering 
  %$1.12402$
  %$1.1123 $
  %$1.0478$
  $1.054 \cdot 10^{-3}$& \centering meV
  \arraybackslash \\  
 \botrule
\end{tabular}
\end{table}

\begin{figure*}
    \centering       
    \includegraphics[width=14cm]{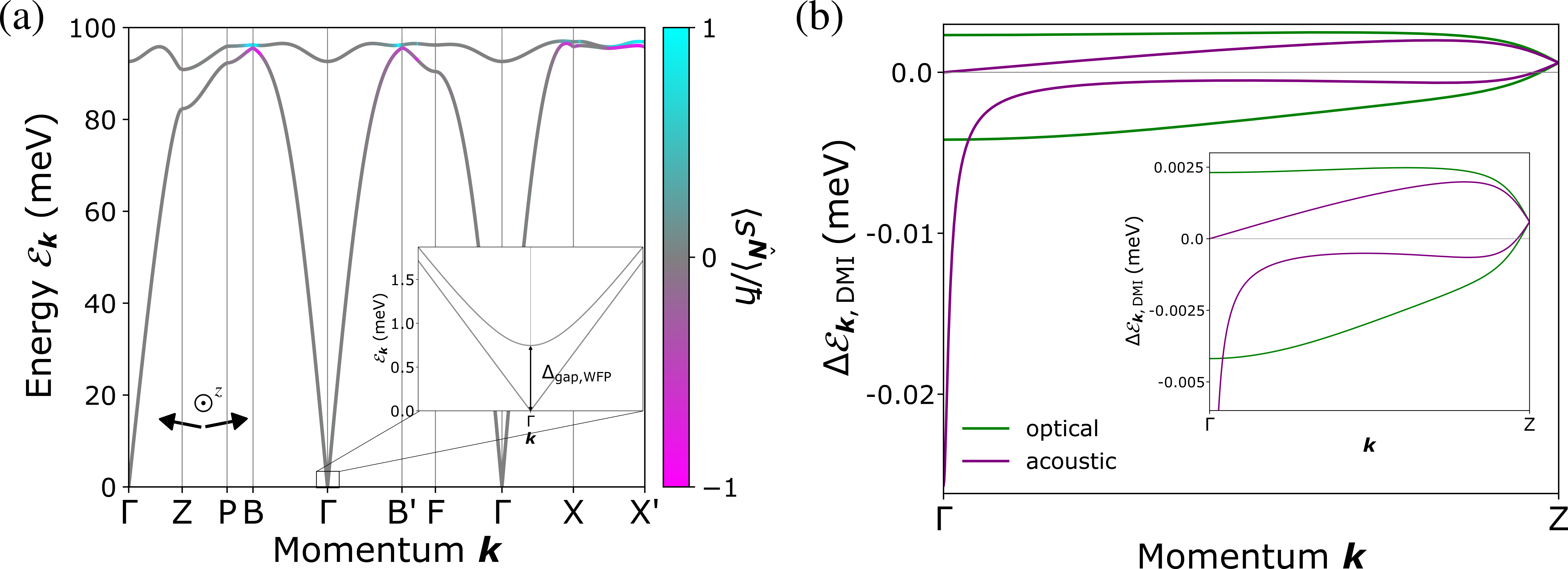}
       \caption{
       (a) Magnon dispersion relation in the weak ferromagnetic phase (WFP) of hematite with Dzyaloshinskii-Moriya interaction (DMI) but without altermagnetism. The inset shows the quasi-ferromagnetic (lower) and quasi-antiferromagnetic (upper) acoustic magnon mode with a gap $\Delta_\text{gap,WFP}$ between them. The color indicates the magnon spin expectation value.  (b) Energy difference $\Delta \varepsilon_{\bm{k},\text{DMI}}$ of the acoustic and optical magnon modes between the cases without and with DMI. %The degeneracies are lifted around $\Gamma$, albeit very weakly.
       We use the parameters in Tab.~\ref{tab:params_nonrel} (but set $\Delta = 0$), Tab.~\ref{tab:dmi} and Tab.~\ref{tab:easy_plane}.
       }
       \label{fig:bands_WFP_no_am}
\end{figure*}

\subsubsection{Influence of easy-plane anisotropy and Dzyaloshinskii-Moriya interaction}
\label{sec:WFP_dmi}
We now introduce the DMI in the same way as in Sec.~\ref{sec:rel}, with the parameters for $D_1$, $D_3$, and $D_4$ in Tab.~\ref{tab:dmi}. The Hamiltonian reads 
\begin{align}
\widetilde{\mathcal{H}}' = \mathcal{H}^{\text{ISO}} + \mathcal{H}^\text{ANISO} + \mathcal{H}^\text{DMI},
\label{eq:ham_WFP_DMI}
\end{align}
with $\mathcal{H}^{\text{ISO}}$ as in Eq.~\eqref{eq:ham_iso}, $\mathcal{H}^\text{ANISO}$ as in Eq.~\eqref{eq:aniso_WFP} and $\mathcal{H}^\text{DMI}$ as in Eq.~\eqref{eq:HDMI}. %We find the following harmonic Hamiltonian $\widetilde{\mathcal{H}}_2' = \widetilde{\mathcal{H}}_2 + \widetilde{\mathcal{H}}_2^\text{D}$, with $\widetilde{\mathcal{H}}_2$ from Eq.~\eqref{eq:harm_ham_wfp}. We define
%where we define
We find
\begin{align}
\widetilde{\mathcal{H}}_2' = \frac{1}{2} \sum_{\bm{k}} \bm{\Psi}_{\bm{k}}^{\dagger} \widetilde{H}_{\bm{k}}' \bm{\Psi}_{\bm{k}}, 
\end{align}
where
\begin{align}
\widetilde{H}_{\bm{k}}' = \widetilde{H}_{\bm{k}} +
	 \widetilde{H}_{\bm{k}}^\text{D}.
     %= \frac{1}{2} \sum_{\bm{k}} \bm{\Psi}_{\bm{k}}^{\dagger} \widetilde{H}_{\bm{k}}^\text{D} \bm{\Psi}_{\bm{k}}.
\end{align}
We take $\widetilde{H}_{\bm{k}}$ as in Eq.~\eqref{eq:ham_kernel_WFP} and the Hamilton kernel including the DMI reads
\begin{widetext}
\begin{align} 
    \widetilde{H}_{\bm{k}}^\text{D} = \begin{pmatrix}
     A_{\bm{k}}^\text{D} & B_{\bm{k}}^\text{D} + \mathrm{i} B'_{\bm{k}}& C_{\bm{k}}^\text{D} + \mathrm{i}C'_{\bm{k}}&  0&  0 & \widetilde{B}_{\bm{k}}^\text{D} - \mathrm{i} \widetilde{B}'_{\bm{k}} & \widetilde{C}_{\bm{k}}^\text{D} - \mathrm{i} \widetilde{C}'_{\bm{k}}& 0 \\[4pt]
     \left(B_{\bm{k}}^\text{D}\right)^* - \mathrm{i} (B'_{\bm{k}})^*& A_{\bm{k}}^\text{D} &  0 & C_{\bm{k}}^\text{D} + \mathrm{i}C'_{\bm{k}} & \left(\widetilde{B}_{\bm{k}}^\text{D}\right)^* - \mathrm{i} (\widetilde{B}'_{\bm{k}})^* & 0 & 0& \widetilde{C}_{\bm{k}}^\text{D} - \mathrm{i} \widetilde{C}'_{\bm{k}}\\[4pt]
      \left(C_{\bm{k}}^\text{D}\right)^* - \mathrm{i}(C'_{\bm{k}})^*& 0  &  A_{\bm{k}}^\text{D} & B_{\bm{k}}^\text{D}  + \mathrm{i} B'_{\bm{k}}& \left(\widetilde{C}_{\bm{k}}^\text{D}\right)^* - \mathrm{i} (\widetilde{C}'_{\bm{k}})^*& 0  & 0 & \widetilde{B}_{\bm{k}}^\text{D} - \mathrm{i} \widetilde{B}'_{\bm{k}}\\[4pt]
      0 & \left(C_{\bm{k}}^\text{D}\right)^* - \mathrm{i}(C'_{\bm{k}})^*& \left(B_{\bm{k}}^\text{D}\right)^* - \mathrm{i} (B'_{\bm{k}})^* &  A_{\bm{k}}^\text{D} & 0 & \left(\widetilde{C}_{\bm{k}}^\text{D}\right)^* - \mathrm{i} (\widetilde{C}'_{\bm{k}})^*& \left(\widetilde{B}_{\bm{k}}^\text{D}\right)^* - \mathrm{i} (\widetilde{B}'_{\bm{k}})^*& 0 \\[4pt]
    0 & \widetilde{B}_{\bm{k}}^\text{D} + \mathrm{i} \widetilde{B}'_{\bm{k}} & \widetilde{C}_{\bm{k}}^\text{D} + \mathrm{i} \widetilde{C}'_{\bm{k}}& 0 & A_{\bm{k}}^\text{D} & B_{\bm{k}}^\text{D} - \mathrm{i} B'_{\bm{k}}& C_{\bm{k}}^\text{D} - \mathrm{i} C'_{\bm{k}} &  0 \\[4pt]
    \left(\widetilde{B}_{\bm{k}}^\text{D}\right)^* + \mathrm{i} (\widetilde{B}'_{\bm{k}})^* & 0 & 0 & \widetilde{C}_{\bm{k}}^\text{D} + \mathrm{i} \widetilde{C}'_{\bm{k}} & \left(B_{\bm{k}}^\text{D}\right)^* + \mathrm{i} (B'_{\bm{k}})^*& A_{\bm{k}}^\text{D} &  0 & C_{\bm{k}}^\text{D}  - \mathrm{i} C'_{\bm{k}} \\[4pt]
     \left(\widetilde{C}_{\bm{k}}^\text{D}\right)^* + \mathrm{i} (\widetilde{C}'_{\bm{k}})^* & 0  & 0 & \widetilde{B}_{\bm{k}}^\text{D} + \mathrm{i} \widetilde{B}'_{\bm{k}} & \left(C_{\bm{k}}^\text{D}\right)^* + \mathrm{i} (C'_{\bm{k}})^* &  0 & A_{\bm{k}}^\text{D} & B_{\bm{k}}^\text{D} - \mathrm{i} B'_{\bm{k}}\\[4pt]
     0 & \left(\widetilde{C}_{\bm{k}}^\text{D}\right)^* + \mathrm{i} (\widetilde{C}'_{\bm{k}})^* & \left(\widetilde{B}_{\bm{k}}^\text{D}\right)^* + \mathrm{i} (\widetilde{B}'_{\bm{k}})^*& 0 &  0 & \left(C_{\bm{k}}^\text{D}\right)^* + \mathrm{i} (C'_{\bm{k}})^* & \left(B_{\bm{k}}^\text{D}\right)^* + \mathrm{i} (B'_{\bm{k}})^*& A_{\bm{k}}^\text{D} 
    \end{pmatrix}.
    \label{eq:ham_kernel_WFP_DMI}
\end{align}
\end{widetext}
\begin{widetext}
    
\end{widetext}
We define 
\begin{subequations}
   \begin{align}
      \bm{f}(\phi_\text{A}, \phi_\text{B}) &= \begin{pmatrix} 0 \\ 0 \\  \sin( \phi_\text{B} - \phi_\text{A} )
     \end{pmatrix}, \\
    \bm{g}(\phi_\text{A}, \phi_\text{B})&=  \begin{pmatrix} \cos \phi_\text{A} + \cos \phi_\text{B} \\ \sin \phi_\text{A} + \sin \phi_\text{B} \\  0 
     \end{pmatrix}, \\
     \bm{h}(\phi_\text{A}, \phi_\text{B}) &=  \begin{pmatrix} \cos \phi_\text{B} -  \cos \phi_\text{A}\\ \sin \phi_\text{B} - \sin \phi_\text{A}  \\  0 
     \end{pmatrix}, 
\end{align}
\end{subequations}
and find the components 
\begin{subequations}
\begin{align}
     A_{\bm{k}}^\text{D} &= -S \bm{f}(\phi_\text{A}, \phi_\text{B})
     \cdot \left[ \bm{D}^{(1)} +  \sum_{j=1}^3 \bm{D}^{(3)}_{\bm{\delta}_{3_j}} +\sum_{j=1}^3 \left(\bm{D}^{(4)}_{\bm{\delta}_{4_j}} + \bm{D}^{(4)}_{-\bm{\delta}_{4_j}} \right)\right], \label{eq:Ak_WFP_DMI} \\
     B_{\bm{k}}^\text{D} &= \frac{S}{2}
     \bm{f}(\phi_\text{A}, \phi_\text{B}) \cdot \left( \bm{D}^{(1)} \mathrm{e}^{\mathrm{i} \bm{k} \cdot \bm{\delta}_1} +  \sum_{j=1}^3 \bm{D}^{(3)}_{\bm{\delta}_{3_j}} \mathrm{e}^{\mathrm{i} \bm{k} \cdot \bm{\delta}_{3_j}} \right), 
    \end{align}
\begin{align}
     B'_{\bm{k}} &= -\frac{S}{2} 
     \bm{g}(\phi_\text{A}, \phi_\text{B}) \cdot \left( \bm{D}^{(1)} \mathrm{e}^{\mathrm{i} \bm{k} \cdot \bm{\delta}_1} +  \sum_{j=1}^3 \bm{D}^{(3)}_{\bm{\delta}_{3_j}} \mathrm{e}^{\mathrm{i} \bm{k} \cdot \bm{\delta}_{3_j}} \right), \\
     C_{\bm{k}}^\text{D} & = 
     \frac{S}{2}\bm{f}(\phi_\text{A}, \phi_\text{B})  \cdot \sum_{j=1}^3 \left(\bm{D}^{(4)}_{\bm{\delta}_{4_j}} \mathrm{e}^{\mathrm{i} \bm{k} \cdot \bm{\delta}_{4_j}} + \bm{D}^{(4)}_{-\bm{\delta}_{4_j}} \mathrm{e}^{-\mathrm{i} \bm{k} \cdot \bm{\delta}_{4_j}} \right) , \\
     C'_{\bm{k}}  &= -\frac{S}{2}
      \bm{g}(\phi_\text{A}, \phi_\text{B}) \cdot \sum_{j=1}^3 \left(\bm{D}^{(4)}_{\bm{\delta}_{4_j}} \mathrm{e}^{\mathrm{i} \bm{k} \cdot \bm{\delta}_{4_j}} + \bm{D}^{(4)}_{-\bm{\delta}_{4_j}} \mathrm{e}^{-\mathrm{i} \bm{k} \cdot \bm{\delta}_{4_j}} \right) , 
\end{align}
\begin{align}
    \widetilde{B}_{\bm{k}}^\text{D} & = 
   % -\frac{S}{2} 
   %  \bm{f}(\phi_\text{A}, \phi_\text{B})  \cdot \left( \bm{D}^{(1)} \mathrm{e}^{\mathrm{i} \bm{k} \cdot \bm{\delta}_1} +  \sum_{j=1}^3 \bm{D}^{(3)}_{\bm{\delta}_{3_j}} \mathrm{e}^{\mathrm{i} \bm{k} \cdot \bm{\delta}_{3_j}} \right) = 
     -  B_{\bm{k}}^\text{D}, \\
    \widetilde{B}'_{\bm{k}} = \, & - \frac{S}{2} \bm{h}(\phi_\text{A}, \phi_\text{B})  \cdot \left( \bm{D}^{(1)} \mathrm{e}^{\mathrm{i} \bm{k} \cdot \bm{\delta}_{1}} + \sum_{j=1}^3 \bm{D}^{(3)}_{\bm{\delta}_{3_j}}\mathrm{e}^{\mathrm{i} \bm{k} \cdot   \bm{\delta}_{3_j}}\right), \\
    \widetilde{C}_{\bm{k}}^\text{D} & = % -\frac{S}{2}   \bm{f}(\phi_\text{A}, \phi_\text{B})  \sum_{j=1}^3 \left(\bm{D}^{(4)}_{\bm{\delta}_{4_j}} \mathrm{e}^{\mathrm{i} \bm{k} \cdot \bm{\delta}_{4_j}} + \bm{D}^{(4)}_{-\bm{\delta}_{4_j}} \mathrm{e}^{-\mathrm{i} \bm{k} \cdot \bm{\delta}_{4_j}} \right) = 
    - C_{\bm{k}}^\text{D} , \\
    \widetilde{C}'_{\bm{k}} = \, &- \frac{S}{2} \bm{h}(\phi_\text{A}, \phi_\text{B}) \cdot \sum_{j=1}^3 \left(\bm{D}^{(4)}_{\bm{\delta}_{4_j}} \mathrm{e}^{\mathrm{i} \bm{k} \cdot \bm{\delta}_{4_j}} + \bm{D}^{(4)}_{-\bm{\delta}_{4_j}} \mathrm{e}^{-\mathrm{i} \bm{k} \cdot \bm{\delta}_{4_j}} \right).
\end{align}
\end{subequations}

As discussed in the beginning of this section, the DMI is able to induce a small canting between the sublattices A (D) and B (C), thus giving rise to a very small magnetization $\bm{M} = \bm{M}_\text{A} + \bm{M}_\text{B}$. Thus the name ``weak ferromagnetic'' phase.
We compute this canting by minimizing the classical energy $E_0$ in the weak ferromagnetic phase
\begin{align}
    \frac{E_0}{S^2 N} &=  2 \left\{ 
    \cos (\phi_\text{A} - \phi_\text{B} )   \left( J_1 + 3 J_3 + 6 J_4\right) \right. \nonumber \\
        & \left. \quad   +3 J_2 + J_5 + 6 J_{13} \right. \nonumber \\
        & \left. \quad 
       % + d_2 \left( \cos^2 \theta_\text{A} + \cos^2 \theta_\text{B} \right) 
        + \bm{f}(\phi_\text{A}, \phi_\text{B})  \cdot \left[ \bm{D}^{(1)} +  \sum_{j=1}^3 \bm{D}^{(3)}_{\bm{\delta}_{3_j}} +\sum_{j=1}^3 \left(\bm{D}^{(4)}_{\bm{\delta}_{4_j}} + \bm{D}^{(4)}_{-\bm{\delta}_{4_j}} \right) \right]\right\}
        \label{eq:classicalenergy}
\end{align}
with respect to the azimuthal angles $\phi_\text{A}$ and $\phi_\text{B}$. We chose the DMI such that we find the canting angle $\delta\phi /2 = (\phi_\text{B} - \phi_\text{A} + \pi)/2 \approx 0.00098$ rad $= 0.0561^{\circ}$, which is in accordance with experiments \cite{morrishCantedAntiferromagnetismHematite1994,hillNeutronDiffractionStudy2008}.

Figure~\ref{fig:bands_WFP_no_am}(a) displays the dispersion relation of hematite in the weak ferromagnetic phase with DMI, but without altermagnetic splitting. 
The Goldstone mode of the previous section now becomes a \textit{pseudo}-Goldstone mode. This is because the Hamiltonian does not hold any continuous spin rotation symmetries due to the presence of the DMI, but the classical energy in Eq.~\eqref{eq:classicalenergy} does: $E_0$ only depends on the relative angle $\phi_\text{A} - \phi_\text{B}$ but not on the absolute angle. Therefore, the harmonic theory predicts a Goldstone mode where there should be none. Fluctuations are expected to lift this pseudo-Goldstone mode \cite{Rau2018_pseudoGoldstone}, as they cause the order-by-disorder phenomenon \cite{Green_2018}. In short, fluctuations create an absolute angle dependence of the ground state energy (see Appendix \ref{sec:order_by_disorder} for the case of quantum fluctuations), which is then expected to lift the Goldstone mode because of magnon-magnon interactions \cite{Rau2018_pseudoGoldstone}. We emphasize, however, that inclusion of a triaxial basal plane magnetocrystalline anisotropy would do the same already at the level of noninteracting magnons  \cite{besserMagnetocrystallineAnisotropyPure1967,flandersAnisotropyBasalPlane1964}, and therefore do not explore this point in greater detail.

\begin{figure*}
   \centering       
   \includegraphics[width=\textwidth]{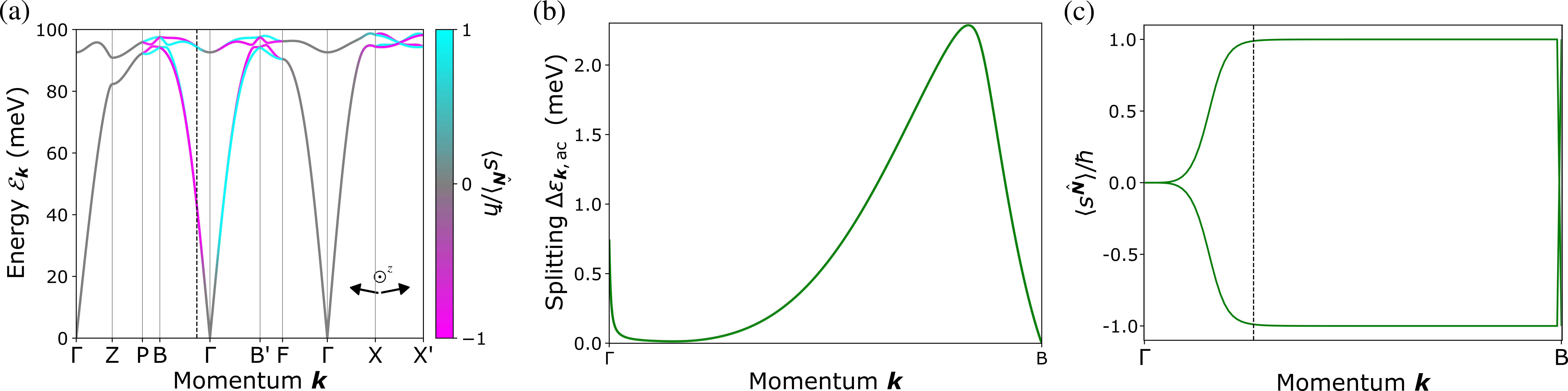}
      \caption{
      (a) 
      %\pps{suggestion: in (a) make the grid lines solid (but still thin and maybe gray even). Then in panel (c) there is the dashed line, and you can indicate this in panel (a) as well. Also, for all panels (in all figures), I would recommend you do not buffer the ends with space. For example here in panel (b) the tiny spacing at the end is needless. It makes it look weird almost like I am missing something. It is also very weird when the bottom is buffered (as negative magnon energies make no sense, so when you buffer the bottom with a bit of space it looks a bit weird, for example in fig 10 this is more visible). Also try to avoid grid lines on top of the plot frame (I see grid lines on top of the frame edge) } 
      Dispersion relation of hematite in the weak ferromagnetic phase. The color reflects the $\langle s^{\hat{\bm{N}}} \rangle$ spin expectation value.  (b) Splitting of the acoustic magnon modes $\Delta \varepsilon_{\bm{k},\text{ac}}$ along the path $\Gamma$-B. The splitting at $\Gamma$ comes from the effective easy-plane anisotropy.  (c) Spin expectation value of the acoustic magnon bands along path $\Gamma$-B. Spin-orbit coupling is responsible for the bands not being spin polarized along the Néel vector direction at $\Gamma$. The dashed line in (a,c) indicates where the effects of spin-orbit coupling become negligible going from $\Gamma$ to B. The parameters are shown in Tabs.~\ref{tab:params_nonrel}, \ref{tab:dmi} and ~\ref{tab:easy_plane}.
      }
      \label{fig:bands_WFP}
\end{figure*}

To clarify the difference between the cases without and with DMI, we show the energy difference $\Delta \varepsilon_{\bm{k},\text{DMI}}$ as in Eq.~\eqref{eq:dmi_diff} of the optical and acoustic magnon modes along the path $\Gamma$-Z in Fig.~\ref{fig:bands_WFP_no_am}(b). 
%The degeneracies of Fig.~\ref{fig:bands_WFP_no_dmi_no_am} are now partly lifted \am{Which degeneracies are you talking about here? Is the $\Gamma$-to-$Z$ direction the one with the larges DMI-induced renormalization?}, as the optical and acoustic modes split into two each (except at the Z point in this figure).
The largest difference is seen at the $\Gamma$ point, with the magnitude $\nu \approx 0.025$ meV, quickly reducing as we increase $\bm{k}$ (as it was also the case in the easy-axis phase in Sec.~\ref{sec:EAP}). This finding emphasizes the rapid decrease in importance of relativistic effects away from $\Gamma$.

\subsubsection{Complete model for the weak ferromagnetic phase with easy-plane anistropy, Dzyaloshinskii-Moriya interaction and altermagnetism}
\label{sec:WFP_full}
We add altermagnetism to the system, arriving at the full Hamiltonian
$\widetilde{\mathcal{H}}'' = \widetilde{\mathcal{H}}' + \mathcal{H}_\text{AM}$, with $\widetilde{\mathcal{H}}'$ as in Eq.~\eqref{eq:ham_WFP_DMI} and $\mathcal{H}_\text{AM}$ as in Eq.~\eqref{eq:ham_am}, where the component $\Delta_{\bm{k}}$ is defined in  Eq.~\eqref{eq:delta}.
We incorporate the Hamilton kernel $H_{\bm{k}}^\text{AM}$ from Eq.~\eqref{eq:ham_kernel_nonrel} into the Hamilton kernel $\widetilde{H}_{\bm{k}}'$in Eq.~\eqref{eq:ham_kernel_WFP_DMI} and Bogoliubov diagonalize.

Figure~\ref{fig:bands_WFP}(a) shows the magnon dispersion including effective in-plane anisotropy, DMI, and altermagnetic splitting. 
The splitting $\Delta \varepsilon_{\bm{k},\text{ac}}$ as in Eq.~\eqref{eq:splitting_ac} is displayed in Fig.~\ref{fig:bands_WFP}(b) for the acoustic magnon bands. Directly at $\Gamma$ the splitting is caused by SOC, specifically by the effective easy-plane anisotropy, as discussed in Sec.~\ref{sec:WFP_anisotropy}. Further away from, but still close to $\Gamma$ the SOC-induced splitting makes room for the altermagnetic spin splitting proportional to $\bm{k}^4$. The spin polarization of the magnon bands along the Néel vector direction is now recovered along spin-split paths in the BZ, as the color shows in Fig.~\ref{fig:bands_WFP}(a) along the paths B-$\Gamma$-B$'$, and X-X$'$. We confirm this fact by plotting the spin expectation value $\langle s^{\hat{\bm{N}}} \rangle$ in Fig.~\ref{fig:bands_WFP}(c) along the path $\Gamma$-B. From $\Gamma$ to the dashed line the influence of SOC dominates the magnon band structure, whereas from the dashed line to the vicinity of B the altermagnetic properties dominate over the SOC. Since the dashed line is still relatively close to $\Gamma$, it is safe to say that the effects of SOC do not obscure the altermagnetic properties of hematite.

%%%%%%%%%%%%%%%%%%%%%%%%%%%%%%%%%%%%%%%%%%%%%%%%%%%%%%%%%%%%%%%%%%%%
%
% Symmetry analysis
%
%%%%%%%%%%%%%%%%%%%%%%%%%%%%%%%%%%%%%%%%%%%%%%%%%%%%%%%%%%%%%%%%%%%%
\section{Experimental implications and outlook}
\label{sec:symmetry}

We conclude from our microscopic analysis that the characteristic spin (or chirality) splitting of magnons in the $g$-wave altermagnet candidate hematite is not obscured by relativistic corrections. This finding implies that---similarly to the case of MnTe \cite{Liu2024MnTeMagnonSplit}---inelastic neutron scattering will be able to resolve the splitting. If necessary, finer resolution is provided by neutron spin echo spectroscopy \cite{kellerNeutronSpinEchoInstrumentation2022}, which may even be able to resolve the anisotropic magnon lifetime due to magnon decays \cite{GarciaGaitan2025, Eto2025MagnonDecayAltermagnet, Cichutek2025MagnonDecayAltermagnet}. As proposed in Ref.~\cite{McClartyAltermagNeutrons2025}, polarized neutron scattering can directly resolve the spin polarization of the magnon bands in altermagnets; similarly to what was done in Ref.~\cite{Nambu2020} for the magnons of the ferrimagnet yttrium iron garnet. According to our analysis, hematite fits the requirements for this method; the splitting is sufficiently large and the interactions breaking spin conservation in the easy-axis and the easy-plane phase are perturbatively small. Inelastic resonant x-ray scattering is yet another method to probe the magnon spin polarization \cite{Takegami2025RIXS, Jost2025RIXS, Biniskos2025RIXS}. 

Beyond revealing the magnon spin splitting in spectroscopy, the time-reversal symmetry breaking of hematite can be probed by transport experiments. Metallic altermagnets are known to exhibit the anomalous Hall effect in the presence of spin-orbit coupling \cite{Smejkal2020CrystalHall, Feng2022}, and its thermal counterparts \cite{Zhou2024}. Similarly, insulating altermagnets exhibit the thermal Hall effect caused by charge-neutral quasiparticles such as magnons \cite{Hoyer2025THEAltermagnet}. The thermal Hall effect describes a heat current response in a transverse direction to an applied temperature gradient. These effects can only be present if the magnetic point group (MPG) is compatible with ferromagnetism (e.g., see Ref.~\cite{Smejkal2022AHEAntiferro}). 

Table~\ref{tab:symmetries_THC} shows the MPG of hematite when we align the Néel vector $\hat{\bm{N}}$ and magnetization $\hat{\bm{M}}$ along certain (high-)symmetry axes: 

(1) In the easy-axis phase the Néel vector aligns out-of-plane and the MPG is $\overline{3}$m, where the three unitary glide-mirrors $\mathcal{M}_{0\overline{1}1, \bm{\tau}}$, $\mathcal{M}_{\overline{1}01, \bm{\tau}}$, and $\mathcal{M}_{\overline{1}10, \bm{\tau}}$, with $\bm{\tau} = \left(\frac{1}{2}, \frac{1}{2}, \frac{1}{2} \right) $, are present. This MPG is \textit{not} compatible with ferromagnetism, hence any Hall-type transport is forbidden.

\begin{figure}
   \centering       
   \includegraphics[width=\columnwidth]{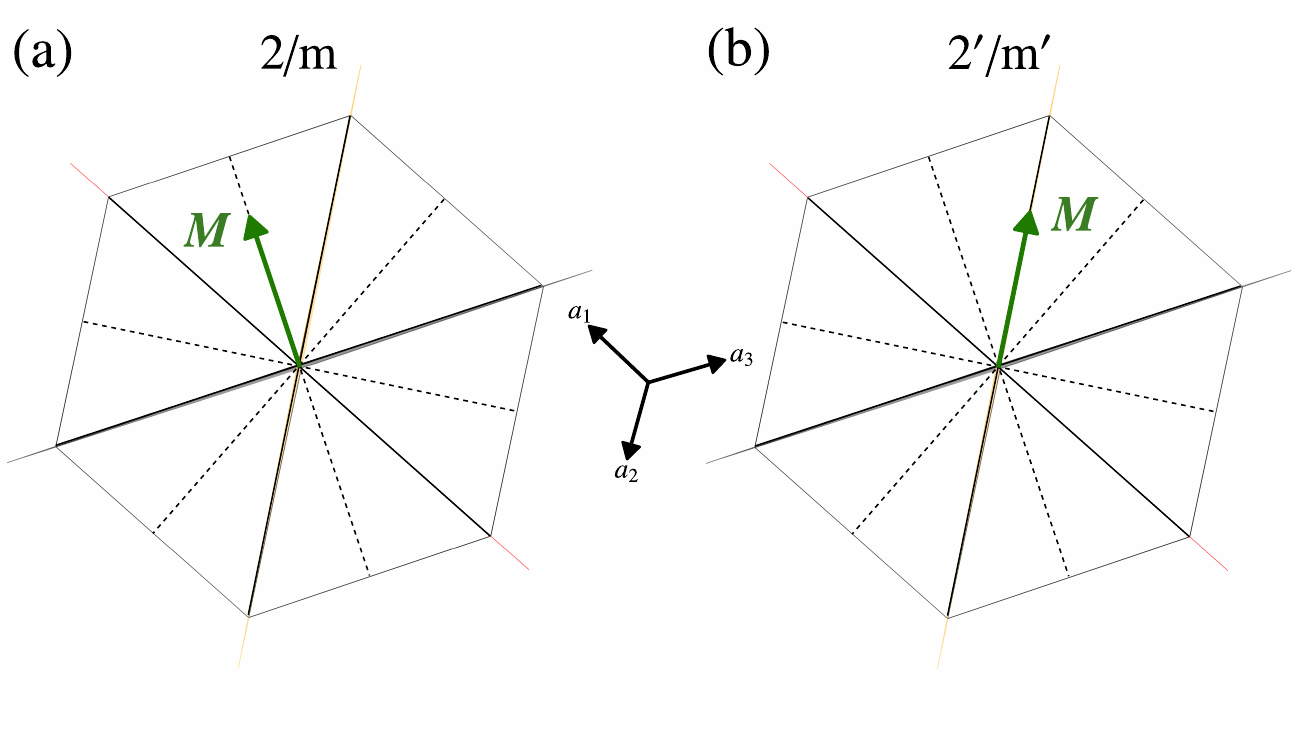}
      \caption{
      Magnetic point groups (MPG) of hematite in the weak ferromagnetic phase. (a) The MPG is 2/m when the the weak ferromagnetic moment is along the rotation axis of a mirror, that is, along one of the dashed lines.  (b) The MPG is 2$'$/m$'$ when the magnetization lies perpendicular to the rotation axis of the mirror, i.e., in the mirror plane along one of the solid lines.
      }
      \label{fig:MPG}
\end{figure}

(2) In the weak ferromagnetic phase, the Néel vector is fully in-plane. When we align the Néel vector along a general non-high-symmetry direction in-plane, all three mirrors are broken, the MPG is compatible with ferromagnetism, and there are no restrictions on the thermal conductivity tensor. Hence, we expect a thermal Hall effect in any geometry, that is, for any two orthogonal directions spanned by the temperature gradient and the transverse heat current.
If the magnetization is parallel to the rotation axis of one of the mirrors, i.e., $\hat{\bm{M}} \parallel [ 0\overline{1}1 ]$, $\hat{\bm{M}}  \parallel [ \overline{1}01 ]$, and $\hat{\bm{M}} \parallel [ \overline{1}10 ]$
%$\hat{\bm{M}} \parallel \mathcal{M}_{0\overline{1}1, \bm{\tau}}$, $\hat{\bm{M}}  \parallel \mathcal{M}_{\overline{1}01, \bm{\tau}}$, and $\hat{\bm{M}} \parallel \mathcal{M}_{\overline{1}10, \bm{\tau}}$
, the MPG is 2/m [as shown in Fig.~\ref{fig:MPG}(a)]. This MPG is also compatible with ferromagnetism. The two-fold rotational axis is also along the direction of the weak ferromagnetic moment ($\hat{\bm{M}}  \parallel 2$) and the Hall effect can occur in the mirror plane.
If the magnetization is, however, perpendicular to this two-fold rotation axis, that is $\hat{\bm{M}} \perp 2$, the MPG is 2$'$/m$'$ [as shown in Fig.~\ref{fig:MPG}(b)], and the Hall effect can occur parallel to the rotation axis of the mirror.
%If the Néel vector lies along the in-plane component of one of the lattice vectors, $\hat{\bm{N}} \paralll \bm{u}_i$, the magnetic point group is 2/m, which is also compatible with ferromagnetism. Here, the two-fold rotational axis is along the direction of the weak ferromagnetic moment and the Hall effect can occur in the mirror plane. Therefore, by rotating the magnetization in-plane one will encounter the 2/m symmetric case six times, which only supports the thermal Hall effect in the respective mirror plane. In between the 2/m symmetric cases, any plane can support a Hall effect. %For example, when measuring the thermal Hall effect in the $ab$ plane upon rotation of the Néel vector, the thermal Hall conductivity would modulate like $\sin(3\varphi)$, where $\varphi$ is the rotation angle, with $\varphi = 0$ corresponding to the case where the Néel vector points along one of the $\bm{u}_i$.

The relativistic corrections to the magnon spectrum are small. In particular, the Dzyaloshinskii-Moriya interaction is the only term that could cause a magnon thermal Hall effect. It is therefore likely that the magnonic contribution to the thermal Hall effect is small as well. However, a phononic contribution to the thermal Hall effect is possible; estimating its magnitude does, however, require a separate microscopic theory.

%%%%%%%%%%%%%%%%%%%%%%%%%%
\begin{table}[t]
\centering
\caption{Magnetic point group symmetries in hematite depending on the
Néel vector $\hat{\bm{N}}$ orientation, or the orientation of the magnetization $\hat{\bm{M}} $, where $\hat{\bm{N}} \perp \hat{\bm{M}}$. Here, $\hat{\bm{N}} \perp \hat{\bm{z}}$ means that $\hat{\bm{N}}$ points in a general direction in the $xy$ (or $ab$) plane. The vectors $\bm{u}_i$ lie along the in-plane components of the lattice vectors $\bm{a}_i$, that is inside the mirror plane. 
}
\label{tab:symmetries_THC}
\begin{tabular}{ p{2.2cm}|p{1.5cm}|p{3.cm}} % |p{4.5cm}
 \toprule
 \centering  Orientation of Néel vector and magnetization & \centering Magnetic point group & \centering Mirrors present 
  \arraybackslash \\
 \hline
 \centering $\hat{\bm{N}} \parallel \hat{\bm{z}}$, \,$\hat{\bm{M}} = \bm{0}$ & \centering $\overline{3}$m & \centering $\mathcal{M}_{0\overline{1}1, \bm{\tau}}$, $\mathcal{M}_{\overline{1}01, \bm{\tau}}$, $\mathcal{M}_{\overline{1}10, \bm{\tau}}$ 
   \arraybackslash  \\
   
  \centering
 $\hat{\bm{N}},\hat{\bm{M}} \perp \hat{\bm{z}}$ & \centering $\overline{1}$ & \centering none  
  \arraybackslash \\

\centering  $\hat{\bm{N}} \parallel \bm{u}_1 $,\, $\hat{\bm{M}} \perp \bm{u}_1$ & \centering 2/m & \centering $\mathcal{M}_{0\overline{1}1, \bm{\tau}}$ 
  \arraybackslash \\
 
 \centering $\hat{\bm{N}} \parallel \bm{u}_2 $,\, $\hat{\bm{M}} \perp \bm{u}_2$ & \centering 2/m  & \centering  $\mathcal{M}_{\overline{1}01, \bm{\tau}}$ 
   \arraybackslash \\
 
 \centering $\hat{\bm{N}} \parallel \bm{u}_3 $,\, $\hat{\bm{M}} \perp \bm{u}_3$ & \centering 2/m & \centering  $\mathcal{M}_{\overline{1}10, \bm{\tau}}$ 
  \arraybackslash  \\

  \centering  $\hat{\bm{N}} \perp \bm{u}_1 $, \,$\hat{\bm{M}}  \parallel \bm{u}_1 $ & \centering 2$'$/m$'$ & \centering $\mathcal{M}_{0\overline{1}1, \bm{\tau}}'$ 
  \arraybackslash \\
 
 \centering $\hat{\bm{N}} \perp \bm{u}_2 $,\, $\hat{\bm{M}} \parallel \bm{u}_2 $ & \centering 2$'$/m$'$  & \centering   $\mathcal{M}_{\overline{1}10, \bm{\tau}}'$
   \arraybackslash \\
 
 \centering $\hat{\bm{N}} \perp \bm{u}_3 $,\, $\hat{\bm{M}}  \parallel \bm{u}_3 $ & \centering 2$'$/m$'$ & \centering   $\mathcal{M}_{\overline{1}01, \bm{\tau}}'$ 
  \arraybackslash  \\
  
 \botrule
\end{tabular}
\end{table}

The smallness of relativistic corrections indicates the need for a nonrelativistic transport probe of $g$-wave altermagnetism. We recall the spin splitter effects in $d$-wave magnets, where an applied force (such as an electric field or temperature gradient) induces a transverse spin current \cite{Naka2019, GonzalezHernandez2021, SmejkalEmerging2022, Smejkal2022GMR}. The spin splitter effect also exists for magnons \cite{Naka2019, Cui2023spinsplittermagnon, Weissenhofer2024}. In contrast to relativistic spin Hall and spin Nernst effects, whose intrinsic contribution arises from interband effects, the spin splitter effect is dissipative and does not require spin-orbit coupling. The usual spin splitter effect is a linear effect and absent in $g$-wave altermagnets. However, recent work has highlighted the \textit{nonlinear} generation of electronic spin currents in magnets with beyond-$s$-wave splitting \cite{Ezawa2025}. These concepts can be extended to magnons in insulating altermagnets, where---relying on the arguments of Ref.~\cite{Ezawa2025}---we expect a third-order magnon spin splitter effect to emerge in hematite (and other $g$-wave altermagnets such as MnTe).

Specifically, let us assume that the $y$ direction of the laboratory frame is aligned with one of the $\hat{\vec{u}}_i$ directions in the $ab$ plane. Put simply, one of the crystallographic mirrors coincides with the $yz$ plane in the laboratory frame.
In this case, the low-energy magnon dispersion of a $g$-wave altermagnet is approximately given by
\begin{align}
    \varepsilon_{\pm,\bm{k}} \approx v k + \kappa k^3 \pm \xi k_z k_x \left( k_x^2 - 3 k_y^2\right),
    \label{eq:dispersion-expansion}
\end{align}
where $v>0$ is the long-wavelength magnon velocity, and $\kappa$ the isotropic part of the cubic nonlinearity. In this expansion, we have neglected relativistic corrections and included crystallographic symmetries only by the altermagnetic spin splitting, which is parametrized by $\xi$. We further assume that a temperature gradient $\bm{\nabla}T$ is applied in the $xy$ plane (the $ab$ basal plane of hematite). The resulting nonequilibrium magnon distribution function $\rho_{\pm,\vec{k}} = \rho(\varepsilon_{\pm,\bm{k}})$ follows from the Boltzmann transport equation within relaxation time approximation,
\begin{align}
    \vec{v}_{\pm,\vec{k}} \cdot \vec{\nabla} \rho_{\pm,\vec{k}} = \frac{\rho_{\pm,\vec{k}}^{(0)} - \rho_{\pm,\vec{k}}}{\tau_{\pm,\vec{k}}}.
\end{align}
Here, we have introduced the magnon group velocity $\vec{v}_{\pm,\vec{k}} = \hbar^{-1} \partial \varepsilon_{\pm,\bm{k}}/ \partial \bm{k} $, the magnon relaxation time $\tau_{\pm,\vec{k}}$, for which we will assume $\tau_{+,\vec{k}} = \tau_{-,\vec{k}} = \tau$ for simplicity, and the magnon equilibrium Bose distribution function $\rho_{\pm,\vec{k}}^{(0)} = \rho^{(0)}(\varepsilon_{\pm,\bm{k}}) = (\mathrm{e}^{\varepsilon_{\pm,\bm{k}}/(k_\mathrm{B}T)} -1)^{-1}$ with Boltzmann's constant $k_\mathrm{B}$.
We expand the distribution function as $\rho_{\pm,\vec{k}} = \sum_{n=0}^\infty \rho^{(n)}_{\pm,\vec{k}}$, where $\rho^{(n)}_{\pm,\vec{k}} \propto (\nabla T)^n$. With this ansatz, we obtain
\begin{align}
    -\tau \bm{v}_{\pm,\vec{k}} \cdot \vec{\nabla} \left( \rho^{(0)}_{\pm,\vec{k}} + \rho^{(1)}_{\pm,\vec{k}} + \rho^{(2)}_{\pm,\vec{k}} + \ldots \right) = \rho^{(1)}_{\pm,\vec{k}} + \rho^{(2)}_{\pm,\vec{k}} + \ldots , 
\end{align}
with the iterative solution
\begin{align}
    \rho^{(n+1)}_{\pm,\vec{k}} = -\tau \bm{v}_{\pm,\vec{k}} \cdot \vec{\nabla} \rho^{(n)}_{\pm,\vec{k}}.
    \label{eq:expansion-of-rho}
\end{align}
Explicitly, for a strictly linear temperature profile, we find
\begin{subequations}
\begin{align}
    \rho^{(1)}_{\pm,\vec{k}} 
    &=
    -\tau \sum_{\alpha} v^\alpha_{\pm,\vec{k}}
     \frac{\partial T}{\partial r_\alpha} 
    \frac{\partial \rho^{(0)}_{\pm,\vec{k}}}{\partial T} ,
    \\
    \rho^{(2)}_{\pm,\vec{k}} 
    &=
    \tau^2 \sum_{\alpha, \beta}
    v^\alpha_{\pm,\vec{k}} v^\beta_{\pm,\vec{k}}
    \frac{\partial T}{\partial r_\alpha}
    \frac{\partial T}{\partial r_\beta} 
    \frac{\partial^2 \rho^{(0)}_{\pm,\vec{k}}}{\partial T^2}
    \\
    \rho^{(3)}_{\pm,\vec{k}} 
    &= 
    -\tau^3 \sum_{\alpha, \beta, \gamma} v^\alpha_{\pm,\vec{k}} v^\beta_{\pm,\vec{k}} v^\gamma_{\pm,\vec{k}} 
    \frac{\partial T}{\partial r_\alpha}
    \frac{\partial T}{\partial r_\beta} 
    \frac{\partial T}{\partial r_\gamma}
    \frac{\partial^3 \rho^{(0)}_{\pm,\vec{k}}}{\partial T^3}.
\end{align}
\end{subequations}

As a result, a nonequilibrium magnon spin current density propagating in the $\mu$ direction in response to the temperature gradient can be expanded as
\begin{align}
    j_\mu 
    &= 
    \frac{\hbar}{V} \sum_{\vec{k}} \left( v^\mu_{+,\vec{k}} \rho_{+,\vec{k}} - v^\mu_{-,\vec{k}} \rho_{-,\vec{k}} \right)
    \nonumber \\
    &\approx 
    \frac{\hbar}{V} \sum_{\vec{k}} \left( v^\mu_{+,\vec{k}} \rho^{(1)}_{+,\vec{k}} - v^\mu_{-,\vec{k}} \rho^{(1)}_{-,\vec{k}} 
    \right)\nonumber \\
    &\quad 
    +
    \frac{\hbar}{V} \sum_{\vec{k}} \left( v^\mu_{+,\vec{k}} \rho^{(2)}_{+,\vec{k}} - v^\mu_{-,\vec{k}} \rho^{(2)}_{-,\vec{k}} 
    \right)\nonumber \\
    &\quad
    +
    \frac{\hbar}{V} \sum_{\vec{k}} \left( v^\mu_{+,\vec{k}} \rho^{(3)}_{+,\vec{k}} - v^\mu_{-,\vec{k}} \rho^{(3)}_{-,\vec{k}} 
    \right). \label{eq:expansion-of-j}
\end{align}
Plugging Eq.~\eqref{eq:expansion-of-rho} into Eq.~\eqref{eq:expansion-of-j} and comparing the result with the constitutive equation 
\begin{align}
    j_\mu \approx \alpha_{\mu,\nu}^{(1)} (- \nabla_\nu T)
    + \alpha_{\mu,\nu}^{(2)} (- \nabla_\nu T)^2
    + \alpha_{\mu,\nu}^{(3)} (- \nabla_\nu T)^3,
\end{align}
where $\alpha_{\mu,\nu}^{(n)}$ is the $n$th-order thermal spin conductivity, we find
\begin{subequations}
\begin{align}
    \alpha_{\mu,\nu}^{(1)}
    &=
    \frac{\hbar \tau}{V} \sum_{\vec{k}} \left( v^\mu_{+,\vec{k}} v^\nu_{+,\vec{k}}
    \frac{\partial \rho^{(0)}_{+,\vec{k}}}{\partial T} 
    -
    v^\mu_{-,\vec{k}} v^\nu_{-,\vec{k}} 
    \frac{\partial \rho^{(0)}_{-,\vec{k}}}{\partial T} \right),
    \\
    \alpha_{\mu,\nu}^{(2)}
    &=
    \frac{\hbar \tau^2}{V} \sum_{\vec{k}} \left[ v^\mu_{+,\vec{k}} \left(v^\nu_{+,\vec{k}}\right)^2
    \frac{\partial^2 \rho^{(0)}_{+,\vec{k}}}{\partial T^2} 
    -
    v^\mu_{-,\vec{k}} \left( v^\nu_{-,\vec{k}} \right)^2
    \frac{\partial^2 \rho^{(0)}_{-,\vec{k}}}{\partial T^2} \right],
    \\
    \alpha_{\mu,\nu}^{(3)}
    &=
    \frac{\hbar \tau^3}{V} \sum_{\vec{k}} \left[ v^\mu_{+,\vec{k}} \left(v^\nu_{+,\vec{k}}\right)^3
    \frac{\partial^3 \rho^{(0)}_{+,\vec{k}}}{\partial T^3} 
    -
    v^\mu_{-,\vec{k}} \left( v^\nu_{-,\vec{k}} \right)^3
    \frac{\partial^3 \rho^{(0)}_{-,\vec{k}}}{\partial T^3} \right].
\end{align}
\end{subequations}
As can be confirmed by plugging Eq.~\eqref{eq:dispersion-expansion} into the above expressions, applying a temperature gradient along a basal plane direction that does \textit{not} coincide with a mirror plane (e.g., the $x$ direction), results in a transverse out-of-plane spin current along the $z$ direction (or $c$ axis) because $\alpha_{z,x}^{(3)} \ne 0$. Applying the temperature gradient along a direction that does coincide with a mirror (e.g., the $y$ direction) does not generate a spin current because $\alpha_{z,y}^{(3)}=0$. This third-order thermal spin splitter effect of magnons is the lowest-order effect, i.e., $\alpha_{\mu,\nu}^{(1)} = \alpha_{\mu,\nu}^{(2)} = 0$, and is symmetry-allowed in any insulating $g$-wave altermagnet (e.g., also in MnTe). It is very similar in spirit to the nonlinear electronic spin splitter effect introduced in Ref.~\cite{Ezawa2025}. Since the spin current polarization is set by the magnetic texture, opposite magnetic domains cause opposite spin currents. Therefore, an experimental detection either requires sub-domain resolution or samples with an unequal distribution of opposite domains.

%%%%%%%%%%%%%%%%%%%%%%%%%%%%%%%%%%%%%%%%%%%%%%%%%%%%%%%%%%%%%%%%%%%%
%
% Conclusion
%
%%%%%%%%%%%%%%%%%%%%%%%%%%%%%%%%%%%%%%%%%%%%%%%%%%%%%%%%%%%%%%%%%%%%
\section{Conclusion}
\label{sec:conclusion}
We have developed a four-sublattice spin-wave theory to describe the magnon spectrum of hematite. The characteristic nonrelativistic spin (or chirality) splitting of magnons in altermagnets is captured by including inequivalent bonds up to the 13\textsuperscript{th} neighbor shell, resulting in a splitting of approximately $2\,$meV. Incorporating relativistic effects, such as magnetocrystalline anisotropies and the Dzyaloshinskii-Moriya interaction, has little impact on the magnon dispersion. The most significant changes occur at the $\Gamma$ point but rapidly diminish away from it. This suggests that the nonrelativistic altermagnetic properties of magnons remain robust against relativistic corrections. This finding is in agreement with Ref.~\cite{Verbeek2024}, where both the electronic spin splitting and the ferroically ordered magnetic triakontadipoles, which were associated with the $g$-wave spin splitting, persist in the weakly ferromagnetic phase with spin-orbit coupling. Thus, hematite is an ideal candidate for detecting spin-split altermagnetic magnons, for instance, via inelastic neutron scattering. Furthermore, we predict a nonlinear thermal spin-splitter effect, where a temperature gradient in the basal plane generates a nonlinear spin current along the $c$ direction.

Looking ahead, intriguing directions include studying the effects of spontaneous strain and substrate clamping, which are known to play an important role in the weak ferromagnetic phase of hematite \cite{Urquhart1956MagnetostrictiveEffects,Voskanyan1968MagnetoelasticPropertiesHematite,wittmannRoleSubstrateClamping2022}, investigating magnon topology \cite{karaki2023materialSearch}, and exploring magnetic excitations beyond conventional magnons, which correspond to $\Delta S = 1$ spin-flip processes. In hematite, the large spin quantum number $S=5/2$ allows for the existence of ``heavy'' or ``multipolar'' magnons---excitations involving local spin flips of $\Delta S = 2,3,4$ and $5$. These excitations produce distinct signatures in resonant inelastic x-ray scattering at approximately $\Delta S \times 100\,$meV \cite{Li2023hematite, Elnaggar2023hematite}. Such excitations can be described by a generalized spin-wave theory \cite{Muniz2014}, as demonstrated in other materials like FeI$_2$ \cite{Bai2021} and FePS$_3$ \cite{Wyzula2022}. An intriguing question for future research is whether the larger spin angular momentum of these excitations enhances their altermagnetic splitting.

%%%%%%%%%%%%%%%%%%%%%%%%%%%%%%%%%%%%%%%%%%%%%%%%%%%%%%%%%%%%%%%%%%%%
%
% Data availability
%
%%%%%%%%%%%%%%%%%%%%%%%%%%%%%%%%%%%%%%%%%%%%%%%%%%%%%%%%%%%%%%%%%%%%
\section{Data availability}
Upon reasonable request, the data and code from the work are available on Zenodo \cite{Hoyer2025hematite_zenodo}.

%%%%%%%%%%%%%%%%%%%%%%%%%%%%%%%%%%%%%%%%%%%%%%%%%%%%%%%%%%%%%%%%%%%%
%
% Acknowledgments
%
%%%%%%%%%%%%%%%%%%%%%%%%%%%%%%%%%%%%%%%%%%%%%%%%%%%%%%%%%%%%%%%%%%%%
\begin{acknowledgments}
%\paragraph{Acknowledgments.}
We are grateful to Peng Rao, Johannes Knolle, Olena Gomonay and Xanthe Verbeek for helpful discussions.
This work was funded by the Deutsche Forschungsgemeinschaft (DFG, German Research Foundation) -- Project No.~504261060 (Emmy Noether Programme). L.Š. acknowledges support from the ERC Starting Grant No. 101165122. P.P.S., A.R. and R.V. acknowledge support by the Deutsche
Forschungsgemeinschaft (DFG, German Research Foundation) for funding through  TRR 288 -- 422213477 (projects A05, B05) and through Va 117/23-1 -- 509751747.
\end{acknowledgments} 

%%%%%%%%%%%%%%%%%%%%%%%%%%%%%%%%%%%%%%%%%%%%%%%%%%%%%%%%%%%%%%%%%%%%
%
% appendix
%
%%%%%%%%%%%%%%%%%%%%%%%%%%%%%%%%%%%%%%%%%%%%%%%%%%%%%%%%%%%%%%%%%%%%
\appendix
%\label{ap:appendix}

\section{Symmetries of the space group \texorpdfstring{R$\overline{3}$c}{} modulo Bravais lattice translations}
\label{ap:symm}
In Tab.~\ref{tab:symmetries}, we list the symmetries of the space group R$\overline{3}$c of hematite modulo Bravais lattice translations.
%%%%%%%%%%%%%%%%%%%%%%%%%%
\begin{table}[H]
\centering
\caption{Symmetries of the space group R$\overline{3}$c (No. 167) modulo Bravais lattice translations \cite{brock2016internationalTabCryst}. The fractional translation reads $\bm{\tau}=(\frac{1}{2}, \frac{1}{2}, \frac{1}{2})$.}
\label{tab:symmetries}
\begin{tabular}{ p{1.5cm}|p{2.5cm}|p{3.cm}|p{1.2cm}} 
 \toprule
 \centering Symmetry & \centering matrix representation in lattice coords. & \centering  matrix representation in Cartesian coords. & \centering rotation axis
  \arraybackslash \\
 \hline
 \centering Identity $\mathcal{E}$ %\pps{suggestion $\mathcal{E}$}
 & \centering $\begin{pNiceMatrix}[columns-width = 0.55cm]
 $1$ & $0$ & $0$\\
 $0$ & $1$ & $0$\\
 $0$ & $0$ & $1$\\
 \end{pNiceMatrix}$  
 & \centering $\begin{pNiceMatrix}[columns-width = 0.55cm]
$1$ & $0$ & $0$\\
 $0$ & $1$ & $0$\\
 $0$ & $0$ & $1$\\
 \end{pNiceMatrix}$
  & \centering  - %$\langle 100 \rangle$, $\langle 010 \rangle$, $\langle 001 \rangle$
\arraybackslash  \\
  \centering Inversion $\mathcal{I}$  %\pps{suggestion $\mathcal{I}$}
 & \centering $\begin{pNiceMatrix}[columns-width = 0.55cm]
 -1 & $0$ & $0$\\
 $0$ & -1 & $0$\\
 $0$ & $0$ & -1\\
 \end{pNiceMatrix}$  
 & \centering $\begin{pNiceMatrix}[columns-width = 0.55cm]
-1 & $0$ & $0$\\
 $0$ & -1 & $0$\\
 $0$ & $0$ & -1\\
 \end{pNiceMatrix}$
  & \centering  - %$\langle 100 \rangle$, $\langle 010 \rangle$, $\langle 001 \rangle$
\arraybackslash  \\
   \centering  $\mathcal{C}_{3}$ %\pps{suggestion $\mathcal{C}_{3}$}
 & \centering $\begin{pNiceMatrix}[columns-width = 0.55cm]
 0 & 0 & 1\\
 1 & 0 & 0\\
 0 & 1 & 0\\
 \end{pNiceMatrix}$  
 & \centering $\begin{pNiceMatrix}[columns-width = 0.55cm]
-1/2 & -\sqrt{3}/2 & 0\\
 \sqrt{3}/2 & -1/2 & 0\\
0 & 0 & 1\\
 \end{pNiceMatrix}$
  & \centering $\langle 111 \rangle$
\arraybackslash  \\
\centering $\mathcal{C}_{3}^{-1}$   %\pps{suggestion $\mathcal{C}_{3}^{-1}$}
 & \centering $\begin{pNiceMatrix}[columns-width = 0.55cm]
 0 & 1 & 0\\
 0 & 0 & 1\\
 1 & 0 & 0\\
 \end{pNiceMatrix}$  
 & \centering $\begin{pNiceMatrix}[columns-width = 0.55cm]
-1/2 & \sqrt{3}/2 & 0\\
 -\sqrt{3}/2 & -1/2 & 0\\
0 & 0 & 1\\
 \end{pNiceMatrix}$
  & \centering $\langle 111 \rangle$
   \arraybackslash  \\
   \centering  $\mathcal{S}_6^{-1}=\mathcal{I}\mathcal{C}_{3}$
   %$\mathcal{C}_{3_1}$ + inv.  \pps{suggestion $\mathcal{S}_6^{-1}=\mathcal{I}\mathcal{C}_{3}$}
 & \centering $\begin{pNiceMatrix}[columns-width = 0.55cm]
 0 & 0 & -1\\
 -1 & 0 & 0\\
 0 & -1 & 0\\
 \end{pNiceMatrix}$  
 & \centering $\begin{pNiceMatrix}[columns-width = 0.55cm]
1/2 & \sqrt{3}/2 & 0\\
- \sqrt{3}/2 & 1/2 & 0\\
0 & 0 & -1\\
 \end{pNiceMatrix}$
  & \centering $\langle 111 \rangle$
\arraybackslash  \\
\centering $\mathcal{S}_6$
%$\mathcal{C}_{3_2}$ + inv. \pps{suggestion $\mathcal{S}_6$}
 & \centering $\begin{pNiceMatrix}[columns-width = 0.55cm]
 0 & -1 & 0\\
 0 & 0 & -1\\
 -1 & 0 & 0\\
 \end{pNiceMatrix}$  
 & \centering $\begin{pNiceMatrix}[columns-width = 0.55cm]
1/2 & -\sqrt{3}/2 & 0\\
 \sqrt{3}/2 & 1/2 & 0\\
0 & 0 & -1\\
 \end{pNiceMatrix}$
  & \centering $\langle 111 \rangle$
\arraybackslash  \\
   \centering $\mathcal{C}_{2\langle 0\overline{1}1 \rangle, \bm{\tau}}$ %\pps{suggestion $\mathcal{C}_{2\langle 0\overline{1}1 \rangle}$}
 & \centering $\begin{pNiceMatrix}[columns-width = 0.55cm]
  -1 & 0 & 0\\
 0 & 0 & -1\\
 0 & -1 & 0\\
 \end{pNiceMatrix}$  
 & \centering $\begin{pNiceMatrix}[columns-width = 0.55cm]
 0 & 1 & 0\\
 1 & 0 & 0\\
 0 & 0 & -1\\
 \end{pNiceMatrix}$
  & \centering $\langle 0\overline{1}1 \rangle$
   \arraybackslash  \\
    \centering $\mathcal{C}_{2\langle  \overline{1}01 \rangle, \bm{\tau}}$
    %$\mathcal{C}_{2_2}$ 
 & \centering $\begin{pNiceMatrix}[columns-width = 0.55cm]
  0 & 0 & -1\\
 0 & -1 & 0\\
 -1 & 0 & 0\\
 \end{pNiceMatrix}$  
 & \centering $\begin{pNiceMatrix}[columns-width = 0.55cm]
 \sqrt{3}/2 & -1/2 & 0\\
 -1/2 & -\sqrt{3}/2 & 0\\
 0 & 0 & -1\\
 \end{pNiceMatrix}$
  & \centering $\langle \overline{1}01 \rangle$
   \arraybackslash  \\
  \centering $\mathcal{C}_{2\langle  \overline{1}10 \rangle, \bm{\tau}}$
  %$\mathcal{C}_{2_3}$ 
 & \centering $\begin{pNiceMatrix}[columns-width = 0.55cm]
  0 & -1 & 0\\
 -1 & 0 & 0\\
 0 & 0 & -1\\
 \end{pNiceMatrix}$  
 & \centering $\begin{pNiceMatrix}[columns-width = 0.55cm]
 -\sqrt{3}/2 & -1/2 & 0\\
 -1/2 & \sqrt{3}/2 & 0\\
 0 & 0 & -1\\
 \end{pNiceMatrix}$
  & \centering $\langle \overline{1}10 \rangle$
   \arraybackslash  \\
   \centering $\mathcal{M}_{0\overline{1}1, \bm{\tau}}=\mathcal{I}\mathcal{C}_{2\langle 0\overline{1}1 \rangle, \bm{\tau}}$
   %$\mathcal{C}_{2_1}$ + inv. $=\mathcal{M}_{0\overline{1}1, \bm{\tau}}$ \pps{suggestion $\mathcal{M}_{0\overline{1}1, \bm{\tau}}=\bm{\tau}+\mathcal{I}\mathcal{C}_{2\langle 0\overline{1}1 \rangle}$}
 & \centering $\begin{pNiceMatrix}[columns-width = 0.55cm]
  1 & 0 & 0\\
 0 & 0 & 1\\
 0 & 1 & 0\\
 \end{pNiceMatrix}$  
 & \centering $\begin{pNiceMatrix}[columns-width = 0.55cm]
 0 & -1 & 0\\
 -1 & 0 & 0\\
 0 & 0 & 1\\
 \end{pNiceMatrix}$
  & \centering $\langle 0\overline{1}1 \rangle$
   \arraybackslash  \\
    \centering $\mathcal{M}_{\overline{1}01, \bm{\tau}}$
    %$=\bm{\tau}+\mathcal{I}\mathcal{C}_{2\langle \overline{1}01 \rangle}$
    %$\mathcal{C}_{2_2}$ + inv. $=\mathcal{M}_{\overline{1}01, \bm{\tau}}$
 & \centering $\begin{pNiceMatrix}[columns-width = 0.55cm]
  0 & 0 & 1\\
 0 & 1 & 0\\
 1 & 0 & 0\\
 \end{pNiceMatrix}$  
 & \centering $\begin{pNiceMatrix}[columns-width = 0.55cm]
 -\sqrt{3}/2 & 1/2 & 0\\
 1/2 & \sqrt{3}/2 & 0\\
 0 & 0 & 1\\
 \end{pNiceMatrix}$
  & \centering $\langle \overline{1}01 \rangle$
   \arraybackslash  \\
  \centering $\mathcal{M}_{\overline{1}10, \bm{\tau}}$
  %$=\bm{\tau}+\mathcal{I}\mathcal{C}_{2\langle \overline{1}10 \rangle}$
  %$\mathcal{C}_{2_3}$ + inv. $=\mathcal{M}_{\overline{1}10, \bm{\tau}}$
 & \centering $\begin{pNiceMatrix}[columns-width = 0.55cm]
  0 & 1 & 0\\
 1 & 0 & 0\\
 0 & 0 & 1\\
 \end{pNiceMatrix}$  
 & \centering $\begin{pNiceMatrix}[columns-width = 0.55cm]
 \sqrt{3}/2 & 1/2 & 0\\
 1/2 & -\sqrt{3}/2 & 0\\
 0 & 0 & 1\\
 \end{pNiceMatrix}$
  & \centering $\langle \overline{1}10 \rangle$
   \arraybackslash  \\
  
 \botrule
\end{tabular}
\end{table}
%%%%%%%%%%%%%%%%%%%%%%%%%%

\section{Density functional theory}
We use \textit{ab initio} methods to estimate the isotropic magnetic exchange coupling $J_i$ up to the 13\textsuperscript{th} nearest neighbor shell through the method of total energy mapping analysis (TEMA) \cite{GlasbrennerNP2015,GuterdingPRB2016,IqbalPRM2017}. TEMA is carried out in two steps: (\textit{i}) \textit{ab initio} simulations of several spin configurations in an appropriate supercell, and (\textit{ii}) mapping the spin configurations and converged energies onto an $S\!=\!5/2$ isotropic Heisenberg Hamiltonian. The couplings are extracted from the second step by a least square fit. 

For the \textit{ab initio} simulations we use full-potential local-orbital (FPLO) software package~\cite{KoepernikPRB1999,OpahlePRB1991}, version 22.00-62, which implements DFT+U in the atomic limit \cite{YlvisakerPRB2009}. In the FPLO calculations, we use the $3\times2\times1$ supercell (in terms of the primitive unit cell). We simulate a total of $27$ spin configurations, with a $10\times8\times8$ $\bm{k}$-mesh, and an energy convergence criteria of $10^{-8}$ Ha. The Hund's coupling $J_{H}$ is set to 1 eV and the on-site Coulomb $U$ is set to 6 eV for all spin configurations. Variations of $U$ mainly affect the overall magnitude of antiferromagnetic Heisenberg couplings $J$, with a dependence of $J \propto 1/U$. Such variations lead to no qualitative difference in the band dispersion.

\begin{figure}
    \centering
    \includegraphics[width=0.85\linewidth]{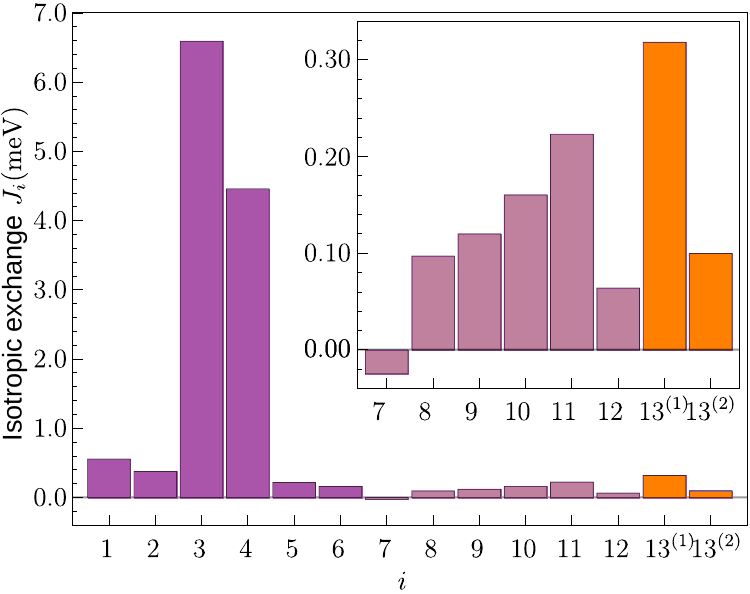}
    \caption{
    Isotropic magnetic exchanges $J_i$ up to 13\textsuperscript{th} nearest neighbor shell from TEMA. The largest couplings are the antiferromagnetic 3\textsuperscript{rd} and 4\textsuperscript{th} nearest neighbors. The inset shows a zoom-in of the longer range couplings. The 13\textsuperscript{th} nearest neighbor shell, responsible for altermagnetism in hematite, is highlighted in orange. As it supports two symmetry-inequivalent bonds, there are two different exchanges.
    }
    \label{fig:TEMA_plot}
\end{figure}

\begin{figure}
    \centering
    \includegraphics[width=0.85\linewidth]{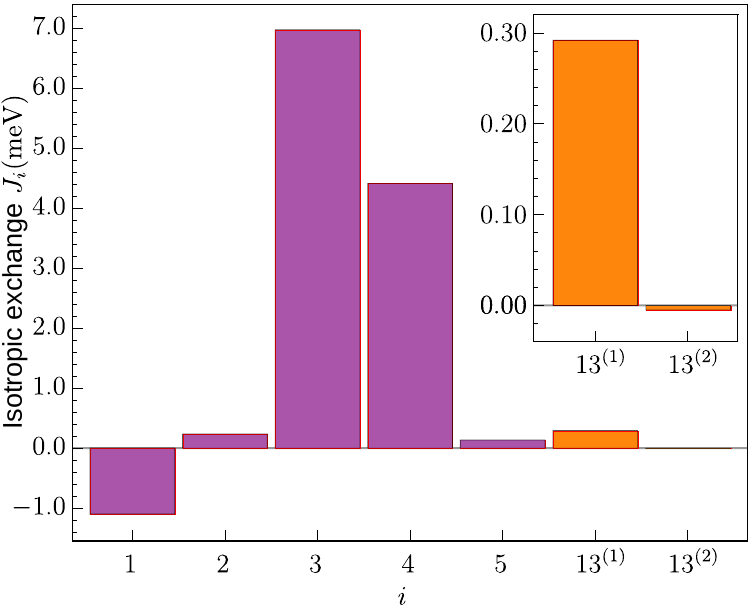}
    \caption{
    Isotropic magnetic exchanges $J_i$, with TEMA carried out for 1\textsuperscript{st} to 5\textsuperscript{th} nearest neighbor shells plus the 13\textsuperscript{th} neighbor shell. Comparing to the full fit (see Fig.~\ref{fig:TEMA_plot}) we see a big difference especially in the 1\textsuperscript{st} nearest neighbor coupling which flips the sign. The overall magnitude of the 13\textsuperscript{th} nearest neighbor shell is unaffected although the difference between $J_{13^{(1)}}$ and $J_{13^{(2)}}$ is more pronounced.
    }
    \label{fig:TEMA_plot_less}
\end{figure}

The results of TEMA are shown in Fig.~\ref{fig:TEMA_plot}. The largest couplings are the antiferromagnetic 3\textsuperscript{rd} and 4\textsuperscript{th} nearest neighbors. The 3\textsuperscript{rd} nearest neighbors couple A-B sublattices and C-D sublattices, while the 4\textsuperscript{th} nearest neighbors couple A-C sublattices and B-D sublattices. The spin pattern in the ground state is of the form $(\text{ABCD})=(\uparrow\downarrow\downarrow\uparrow)$, which is energetically stabilized by antiferromagnetic 3\textsuperscript{rd} and 4\textsuperscript{th} nearest neighbor exchanges, making the ground state very stable. Further neighbors add finer features on-top of the ground state, with altermagnetism appearing because of the 13\textsuperscript{th} nearest neighbor shell having two symmetry inequivalent exchanges $J_{13^{(1)}}=J_{13}+\Delta$, $J_{13^{(2)}}=J_{13}-\Delta$.

Additionally we try fitting TEMA to a reduced set of parameters, keeping only 1\textsuperscript{st} to 5\textsuperscript{th} nearest neighbor shells plus the 13\textsuperscript{th} neighbor shell. The results are shown in Fig.~\ref{fig:TEMA_plot_less}. While the largest couplings are relatively unaffected, we see a huge difference in the 1\textsuperscript{st} neighbor neighbor which now flips its sign to ferromagnetic, as opposed to antiferromagnetic in the full fit. Also the difference between $J_{13^{(1)}}$ and $J_{13^{(2)}}$ is more pronounced, with $J_{13^{(2)}}$ even flipping the sign.

\begin{figure}
    \centering
    \includegraphics[width=0.85\linewidth]{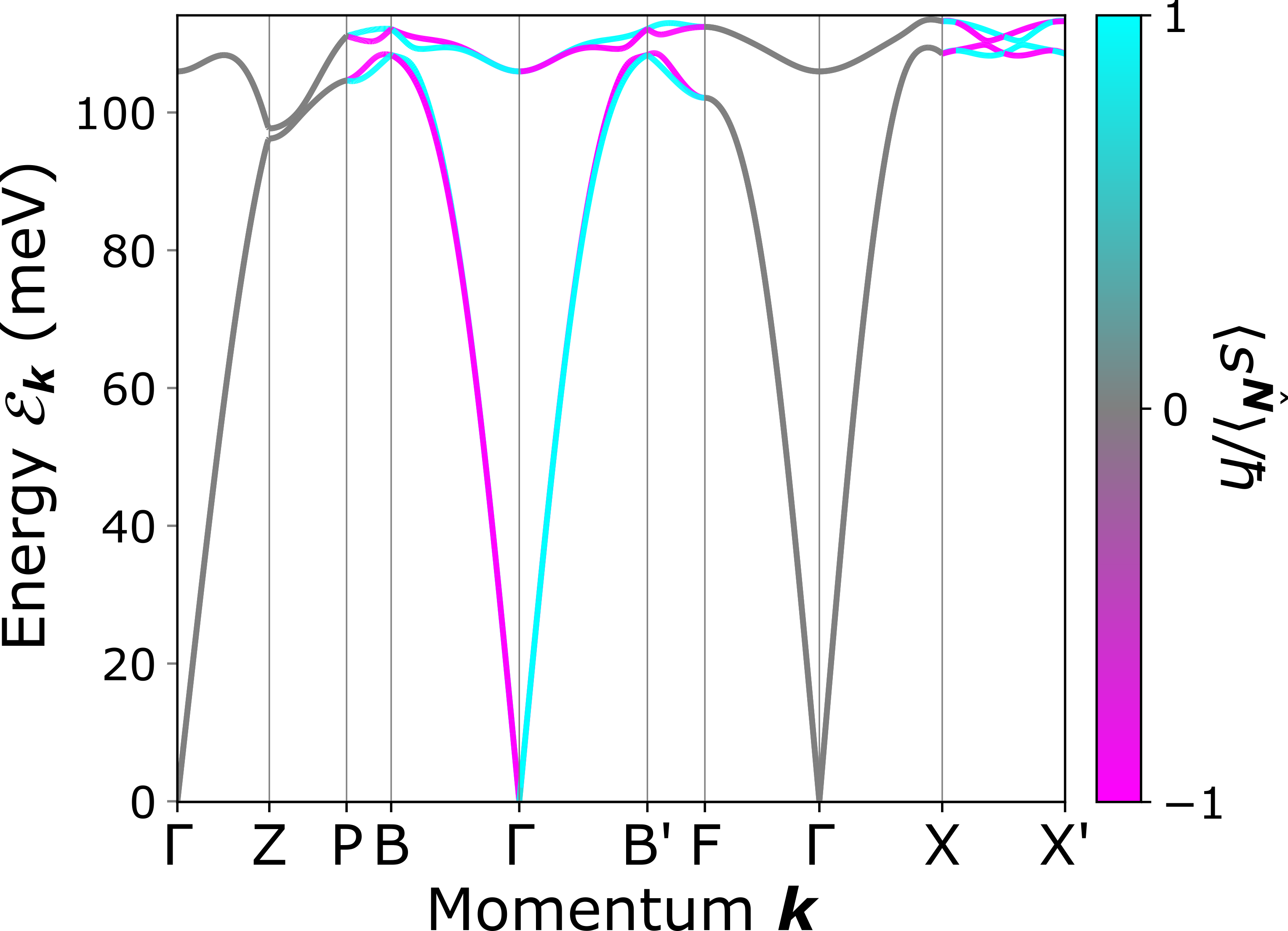}
    \caption{
    Nonrelativistic magnon dispersion relation of hematite with parameters as in Fig.~\ref{fig:TEMA_plot_less}.
    }
    \label{fig:TEMA_bands}
\end{figure}

Figure~\ref{fig:TEMA_bands} shows the nonrelativistic magnon dispersion relation of hematite with the parameters as in Fig.~\ref{fig:TEMA_plot_less}. Compared to Fig.~\ref{fig:disperion_nonrelativistic}(a), which is based on inelastic neutron scattering data, the bandwidth is slightly too large, but the shape of the dispersion relation is in overall good agreement with it.

\label{ap:DFT}

%%%%%%%%%%%%%%%%%%%%%%%%%%%%%%%%%%%%%%%%%%%%%%%%%%%%%%%%%%%%%%%%%%%%
%
% Order-by-quantum disorder
%
%%%%%%%%%%%%%%%%%%%%%%%%%%%%%%%%%%%%%%%%%%%%%%%%%%%%%%%%%%%%%%%%%%%%
\section{Order-by-quantum disorder}
\label{sec:order_by_disorder}
In the weak ferromagnetic phase in the presence of the easy-plane anisotropy, the DMI and the altermagnetic exchange [cf.~Sec.~\ref{sec:WFP_full}], we encounter the case of a pseudo-Goldstone mode. It arises because the classical ground state energy $E_0$ in Eq.~\eqref{eq:classicalenergy} has no absolute angle dependence, resulting in an accidental continuous symmetry. Here, we invoke order-by-quantum disorder arguments to show that quantum fluctuations induce an absolute angle dependence in the ground state energy.

In spin wave theory, the classical ground state energy $E_0$ in Eq.~\eqref{eq:classicalenergy} gets corrected to leading order in $1/S$ by quantum fluctuations 
\begin{align}
    \frac{\delta E_0}{S^2 N}  = \frac{1}{2 S^2 N} \sum_{\bm{k}}^N \sum_{n=1}^4 \left( \varepsilon_{n, \bm{k}} - A_{n, \bm{k}} \right).
\end{align}
We expect an anisotropy, i.e., an absolute angle dependence, from these quantum fluctuations because the energies $\varepsilon_{n, \bm{k}}$ depend on the relative orientation of the Néel and DMI vectors. The term $A_{n, \bm{k}}$ does not [cf.~Eqs.~\eqref{eq:Ak_WFP_1} and \eqref{eq:Ak_WFP_DMI}], and therefore does not contribute to the anisotropy. 

\begin{figure}[t]
    \centering       
    \includegraphics[width=5cm]{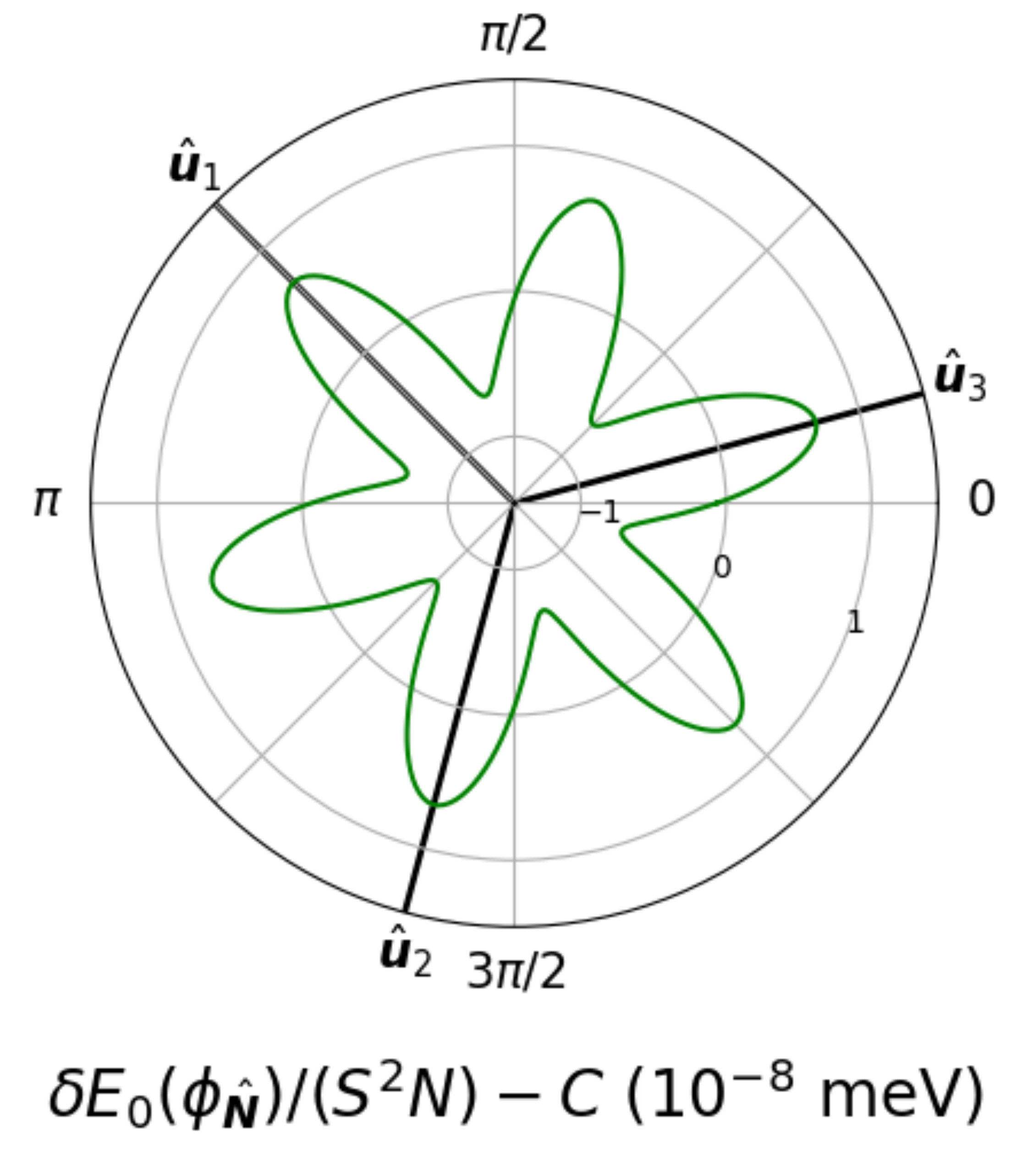}
       \caption{
       The Dzyaloshinskii-Moriya interaction-induced quantum fluctuations $\delta E_0$ in the weak ferromagnetic phase as a function of the azimuthal angle $\phi_{\hat{\bm{N}}}$ of the Néel vector show a triaxial anisotropy. The constant $C \sim -1.167\,$meV.}
       \label{fig:quantum_fluct}
\end{figure}

In Fig.~\ref{fig:quantum_fluct}, we show $\delta E_0/(S^2 N)$ as a function of the azimuthal angle $\phi_{\hat{\vec{N}}}$ of the Néel vector. The in-plane components of the lattice vectors $\hat{\bm{u}}_i$, with $i \in \{1, 2, 3\}$ are marked in black. We find a sixfold modulation of the energy, and therefore an effective triaxial anisotropy. If the Néel vector is aligned with the in-plane components of a lattice vector, i.e., $\bm{N} \parallel \pm \hat{\bm{u}}_i$, the quantum fluctuations are maximized. To minimize the energy, the magnetization $\bm{M}$, which lies perpendicular to the Néel vector, should lie in the direction of the ``petals''.

%Computing $\frac{\delta E_0}{S^2 N}$ as a function of the azimuthal angle $\phi_{\hat{\bm{N}}}$ of the Néel vector in the WFP indicates an effective triaxial anisotropy with a scale of $d=10$ neV in the system [cf. Fig~\ref{fig:quantum_fluct}].

%Such a triaxial anisotropy would cause a gap that comes with exchange enhancement \rh{read about that}. This would lead to a gap $\propto \sqrt{d E_{\text{ex}}}$, where $E_{\text{ex}} = \sum_i z_i J_i$ denotes the exchange field (or total exchange energy) seen by a spin. The parameter $z_i$ counts the number of neighbors in the $i$th shell. According to our band width, this exchange field adds up to $E_{\text{ex}} \sim 100$ meV. Using these numbers, we estimate the gap to be $\sim \sqrt{d E_{\text{ex}}} \sim 0.03$ meV. This corresponds to $7$ GHz, which fits to the gap of the lower (quasi-ferromagnetic) mode in the WFP seen in antiferromagnetic resonance experiments \cite{velikov1969antiferromagnetic}.
%In principle, one could write such a triaxial anisotropy into the Hamiltonian, which would gap out the pseudo-Goldstone mode.

%%%%%%%%%%%%%%%%%%%%%%%%%%%%%%%%%%%%%%%%%%%%%%%%%%%%%%%%%%%%%%%%%%%%
%
% BIBLIOGRAPHY
%
%%%%%%%%%%%%%%%%%%%%%%%%%%%%%%%%%%%%%%%%%%%%%%%%%%%%%%%%%%%%%%%%%%%%
\bibliography{hematite_bib}

\end{document}